\documentclass[fleqn,usenatbib]{mnras}

\usepackage{newtxtext,newtxmath}

\usepackage[T1]{fontenc}

\DeclareRobustCommand{\VAN}[3]{#2}
\let\VANthebibliography\thebibliography
\def\thebibliography{\DeclareRobustCommand{\VAN}[3]{##3}\VANthebibliography}

\usepackage{graphicx}	
\usepackage{amsmath}	
\usepackage{longtable}
\usepackage{threeparttablex}
\usepackage{footnote}
\usepackage{xcolor}
\usepackage[width=.98\textwidth]{caption}
\usepackage{subcaption}

\title[Supernova Remnants in NGC 7793] {Supernova Remnant properties and Luminosity Functions in NGC 7793 using MUSE IFS}

\author[Kopsacheili M. et al.]{
Maria Kopsacheili,$^{1,\, 2}$\thanks{E-mail: kopsacheili@ice.csic.es}
Cristina Jiménez-Palau,$^{1}$
Lluís Galbany,$^{1,\, 2}$
Panayotis Boumis,$^{3}$
Raúl González-Díaz.$^{1,\,4}$
\\
$^{1}$Institute of Space Sciences (ICE, CSIC), Campus UAB, Carrer de Can Magrans, s/n, E-08193 Barcelona, Spain
\\
$^{2}$Institut d’Estudis Espacials de Catalunya (IEEC), E-08034 Barcelona, Spain\\
$^{3}$Institute for Astronomy, Astrophysics, Space Applications
and Remote Sensing, National Observatory of Athens,
15236 Penteli, Greece\\
$^{4}$Instituto Nacional de Astrofísica, Óptica y Electrónica, Luis E. Erro 1, 72840 Tonantzintla, Puebla, México\\
}

\date{Accepted 22 March 2024. Received 4 August 2023; in original form ZZZ}

\pubyear{2015}

\begin{document}
\newcommand{\ha}{H$\rm{\alpha}$}
\newcommand{\hb}{H$\rm{\beta}$}
\newcommand{\oiii}{[\ion{O}{III}]}
\newcommand{\oi}{[\ion{O}{I}]}
\newcommand{\oii}{[\ion{O}{II}]}
\newcommand{\sii}{[\ion{S}{II}]}
\newcommand{\nii}{[\ion{N}{II}]}
\newcommand{\sig}{$\rm \sigma$}
\newcommand{\m}{$\rm \mu$}
\newcommand{\siiratio}{$\rm [\ion{S}{II}]\, 6716$\AA/$6737$\AA}
\newcommand{\lam}{$\rm \lambda$}

\label{firstpage}
\pagerange{\pageref{firstpage}--\pageref{lastpage}}
\maketitle

\begin{abstract}
In this study we use MUSE Integral Field Spectroscopy (IFS), along with multi-line diagnostics, for the optical identification of Supernova Remnants (SNRs) in the galaxy NGC 7793. We find in total 238 SNR candidates, 225 of them new identifications, increasing  significantly the number of known SNRs in this galaxy. The velocity dispersion of the candidate SNRs was calculated, giving a mean value of $\rm 27\, km\, s^{-1}$. We construct the \ha, \sii, \oiii, and \sii\ - \ha\ luminosity functions, and for the first time, the \nii, \nii\ - \ha, \nii\ - \sii, \oiii\ - \sii, and \oiii\ - \nii\ luminosity functions of the candidate SNRs. Shock models, along with the observed multi-line information were used, in order to estimate shock velocities. The $\rm \sim 65\%$ of the SNRs present velocities < 300 $\rm km\, s^{-1}$. There is a clear correlation between shock velocity and \oiii/\hb\ ratio, and a less clear but still evident correlation in the relation between shock velocity and the \sii/\ha, \nii/\ha\ ratios. We also use the \sii 6716/31 ratio of the SNR candidates to calculate their post-shock density, assuming different temperatures. The median value of the density of our sample is $\rm \sim 80\, cm^{-3}$, for a temperature of $\rm T = 10^4\, K$. No correlation between shock velocity and  density, or  density and SNRs with \sii/\ha\ > 0.4 and \sii/\ha\ < 0.4 is observed. 
\end{abstract}




\section{Introduction} \label{intro}
Supernova Remnants (SNRs) are very important ingredients of  galaxies. They enrich the Interstellar Medium (ISM) with heavy elements and they depose to it large amount of mechanical energy that heat and shape it. Their un-biased systematic study can give us information on the massive-star formation rate of a galaxy since they depict massive star's endpoint life.

In order to explore the feedback of SNRs to the ISM and to the host galaxy, detailed study of SNRs and SNR populations in different wavelengths is very important. In the last decades a lot of surveys have been published on Galactic SNRs and SNRs in Magellanic Clouds (MCs) in different wavelengths (e.g.: \citealt{2006ApJ...642..260S}; \citealt{2008A&A...485...63F}; \citealt{2013ApJ...772..134M}; \citealt{2019A&A...631A.127M}; \citealt{2020ApJ...898L..51W}; \citealt{2021ApJ...920...90F}; \citealt{2022MNRAS.512.1658B}; \citealt{2021MNRAS.507..971C},\citeyear{2022MNRAS.513L..83C}; \citealt{2022ApJ...932...26T}; \citealt{2022MNRAS.tmp.1558P}; \citealt{2022MNRAS.514..728A}). Their proximity allows a detailed investigation of their physical properties, morphological and kinematic characteristics, their interaction with their ambient medium, and their progenitor. On the other hand,  SNRs of more distant galaxies provide the opportunity of studying larger samples in various galactic environments, e.g. in galaxies with different morphology, metallicity, density, star formation rate (SFR). Moreover, such studies free of selection effects, in combination with population synthesis models can give an image of the mechanical energy deposited by the SNRs to the host galaxy (e.g. \citealt{2022MNRAS.514.3260K}). Many extragalactic studies (e.g. \citealt{2010ApJ...725..842L}; \citealt{2013MNRAS.429..189L}; \citealt{2011AAS...21725633P}; \citealt{2019MNRAS.488..803M}; \citealt{2020MNRAS.495..479R}; \citealt{2021MNRAS.500.2336Y}; \citealt{2021MNRAS.507.6020K}; \citealt{2021MNRAS.502.1386C}) have presented hundreds of SNRs, revealing properties, such as their size, their shock excitation, and correlations between different wavelengths.

The most traditional tool for the optical identification of SNRs is the  \sii/\ha\ ratio. In the first extragalactic studies they noticed that this ratio is relatively high  in SNRs compared to \ion{H}{II} regions (\citealt{Mathewson&Clarke}; \citeauthor{1978A&A....63...63D}\, \citeyear{1978A&A....63...63D}; \citeyear{1980A&AS...40...67D}). The threshold that finally was established for this ratio is 0.4 (\citealt{1978A&A....63...63D}), while a typical \sii/\ha\ ratio for \ion{H}{II} regions is $\sim 0.1 - 0.2$ . Since then, sources with \sii/\ha\ > 0.4 are considered as SNRs. This is an empirical diagnostic, based on extragalactic studies of SNRs, that seems to efficiently differentiate shock excited regions like SNRs, from photoionized like \ion{H}{II} regions. This criterion has been used a lot the last decades giving large samples of SNRs. Over the years, this threshold has been slightly modified, in order to take into account different densities and metallicities (e.g. \citealt{2013MNRAS.429..189L}), and more emission lines have been proposed as more efficient ones, for example oxygen lines (e.g. \citealt{Fesen1985}). More recently, \citet{2020MNRAS.491..889K} used shock models from \citet{2008ApJS..178...20A} and photoionization models from \citet{2001ApJ...556..121K} and \citet{2010AJ....139..712L}, along with a Support Vector Machine method, in order to develop diagnostics for the identification of SNRs. These diagnostics use the emission-line ratios \oi/\ha, \nii/\ha, \sii/\ha, \oii/\hb, and \oiii/\hb, combined in 2 and 3 dimensions. They seem to identify at least 50\% more SNRs than the traditional \sii/\ha > 0.4 alone. The SNRs missed by using the traditional diagnostic, usually are low-excitation SNRs, e.g. SNRs with low shock velocities.

The new, multi-line diagnostics have been recently used for the identification of extragalactic SNRs. \citet{2021MNRAS.502.1386C} identified 59 SNRs using one of the multi-dimensional diagnostics, instead of 20 SNRs by using the [\ion{S}{II}]/H$\rm \alpha$ > 0.4 criterion, in the galaxy NGC 4030. \citet{2021MNRAS.503.3856G} obtained spectra of the SNR MCSNR J0127-7332 in the Small Magellanic Cloud (SMC). This SNR gives   [\ion{S}{II}]/H$\rm \alpha$ ratios $< 0.24$, far from the traditional threshold. In addition, it satisfies all the multi-dimensional criteria of \citet{2020MNRAS.491..889K}. These diagnostics have been also used in Galactic SNRs as an extra test of their shock-excited nature (\citealt{2022MNRAS.512.1658B}; \citealt{2022MNRAS.tmp.1558P}).

In order to apply these multi-line diagnostics, Integral Field Spectroscopy (IFS) is the ideal approach. It allows the exploration of the spatial distribution of the sources/galaxy of interest (as it has been done with imaging), but providing also spectral information for each pixel, for a wide range of wavelengths. In addition, these data are relieved of some issues that we face when applying the traditional diagnostics with imaging. The most important ones are: i) bad subtraction of the continuum radiation, because in imaging a uniform factor is used for all the stars of the field, which is not correct because the contribution of each star is different; ii) linear pseudo-continuum is usually assumed because the (varying) underlying absorption of \ha\ cannot be distinguished from the overlapping emission; iii) usually the \ha\ filters are wide enough to include the nitrogen doublet emission lines \nii\ $\rm \lambda \lambda 6548,84 $. The problem is that we cannot know a priori the contribution of the \nii\ in the SNRs and hence to correct for this. Sometimes we use a fixed value which comes for the average of the \nii/\ha\ ratios of SNRs in galaxies with similar metallicities, however this is not very  accurate.


In this work, we use IFS in order to identify new optical SNRs using multi-dimensional diagnostics, more efficient in low-excitation SNRs than the commonly used   [\ion{S}{II}]/H$\rm \alpha > 0.4$ diagnostic (\citealt{2020MNRAS.491..889K}). For this purpose, the galaxy NGC 7793 has been selected. This is a flocculent, spiral, almost face-on galaxy, 3.7 Mpc away from us \citep{2011ApJS..195...18R},  and part of Sculptor galaxy group, with an extinction corrected \ha\ star formation rate (SFR) of $\rm 0.51\, M_{\odot}\, yr^{-1}$ \citep{2009ApJ...706..599L}. 

NGC 7793 has been studied before by \citet{Blair1997} who reported 27 optical SNRs, and more recently by \citet{2021MNRAS.507.6020K} who presented an updated list of SNRs, with 42 new identifications. In both surveys, the SNR sample has been selected with the \sii/\ha\ > 0.4  diagnostic and it presents a relatively high number of SNRs. The new, multi-line diagnostics are expected to give an even higher number. The relatively large SNR sample, in combination with the availability of IFS was the main motivation in the selection of this galaxy. In addition, {\textit{Chandra}} X-ray data is available, and hence a multi-wavelength study of SNRs can be performed (Kopsacheili et al., in prep.).

Past studies have also presented radio and X-ray SNRs in NGC7793. 
\citeauthor{2002ApJ...565..966P} (\citeyear{2002ApJ...565..966P}; \citeyear{2011AJ....142...20P}) reported 5 new radio SNRs (not yet identified in optical wavelengths) and 2 radio and X-ray counterparts to optical ones.  More recently, \citet{2014Ap&SS.353..603G} presented
a catalogue of 14 radio SNRs, 9 of them being new identifications.



The structure of the paper is the following: In \S \ref{data} we describe the data and the analysis performed, in order to make the final maps.  In \S \ref{det_phot}  the detection and the photometry procedure is presented and in \S \ref{diag} the diagnostics  used for the SNR identification. In \S \ref{results} the sample of the candidate SNRs and their properties are performed. The discussion of our results is presented in  \S \ref{discussion}, along with comparison with other studies, luminosity functions,  velocity and density distributions. We also construct the density maps,  exploring possible correlations between those and properties of the SNR candidates. Finally, our conclusions are presented in  \S \ref{conclusions}.


\section{Data description and Analysis} \label{data}

\subsection{Integral Field Spectroscopy}
\begin{figure}
\centering
 \includegraphics[width=0.5\textwidth]{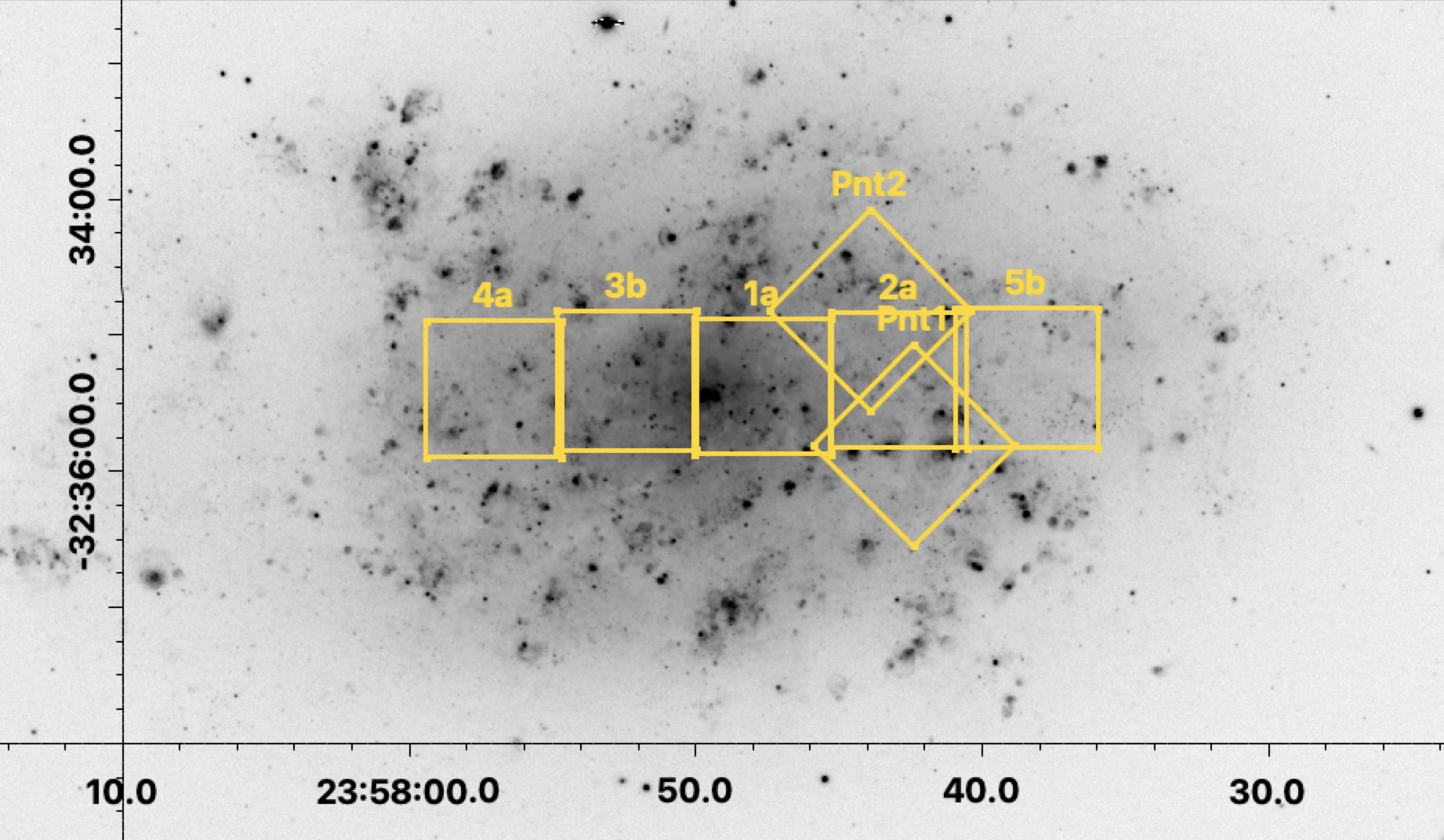}
\caption{\label{fig:ifu_fields} \ha\ + \nii\ image of NGC 7793 obtained by 4m Blanco telescope at CTIO. The yellow boxes show the regions covered by the MUSE instrument and their size is $\rm 60\times 60\, arcsec^2$.}
\end{figure}

For our study we have used Multi Unit Spectroscopic Explorer (MUSE) archival data. MUSE is an integral field spectrograph, installed in the Very Large Telescope (VLT). Its nominal wavelength range is 4800-9300 \AA. The spectral resolution is 1770 at 4650 \AA\ and 3590 at 9300 \AA. The field of view is $\rm 60\times60\, arcsec^2$ and the pixel scale is $\rm 0.2\,arcsec\,pixel^{-1}$.  MUSE observations of NGC 7793 available on the ESO Archive Science Portal, do not cover the full area of the galaxy but  mostly central regions (see \autoref{fig:ifu_fields}). The details of the observations are listed in \autoref{table:muse_info}.

\begin{table*}
	\centering
		\caption{The information of the MUSE observations, obtained from the ESO archival page} 
	\begin{tabular}{llcccclc} 
		\hline
		Object & Program Id  & Exp.  & PI & RA & DEC & Obs Date & Seeing (FWHM)\\
		     &  & (s)  &    & (J2000) & (J2000) &  & (arcsec)\\
		\hline
		NGC7793\_1a   & 097.B-0899 & 3500 & Ibar, Edo     & 23:57:47.5 & -32:35:19.7 & 2016-05-31 & 1.20\\
		NGC7793\_3b   & 097.B-0899 & 5250 & Ibar, Edo     & 23:57:52.0 & -32:35:20.5 & 2016-07-01 & 0.81\\
		NGC7793\_Pnt2 & 60.A-9188  & 2100 & Adamo, Angela & 23:57:43.8 & -32:34:50.8 & 2017-08-15 & 0.62\\
		NGC7793\_Pnt1 & 60.A-9188  & 2100 & Adamo, Angela & 23:57:42.3 & -32:35:48.9 & 2017-08-15 & 0.56\\
  		NGC7793\_2a   & 097.B-0899 & 1750 & Ibar, Edo     & 23:57:42.8 & -32:35:22.8 & 2016-07-01 & 0.74\\
		NGC7793\_4a   & 097.B-0899 & 1750 & Ibar, Edo     & 23:57:49.8 & -32:35:29.5 & 2016-07-01 & 0.87\\
            NGC7793\_5b   & 097.B-0899 & 3500 & Ibar, Edo     & 23:57:49.8 & -32:35:29.5 & 2016-07-01 & 0.59\\
		\hline
	\end{tabular}
	\label{table:muse_info}
	
\end{table*}

 In order to obtain the final maps of the emission lines we are interested in from  archival data cubes, we have followed some specific steps. First,  the data were corrected for Milky Way extinction using the formula of \citet{1999PASP..111...63F}, flux reddening, and redshift, and  a new cube was produced. The spectrum of each pixel consists of stellar and nebular emission. For this study, maps of the latter should be created. This was achieved by fitting to each emission line of interest (i.e \ha, \sii\ 6717, \sii\ 6731, \nii\ 6583, \hb, and  \oiii\ 5007)  a Gaussian using the mpfit code\footnote[1]{https://github.com/segasai/astrolibpy/tree/master/mpfit} and integrating over it. The integration was done above the stellar continuum level. This way, we get rid of the stellar continuum radiation and the emission-line maps contain only the nebular emission. Following this procedure, we do not account for the stellar absorption to which the \ha\ and \hb\ lines are superimposed. For this reason, later on, in the photometry, we considered an extra uncertainty in the \ha\ and \hb\ fluxes, which comes from the comparison between the spectra we worked on, and those for which we were able to estimate the absorption lines (even with high uncertainties), using the spectral synthesis code \texttt{STARLIGHT} \citep{2005MNRAS.358..363C}. The fluxes that were used for this comparison come from the integration of the fitted Gaussian to the spectra of the initial cube (i.e after the correction of the Milky Way extinction, flux reddening and redshift), and to those of the cube created after running the \texttt{STARLIGHT} code.


The emission lines that were used in order to apply the multi-line diagnostics are: \hb\ \lam 4861, \ha\ \lam 6563,  \oiii\ \lam 5007, \nii\ \lam 6584, \sii\ \lam 6717, and \sii\ \lam 6731.

\section{Detection and Photometry} \label{det_phot}

For the source detection we used the SExtractor\footnote{https://www.astromatic.net/software/sextractor} (\citealt{sextractor}) on the stacked \ha\ - \sii\ - \oiii\ 2D maps of each field, in order to increase the signal to noise (S/N) of the sources and take into account sources that appear in every of these wavelengths. The input parameters for the SExtractor were selected such as all the sources of the field, but not random noise, to be detected. For this reason we tried various combinations of these parameters and the best is the one that includes the following parameters: a) detection threshold at 1.5\sig\ above the background, b) the minimum
number of source pixels equal to 5, and c) the background mesh size
at 6 pixels in order to account for small-scale background variations. Those variations would not be detected if this parameter were too large, while if it were too small, the output list of sources would be contaminated by random noise. The fact that NGC 7793 is a nearby galaxy (3.7 Mpc) in combination with the high spatial resolution of the MUSE instrument, allows the view of some SNRs as extended sources that present some structure, rather than only point-like sources. For this reason, we checked the detected sources and we adjusted the aperture for each of them properly. Then, the sources were eye-inspected, making sure that they are not just random background enhancements or parts of large filaments.

The next step was to apply astrometric corrections on the maps. We used the {\texttt{ccmap}} package  of \texttt{IRAF} \citep{1993ASPC...52..173T} with a second order polynomial along with the 2MASS catalog.
For the photometry we used the {\texttt{phot}} package  of \texttt{PyRAF} (a command language for \texttt{IRAF} based on the Python scripting language). For each source an annulus of 2 pixels width and 2 pixels away from the aperture used for the photometry was considered as background, and it was subtracted from the flux of the source. We then kept the sources for which the \ha\ and \hb\ fluxes were higher than 3\sig\ and we corrected for the extinction based on the Balmer decrement. Finally, we kept those for which also the fluxes of the \sii, \nii, and \oiii\ were higher than 3\sig, as well as, the ratios \sii/\ha, \nii/\ha, and \oiii/\hb. 

\section{Diagnostics for the identification of Supernova Remnants} \label{diag}
Aiming to identify optical SNRs in NGC 7793, we use the multi-line diagnostics by \citet{2020MNRAS.491..889K}, that combine the following emission line ratios: a) \sii \lam\lam 6717,31 / \ha\ - \nii \lam 6584 / \ha, b) \sii \lam\lam 6717,31 / \ha\ - \oiii \lam 5007 / \hb, c) \nii \lam 6584 / \ha\ - \oiii \lam 5007 / \hb, and d) \nii \lam 6584 / \ha\ - \sii \lam\lam 6717,31/ \ha\ - \oiii \lam 5007 / \hb. In order to quantitatively compare the efficiency of the multi-line diagnostics with the traditional one, we also used the \sii/\ha\ > 0.4 criterion. We do not use the diagnostic of \citet{2020MNRAS.491..889K} that contain the \oii \lam\lam 3727,29 and \oi \lam 6300 lines, because for the former case MUSE does not cover the region of those blue wavelengths and in the latter, the \oi\, line, although it is one of the very good indicators for the separation of SNRs from \ion{H}{II} regions (e.g. \citealt{Fesen1985}; \citealt{2020MNRAS.491..889K}), it is very weak so we have very few cases where we detect it with a S/N ratio higher than 3.

In the case of multi-line diagnostics, the separating lines (2D diagnostics) or surface (3D diagnostic) are characterized by a single function $f$ \citep{2020MNRAS.491..889K}:

\noindent{For the \sii/\ha\ - \nii/\ha\ diagnostic:}
\begin{equation} \label{eq:1}
\begin{aligned}
   f(x, y) = & -2.452xy + 0.029xy^2 + 2.244x + 0.175x^2y - 0.257x^2 \\
             & - 0.043x^3 + 1.116y + 1.388y^2 - 0.217y^3 + 2.763,
\end{aligned}
\end{equation}
\noindent{where x = $\rm log_{10}($\sii/\ha$)$ and y = $\rm log_{10}($\nii/\ha$)$}.\\

\noindent{For the \nii/\ha\ - \oiii/\hb\ diagnostic:} \label{eq:2}
\begin{equation}
\begin{aligned}
f(x, y) = 0.939x + 1.0y + 0.469,
\end{aligned}
\end{equation}
\noindent{where x = $\rm log_{10}($\nii/\ha$)$ and y = $\rm log_{10}($\oiii/\hb$)$}.\\

\noindent{For the \sii/\ha\ - \oiii/\hb\ diagnostic:} \label{eq:3}
 \begin{equation}
\begin{aligned}
  f(x, y) = & -0.781xy - 0.318xy^2 + 4.66x + 0.148x^2y - 1.821x^2\\
               & + 0.079x^3 + 4.433y - 0.479y^2 + 0.255y^3 + 3.403,
\end{aligned}
\end{equation}
\noindent{where x = $\rm log_{10}($\sii/\ha$)$ and y = $\rm log_{10}($\oiii/\hb$)$}.\\

\noindent{For the \nii/\ha\ - \sii/\ha\ - \oiii/\hb\ diagnostic:} \label{eq:4}
 \begin{equation}
\begin{aligned}
 f(x, y, z) = & -2.913x^3 + -0.638x^2z + 1.568x^2y + 0.407x^2 \\
                 & - 0.866xz^2 + -2.264xyz + 1.508xz  + 1.753xy^2 \\
                 & - 6.913xy    + 0.001x + 1.463z^3 - 1.325yz^2 \\
                 & - 2.732z^2 + 1.823y^2z + -2.697yz + 4.377z  \\
                 & - 1.585y^3   + 0.770y^2+  1.267y + 2.413,
\end{aligned}
\end{equation}
\noindent{where $x$ = $\rm log_{10}($\nii/\ha$)$,  $y$ = $\rm log_{10}($\sii/\ha$)$, and $z$ = $\rm log_{10}($\oiii/\hb$)$}.\\


\noindent{In every case, in order for a detected source to be a SNR, the $f(x,y)$ or $f(x, y, z)$ should be positive. 
All the diagnostics (included those that have not been used in this study) can be accessed and directly utilized via \href{https://github.com/mariakop21/Diagnostics-for-SNR-identification.git}{this GitHub repository}.



\begin{figure*}
\minipage{1\textwidth}
  \includegraphics[width=0.95\linewidth]{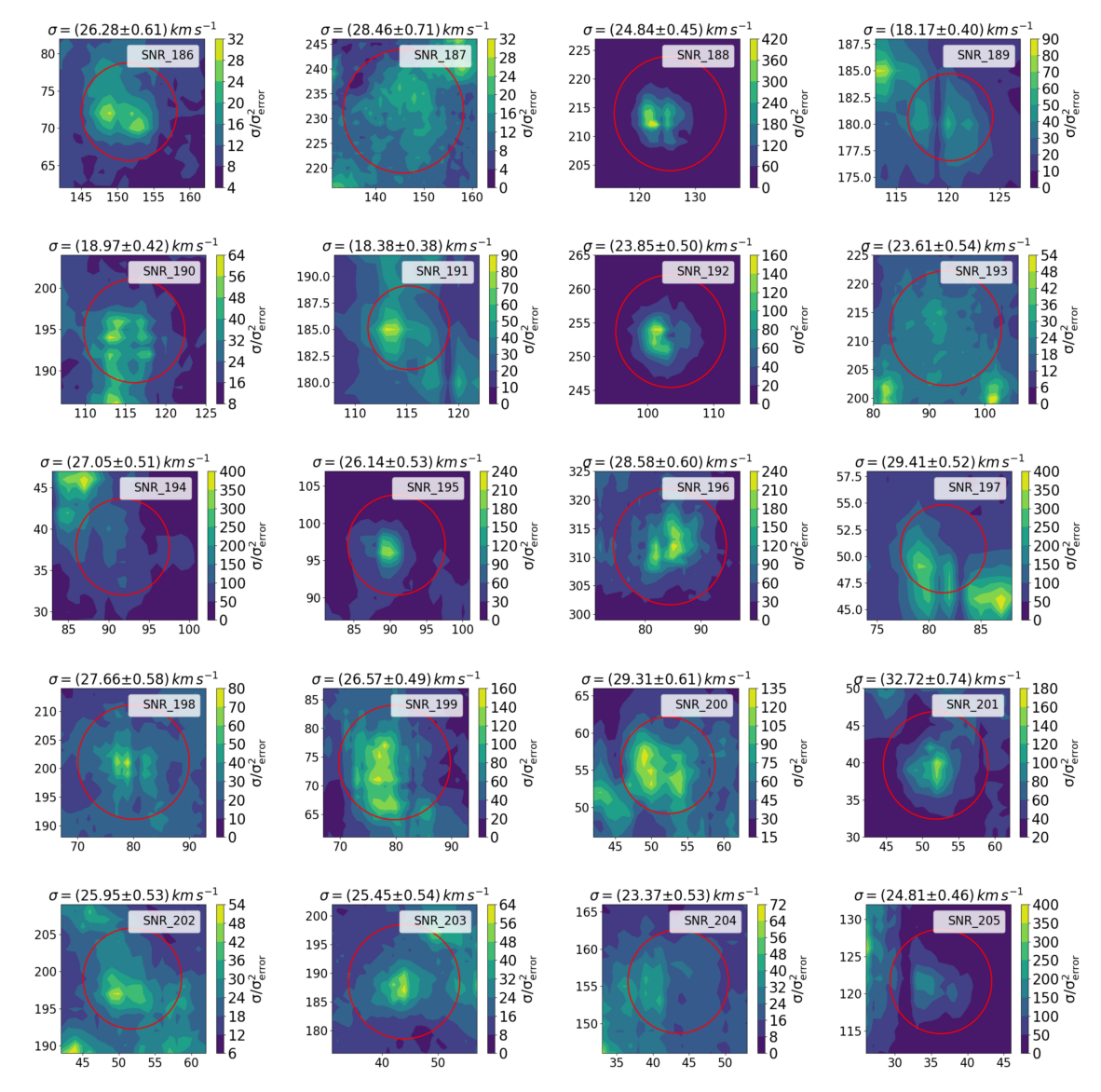}
  
  \caption{The $\rm \sigma/\sigma_{error}^2$ map of the Gaussian fitted to the \ha\ line of each pixel. The maps have been constructed in the regions of the SNR candidates. The red circles indicate the apertures used for the photometry for each SNR candidate. Above each map the velocity dispersion ($\rm \sigma$) of the integrated \ha\ flux of the candidate SNRs is presented. The maps of the whole sample of the SNR candidates can be found on the online version.} \label{fig:VDs}
\endminipage
\end{figure*}

\begin{figure*}
\minipage{0.98\textwidth}
\centering
 \includegraphics[width=\textwidth]{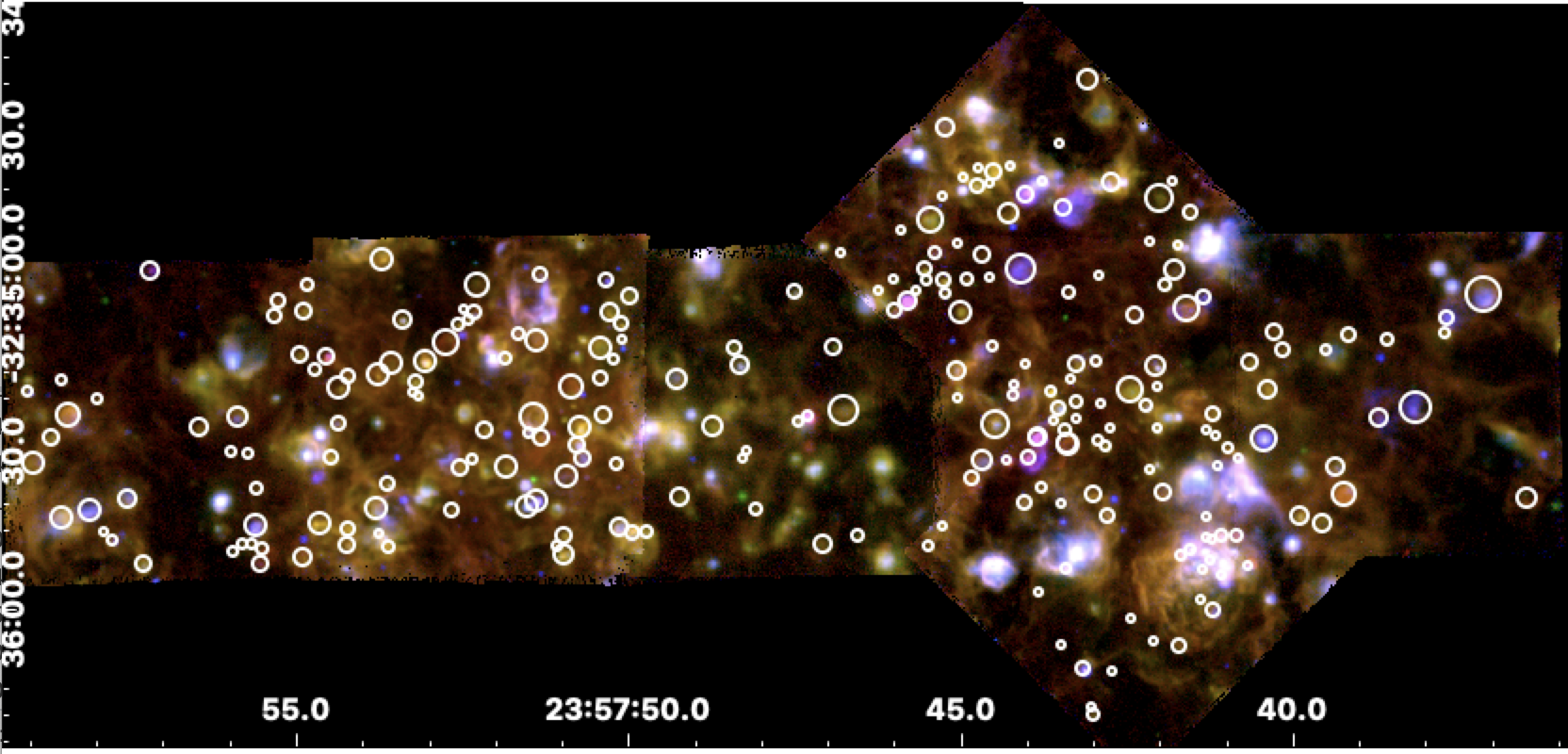}
\caption{\label{fig:RGB_sources} The optical RGB image of the MUSE IFS fields of NGC 7793 consisting of: \sii\ (red color); \ha\  (green color); \oiii\ (blue color). The white circles indicate the position of the identified SNRs and their size, the aperture used for the photometry.}
\endminipage
\end{figure*}


\section{Results} \label{results}
\subsection{Optical Supernova Remnants - Dispersion velocities}
In total, we identified 238 optical SNR candidates using the new, multi-line diagnostics, down to a \ha\ luminosity of $\rm \sim 9 \times 10 ^ {-17}\, erg\, cm^{-2}\, s^{-1}$. In \autoref{table:opt_SNRs} in the Appendix we present these SNR candidates  along with their properties. In the cases where we have two or more cubes overlapping (i.e see \autoref{fig:ifu_fields} Pnt1, Pnt2 with 2a), we keep the photometry of the cube that gives the higher S/N for each source in \ha. It is important to mention that this sample is very probable to contain "false positives", i.e. nebulae, such as \ion{H}{II} regions, that have been incorrectly identified as SNRs. For the moment the quantification of the contamination by \ion{H}{II} regions is limited to the percentages presented in \citet{2020MNRAS.491..889K}  which are: 1.0\%, 1.3\%, 2.7\%,1.5\%, 1.9\% for the diagnostics that include the lines \sii/\ha-\nii/\ha, \sii/\ha-\oiii/\hb, \nii/\ha-\oiii/\hb, \nii/\ha-\sii/\ha-\oiii/\hb\ and \sii/\ha\ respectively. These percentages correspond to the contamination as was calculated by the python module scikit-learn for the Support Vector Machine method, used for the development of the diagnostics. It is defined as the number of
false positives (i.e photoionization models that incorrectly were classified as shock models) over the sum of true and false positives (i.e. all the models that were classified as shock models). More realistic values would be expected if we had a wider ranges of parameters of the shock and photoionization models used for the development of the diagnostics, for example shock velocities that extend to values lower than 100 $\rm km\,s^{-1}$. In addition, a limitation in this kind of models is that they do not take into account SNRs evolving in more complex environments, for example SNRs embedded or interacting with \ion{H}{II} regions or molecular clouds. 

A more conservative sample can be possibly presented by calculating the velocity dispersion (VD) of the candidate SNRs. \citet{2019ApJ...887...66P} noticed a clear separation in the VD between \ion{H}{II} regions and SNRs, where most of SNRs present velocities ($\rm \sigma$ of the fitted to the line Gaussian) around 50 $\rm km\,s^{-1}$ (from $\sim$ 19 $\rm km\,s^{-1}$ to $\sim$ 110 $\rm km\,s^{-1}$), while \ion{H}{II} regions lower than 17.5 $\rm km\,s^{-1}$ (these values are corrected for the instrumental broadening). More recently, \citet{2023MNRAS.524.3623V} introduced the parameter $\rm \xi = \sigma\frac{[S\,II]}{H\alpha}$, where $\rm \sigma$ is the VD (i.e. the sigma value of the Gaussian), as a more effective criterion than using the traditional \sii/\ha\ > 0.4 alone. In this work we calculate the VD (i.e. the $\rm \sigma$ value) of the \ha\ line by fitting a Gaussian distribution to it, and taking into account the uncertainties of the \ha\ flux. We correct for the instrumental broadening by using the formula $\rm \sigma _{cor} = \sqrt{\sigma _{obs} ^2 - \sigma _{inst} ^2}$, where $\rm \sigma _{inst} = 49.3\, km\, s^{-1}$ is the instrumental $\rm \sigma$ VD for \ha\ \citep{2017A&A...608A...5G}. The final $\rm \sigma$ VD of the candidate SNRs are presented in the last column of \autoref{table:opt_SNRs} and they are between $\sim$ 15 $\rm km\,s^{-1}$ and $\sim$  70 $\rm km\,s^{-1}$ (with a mean value of $\sim$  27 $\rm km\,s^{-1}$). There are 9 candidate SNRs  for which the VD is lower that the VD range suggested in \citet{2019ApJ...887...66P} (most of them between $\sim$ 18 $\rm km\,s^{-1}$ and $\sim$  19 $\rm km\,s^{-1}$). However, we should note that in our case these velocities represent an average VD since the emission comes from the entire source (in most of the cases a compact object) and not from individual filaments or part of the shells of the SNRs, and for this reason we do not exclude them. In order to compare the VD of the candidate SNRs with those of the surrounding ionized gas, the $\rm \sigma/\sigma_{error}^2$ map of each source and its surrounding ISM is constructed. In \autoref{fig:VDs} we indicatively present some of those, while the rest of the maps can be found on the online version (\autoref{fig:VDs_all}).  The final sample of the candidate SNRs along with their properties is presented in \autoref{table:opt_SNRs}.

In \autoref{fig:RGB_sources}, the optical RGB image of the MUSE IFS fields of NGC 7793 is presented. The red, green, and blue colors correspond to the \sii, \ha,  and \oiii\ bands, respectively. The white circles indicate the position of the candidate SNRs and their size, the aperture that we used for the photometry. In \autoref{fig:spectra} of the online version the spectra of the SNR candidates are shown.

The number of the candidate SNRs is high compared to that presented in previous studies (e.g. \citealt{2021MNRAS.507.6020K}). However, it is very similar to the number of  known Galactic SNRs (\citeauthor{2019JApA...40...36G} \citeyear{2019JApA...40...36G}; \citeyear{Green2022}).

\section{Discussion} \label{discussion}
\subsection{Multi-line diagnostics}

\begin{figure*}
\minipage{1\textwidth}
  \includegraphics[width=\linewidth]{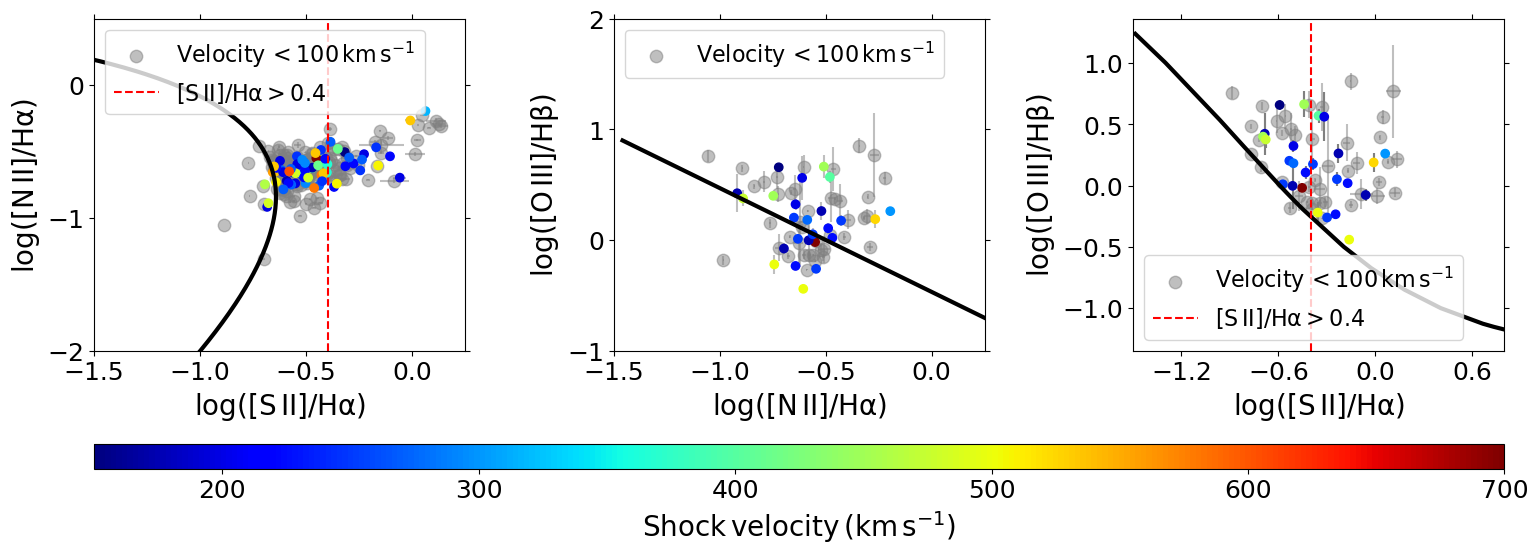}

  \caption{The SNRs identified in this work on the i) log(\sii/\ha) - log(\nii/\ha) plane (left panel), ii) log(\nii/\ha) - log(\oiii/\hb) plane (middle panel), and iii) log(\sii/\ha) - log(\oiii/\hb) plane (right panel). The black lines show the diagnostics that we used for the SNR identification  \citep{2020MNRAS.491..889K}. The red-dashed line is the \sii/\ha\ = 0.4 line and it interprets the traditional diagnostic. The colorbar indicates the shock velocity of the candidate SNRs, that was found interpolating in the \sii/\ha\ - \nii/\ha\ - \oiii/\hb\ - Shock velocity space of the shock models of \citeauthor{2008ApJS..178...20A}  (\citeyear{2008ApJS..178...20A}; \autoref{fig:4dmod}). The grey points are the SNRs for which the shock velocity is lower than 100 $\rm km\, s^{-1}$, since they fall below the interpolation limit of the shock models.}\label{fig:diagnostics}
\endminipage
\end{figure*}

In this work we have used 5 diagnostics in order to identify optical SNRs. These are the traditional \sii/\ha\ > 0.4 criterion, and those from \citet{2020MNRAS.491..889K} that combine the emission-line ratios \sii/\ha, \nii/\ha, \oiii/\hb\ in 2 and 3 dimensions (equations \ref{eq:1} - \ref{eq:4}). We considered as SNRs all the sources that have been identified using each diagnostic individually. As can be seen in \autoref{table:opt_SNRs}, the \sii/\ha\ - \nii/\ha\ diagnostic recovers most of the SNRs of our total sample ($\sim 96\%$), and about 72\% more than the \sii/\ha\ criterion. On the other hand, there is no SNR that has been detected only with the \sii/\ha\ criterion. With the \sii/\ha\ - \oiii/\hb\ criterion we recover $\sim 27\%$ of the sample, with the \nii/\ha\ - \oiii/\hb\ the $\sim 18\%$, and with the \nii/\ha\ - \sii/\ha\ - \oiii/\hb\ diagnostic the $\sim 28\%$. Diagnostics that contain the \oiii\ and \hb\ lines are those that detect the smallest number of SNRs. This happens because only few SNRs present strong \oiii\ and \hb\ emission and with a S/N higher than 3. Those are usually SNRs with high shock velocities (e.g. \citealt{1979ApJS...39....1R}).

In \autoref{fig:diagnostics} we show the 2D diagnostics (black lines) along with the identified SNRs. In the left panel we see the \sii/\ha\ - \nii/\ha\ diagnostic, in the middle panel the \nii/\ha\ - \oiii/\hb\ diagnostic, and in the right panel the \sii/\ha\ - \oiii/\hb\ diagnostic. We remind that in order for a source to be considered as a SNR, it must be above the black line, in the cases of the \nii/\ha\ - \oiii/\hb\ and \sii/\ha\ - \oiii/\hb\ diagnostics or on the right of the black line in the \sii/\ha\ - \nii/\ha\ diagnostic (or in other words, $f(x_0, y_0) > 0$, where $f$ is the function that describes each black line and $ x_0$, $
y_0$ is the logarithm of the emission line ratio of the x and y axes respectively). In each plot we include the SNRs which present emission in all the lines used for the corresponding diagnostic. This means that in reality, there are more candidate SNRs that fall out of the locus of SNRs (i.e. $ f(x_0, y_0) < 0$) but we cannot plot them  in all cases, since no emission has been detected. The red-dashed line is the \sii/\ha\ = 0.4 line (the traditional diagnostic), and the colorbar shows the shock velocity of the SNRs, calculated as described in \S \ref{velocities}.

As we see, although the \sii/\ha\ - \nii/\ha\ criterion identifies most of the SNRs, it misses a $\sim 4\%$ which is recovered by the  \sii/\ha\ - \oiii/\hb\ and \nii/\ha\ - \sii/\ha\ - \oiii/\hb\ diagnostics. These SNRs have relatively high velocities and hence this probably means that the temperature is still high enough to produce strong \sii\ lines. The \nii/\ha\ - \oiii/\hb\ misses $\sim 73\%$ of the SNRs. The main reason for that is the \oiii\ and \hb\ emission which is absent for many of our sources. The strict cutoff that we have set in the emission lines and in the emission line ratios (higher than 3\sig\ in every case) result to even smaller number of candidate SNRs. We should also mention here that this diagnostic presents relatively high incompleteness, compared to the others \citep{2020MNRAS.491..889K}. We see a similar behavior of this diagnostic in its application on spectra of the Galactic SNR G132.7+1.3 \citep{2022MNRAS.512.1658B}. The majority of SNRs that have not been identified using the \sii/\ha\ - \oiii/\hb\ criterion do not present emission in \oiii\ and/or \hb. Among the SNRs missed by this criterion, only 5 have been detected in \oiii\ and \hb\ but with a significance below 3\sig\ in \oiii\ and/or \hb\ and/or \oiii/\hb. We note here that the metallicity of the galaxy could be a factor that affects some emission line ratios, for example the \nii/\ha, and hence their position in the plots of \autoref{fig:diagnostics}. However, the diagnostics used for the identification have been constructed with shock models that take into account solar and twice solar (\citealt{1989GeCoA..53..197A}, \citealt{2005ASPC..336...25A}, \citealt{2005ApJ...619..755D}), Large Magellanic Cloud (LMC) and Small Magellanic Cloud \citep{1992ApJ...384..508R} metallicities. The metallicity of NGC 7793 is  between solar and LMC metallicity \citep{2014AJ....147..131P}, i.e. within the metallicity range considered in the diagnostics, and hence the location of the candidate SNRs above or below the black lines in \autoref{fig:diagnostics}, should not be affected by the metallicity of the galaxy.

\subsection{Comparison with other studies}

There are various studies that report SNRs in this galaxy in optical (\citealt{2021MNRAS.507.6020K}; \citealt{2020A&A...635A.134D}; \citealt{Blair1997}), X-ray (\citealt{2011AAS...21725633P}), and radio wavebands (\citealt{2014Ap&SS.353..603G}; \citealt{2002ApJ...565..966P}). In the optical band, our sources  NGC7793$\_$SNR$\_$206,  NGC7793$\_$SNR$\_$136,  NGC7793$\_$SNR$\_$89,  and NGC7793$\_$SNR$\_$126, have been reported before by \citet{2021MNRAS.507.6020K} as 13, 22, 12, and 23 respectively and by \citet{Blair1997} as S16, S11, S6, and S10 respectively. Our sources NGC7793$\_$SNR$\_$226,  NGC7793$\_$SNR$\_$98, NGC7793$\_$SNR$\_$90, and NGC7793$\_$SNR$\_$91 coincide with the sources 14, 10, 11, and 24 from \citet{2021MNRAS.507.6020K}, while the NGC7793$\_$SNR$\_$9, NGC7793$\_$SNR$\_$233, and NGC7793$\_$SNR$\_$19 with the s3, S22 and the photometric SNR S4 from \citet{Blair1997} respectively. The spectrum of S4 does not satisfy the \sii/\ha\ > 0.4 criterion so finally \citet{Blair1997} do not consider it as a SNR. In addition, the sources NGC7793$\_$SNR$\_$92  and NGC7793$\_$SNR$\_$237, coincide with the SNRs 23 and 30 of the less secure sample of \citet{2021MNRAS.507.6020K} for which \sii/\ha\ - 0.4 < 3\sig.
Our sources NGC7793$\_$SNR$\_$126, NGC7793$\_$SNR$\_$89, and NGC7793$\_$SNR$\_$90 have been also identified by \citet{2020A&A...635A.134D}. In that study, our source NGC7793$\_$SNR$\_$91 is considered as a planetary nebula (PN). This source presents a ring-like structure of a diameter of $\rm \sim 30\, pc$ and hence it cannot be a PN (typically 0.2 - 2 pc; e.g. \citealt{2006agna.book.....O}), despite its interestingly strong \oiii\ emission (which is the reason why \citealt{2020A&A...635A.134D} categorized it as a PN). In total, 13 out of the 238 SNRs identified in this work, have been identified as optical SNRs in previous studies. These 13 SNRs are a sub-sample of those presented in \citet{2021MNRAS.507.6020K} and  \citet{Blair1997}, since only those are in the fields covered by MUSE.

Our source NGC7793$\_$SNR$\_$136 is the optical counterpart of the X-ray and radio SNR  CXOU J235747.2-323523 in the work of \citeauthor{2011AAS...21725633P} (\citeyear{2011AAS...21725633P}; \citeyear{2002ApJ...565..966P}) and \citet{2014Ap&SS.353..603G}. The source NGC7793$\_$SNR$\_$233 coincides with the radio SNR J235758-323523 from \citet{2014Ap&SS.353..603G}.

In this work we identify 238 candidate SNRs, 225 of which are new identifications. The {{main reason}} for which we detect a significant higher number of SNRs compared to previous studies is the new multi-line diagnostics that recover SNR populations, ignored by the traditional \sii/\ha\ criterion, i.e. low-excitation SNRs  (up to 72\% more SNRs).

\subsection{Luminosity Functions}
\subsubsection{1-Dimensional Luminosity Functions}
Having a quite large sample of SNRs we construct 1D and 2D luminosity functions (LFs).
In \autoref{fig:1D_LF} we present the \ha, \nii, \sii, and \oiii\ LFs for our sample of SNRs. In all cases we fit  a Gaussian distribution (black lines). In order to determine the best-fit parameters we used a Markov Chain Monte Carlo (MCMC) method, performing 20000 iterations and excluding the first 500 ones. The best-fit mean ($\rm \mu$) and sigma ($\rm \sigma$) values are presented in \autoref{table:1d_LF_params}.
As can be seen, the mean value of the \oiii\ LF presents high uncertainties. This is probably result of the relatively low number of SNRs that present \oiii\ emission and hence that are used for the construction of the \oiii\ LF.

The \ha\ LF of extra-galactic SNRs has been calculated in various studies (e.g.  \citealt{1997ApJS..112...49M}; \citealt{2013MNRAS.429..189L};  \citealt{2014ApJ...786..130L}; \citeyear{2014ApJ...793..134L};  \citealt{2020MNRAS.491..889K}), while in less studies also the distribution of the \sii\ luminosities is constructed (e.g. \citealt{2014ApJ...786..130L}; \citeyear{2014ApJ...793..134L}; \citealt{2019MNRAS.488..803M}). However, in this work we present for the first time  also the \nii\ and \oiii\ LFs for SNRs. The comparison of LFs of SNRs evolving in different environments can reveal and/or explain populations with different characteristics. Possible trends and correlations can be more evident and accurate when we compare LFs of different wavelengths at the same time.


\begin{figure*}
\minipage{0.9\textwidth}
  \includegraphics[width=0.5\linewidth]{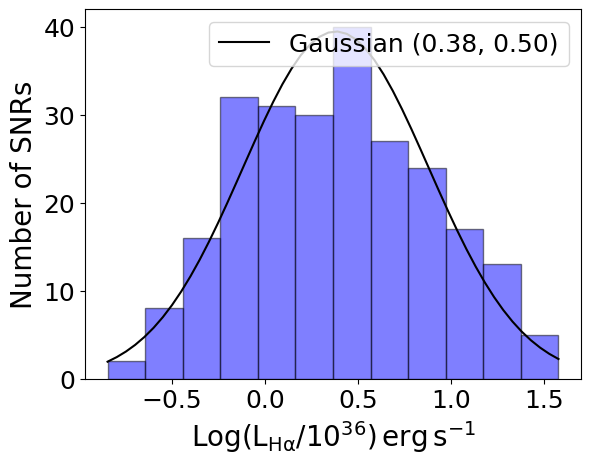}
  \hfil
  \includegraphics[width=0.5\linewidth]{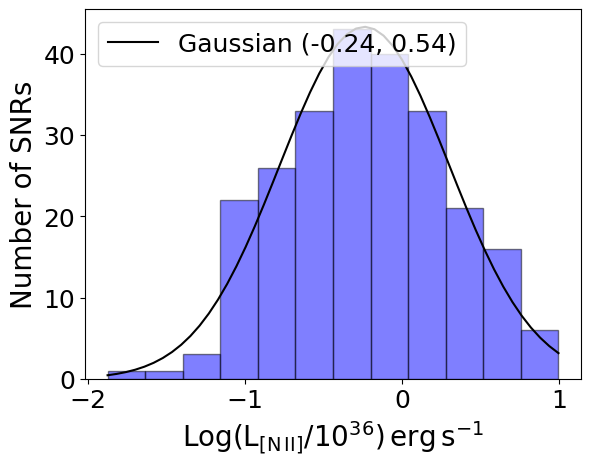}
  
  \includegraphics[width=0.5\linewidth]{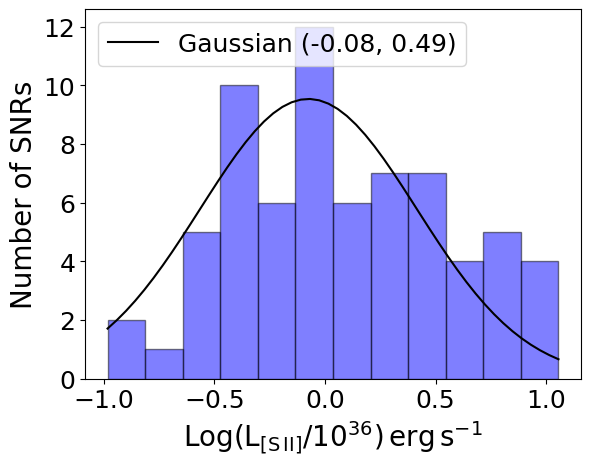}
  \hfil
  \includegraphics[width=0.5\linewidth]{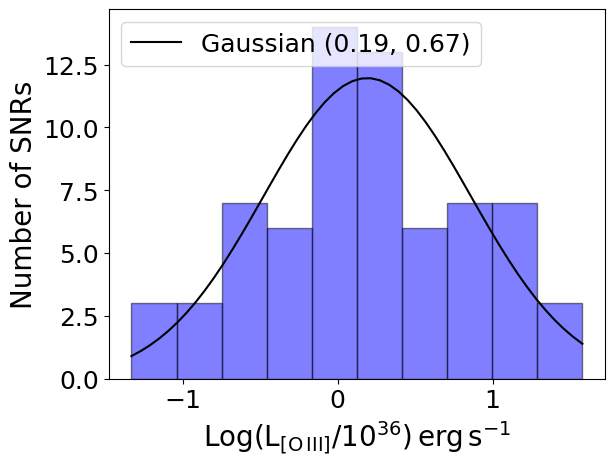}
  
  \caption{Luminosity functions: The blue histogram shows the distribution of the \ha\ (top left), \nii\ (top right), \sii\ (bottom left), and \oiii\ (bottom right) luminosities of the SNRs identified in this study using multi-line diagnostics. The black line is the best-fit Gaussian. The best-fit parameters  \m\ and \sig\ of the Gaussian are shown in the label of each image as: Gaussian (\m, \sig ) and in  \autoref{table:1d_LF_params}.} \label{fig:1D_LF}
\endminipage
\end{figure*}

\subsubsection{2-Dimensional Luminosity Functions}
We also present the joint LFs of: \sii\ - \ha, \nii\ - \ha, \nii\ - \sii, \oiii\ - \sii, and \oiii\ - \nii\ following the method of \citet{2021MNRAS.507.6020K} who presented the 2D \sii\ - \ha\, LF of SNRs in the same galaxy. In order to construct them we proceed as follows: First we find the best-fit line of the candidate SNRs on the plane i) \sii\ - \ha,  ii) \nii\ - \ha,   iii) \nii\ - \sii, iv) \oiii\ - \sii,  and v) \oiii\ - \nii, (black line in the top left panel of Figures \ref{fig:SIIHa_LF} - \ref{fig:OIIINII_LF}). Then, we calculate the distance of each point (blue circles) from the best-fit line (the distance along the vertical axes, as indicated by the dashed lines in the top left panel). This will give us the width of the 2D LF along the best-fit line. The distribution of these distances is presented with the blue histogram in the top-middle panel. To each distribution we fit a Gaussian, the $\rm \sigma$ of which will give us the width of the 2D LF. In order to determine the best-fit parameters of the Gaussian, we used MCMC method with 20000 iterations and excluding the first 500 ones. The best-fit $\rm \sigma$ of every distribution is presented in the 3rd column of \autoref{table:2d_LF_params}, indicated as "Width", and the best-fit Gaussian is depicted with the black line in the middle top panel. We then project each point to the best-fit line (along the dashed lines in the top left panel) and at this point all the points are located on the best-fit line (top right panel). The distribution of the points along the best-fit line, will give us the shape of the 2D LF. This distribution is presented with a green histogram in the bottom-left panel of Figures \ref{fig:SIIHa_LF} - \ref{fig:OIIINII_LF}. We fit to each one a Gaussian and we find the best-fit parameters using again MCMC method with 20000 iterations and 500 burns. Those are presented in \autoref{table:2d_LF_params}. The black line in the bottom-left panel of Figures \ref{fig:SIIHa_LF} - \ref{fig:OIIINII_LF} shows the best-fit Gaussian. Finally, we have the shape of the 2D LF and its width and we present it in 3D in the bottom right panel of Figures \ref{fig:SIIHa_LF} - \ref{fig:OIIINII_LF}.

The joint \sii-\ha\ LF has been presented in the past by \citet{2021MNRAS.507.6020K}. In this work the 2D \nii\ - \ha, \nii\ - \sii, \oiii\ - \sii, and \oiii\ - \nii\ LFs are presented too. In the joint LFs that contain the \oiii\ line we see a large scatter around the best-fit line (top left panel in Figures \ref{fig:OIIISII_LF} and \ref{fig:OIIINII_LF}). This is probably an indication that the \oiii\ emission comes from a different region of the shock than  the \sii\ and \nii\ emission where we see a more tight relation (a width of 0.14; \autoref{fig:NIISII_LF}), with probably different conditions, such as velocity (e.g. \citealt{2022MNRAS.512.1658B}).

1D and 2D multi-line LFs can be also compared with theoretical analyses that predict luminosity functions combining models for the evolutionary phase of SNRs and shock models. A first attempt to this direction has been done by \citet{2022MNRAS.514.3260K},  who assumed and age and a density distribution of SNRs and based on those they predicted their evolutionary phase using the relations of \citet{1988ApJ...334..252C}, which among others give the velocity of a SNR. Then, interpolating in the space defined by velocity - density - emission line in the shock models of \citet{2008ApJS..178...20A}, they find the \ha\ and the \sii\ luminosities and they construct theoretical LF which they compare with the observational ones from the work of \citet{2021MNRAS.507.6020K}. As it is mentioned in the work of \citet{2022MNRAS.514.3260K}, more dimensions, i.e. more emission-line ratios can predict more accurate theoretical LFs since the degeneracy of the shock models is significantly reduced. 

We note that unlike the work of \citet{2021MNRAS.507.6020K}, we do not apply incompleteness corrections in the LFs of our sample. In this study we identify  point-like  sources, but also sources that present some extended structure. \citet{2021MNRAS.507.6020K}, added on their data artificial, point-like sources, that can be considered as SNRs, since their sample consists of only point-like SNRs, and based on those they calculated the incompleteness. In order to estimated the incompleteness of sources for which we observe a structure, we would need to add on our data artificial objects that also present some structure. However,  the morphology of SNRs, can vary a lot, in size and in shape for example, that makes very difficult the construction of such artificial nebulae.

\begin{figure*}
\minipage{0.9\textwidth}
  \includegraphics[width=1\linewidth]{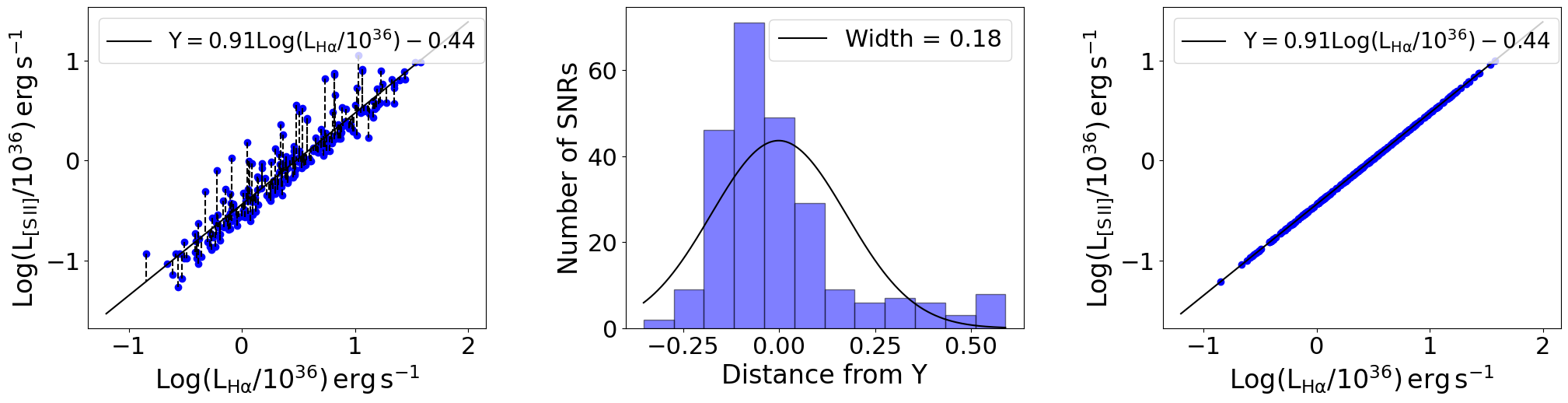}
  \includegraphics[width=0.5\linewidth]{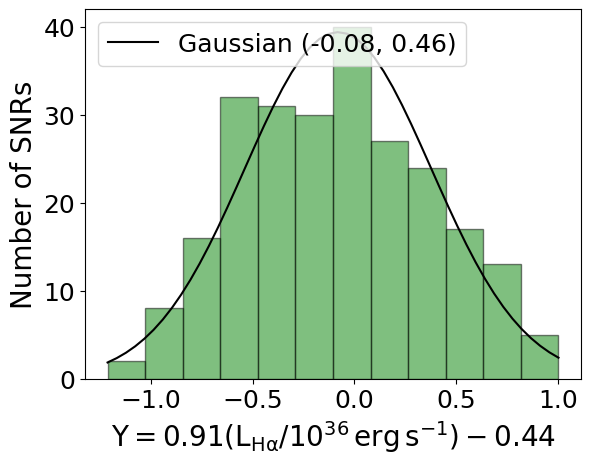}
\hfil
  \includegraphics[width=0.5\linewidth]{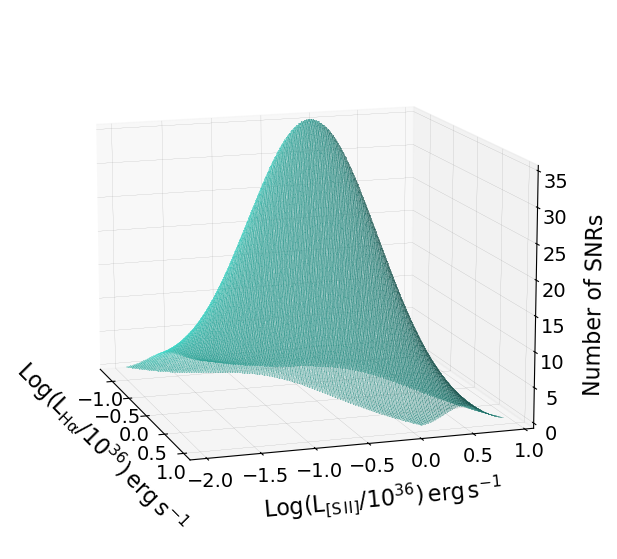}
  \caption{The joint \sii\ - \ha\ luminosity function. {\textit{Top left}}: the identified SNRs (blue circles) and the best-fit line (black line; it is given by the function of the label) on the \sii\ - \ha\ plane. The dashed lines show the distance of each circle to the best-fit line along the \sii\ axis. {\textit{Top middle}}: The distribution of the distances of each circle to the best-fit line along the \sii\ axis. The black line is the best-fit Gaussian, the \sig\ of which will give the width of the 2D LF, and it is presented in the label. {\textit{Top right}}: The "projection" of each circle of the top left panel on the best-fit line. It is the "projection" along the \sii\ axis (as indicated by the dashed lines in the top left panel) and not the vertical projection. {\textit{Bottom left}}: The distribution of the sources (blue circle in the top left and top right panels) along the best-fit line. We have fitted a Gaussian to this distribution, the \m\ and \sig\ of which appear in the label as: Gaussian (\m, \sig ). This Gaussian will give the shape of the 2D LF. {\textit{Bottom right}}: The 2D \sii\ - \ha\ luminosity function. The shape of the 2D LF is given by the Gaussian of the bottom left panel, and the width from the Gaussian of the top middle panel. All the best-fit parameters are shown in \autoref{table:2d_LF_params}.}\label{fig:SIIHa_LF}
\endminipage
\end{figure*}

\begin{figure*}
\minipage{0.9\textwidth}
  \includegraphics[width=1\linewidth]{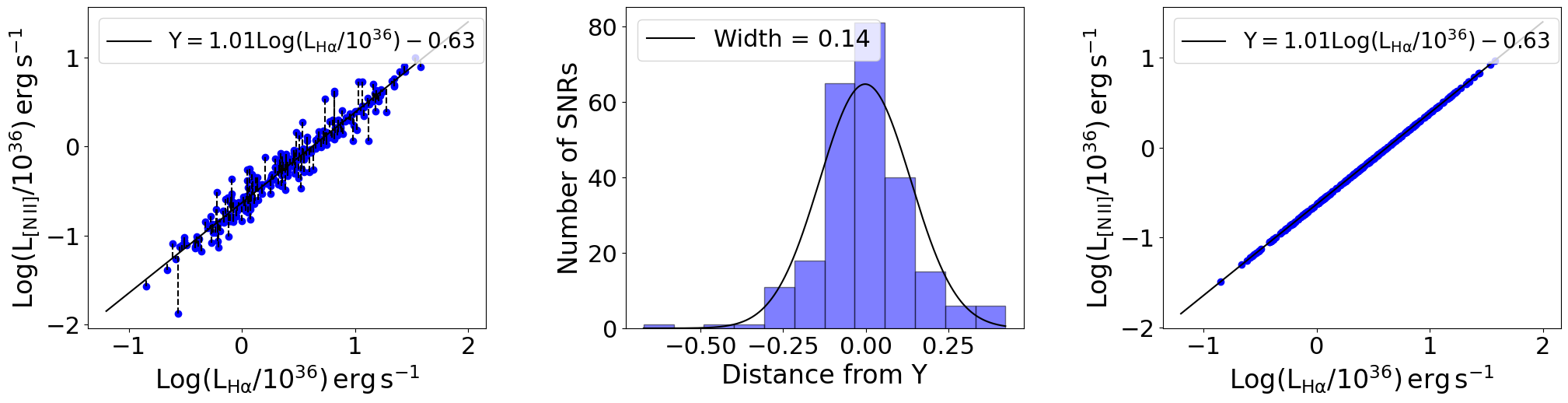}
  \includegraphics[width=0.5\linewidth]{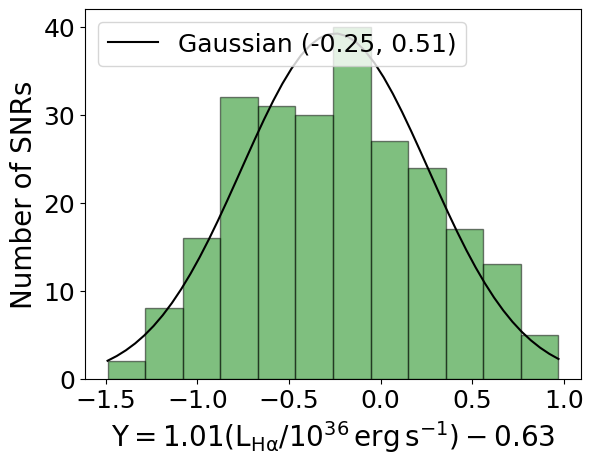}
\hfil
  \includegraphics[width=0.5\linewidth]{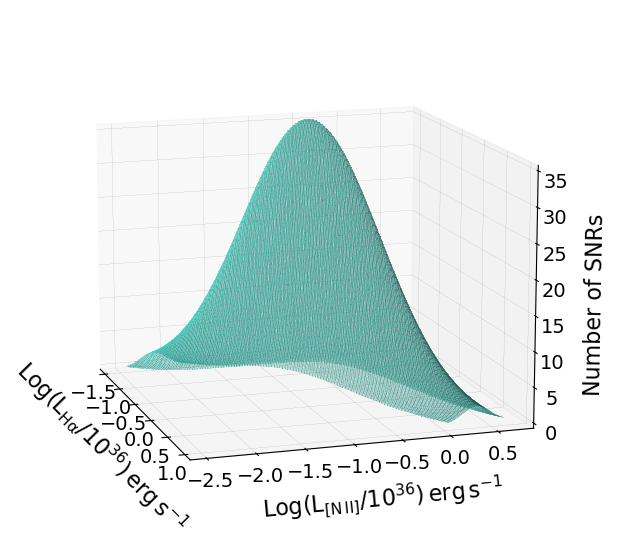}
  \caption{Same as \autoref{fig:SIIHa_LF} but for the joint \nii\ - \ha\ luminosity function.}\label{fig:NIIHa_LF}
\endminipage
\end{figure*}

\begin{figure*}
\minipage{0.9\textwidth}
  \includegraphics[width=1\linewidth]{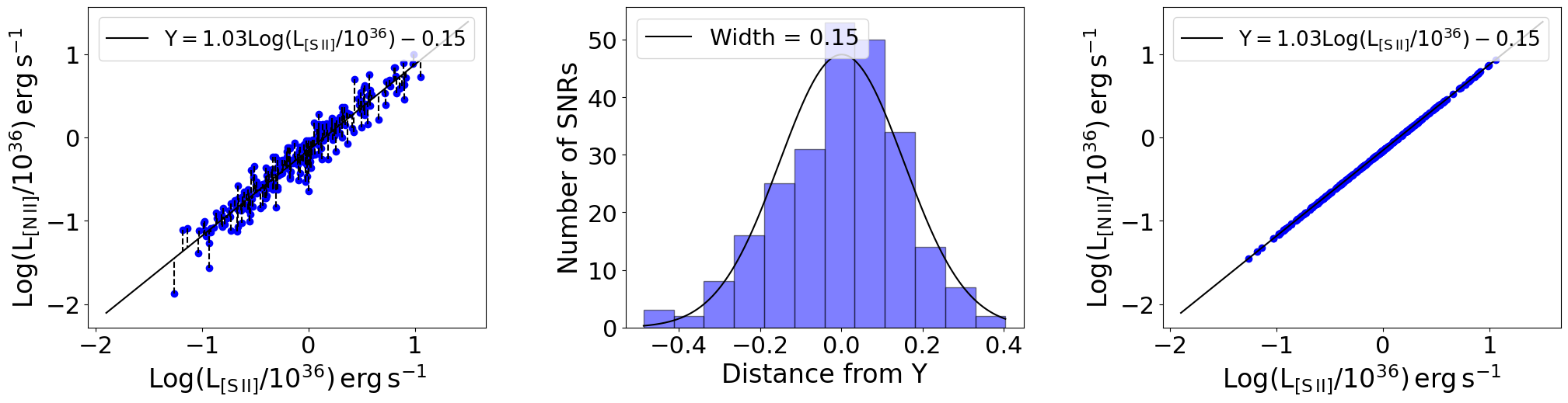}
  \includegraphics[width=0.5\linewidth]{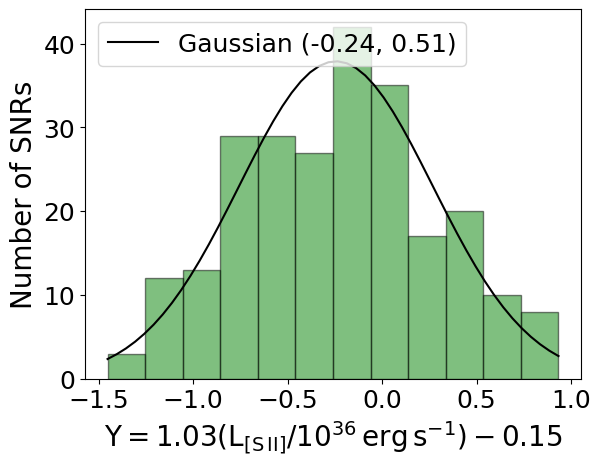}
\hfil
  \includegraphics[width=0.5\linewidth]{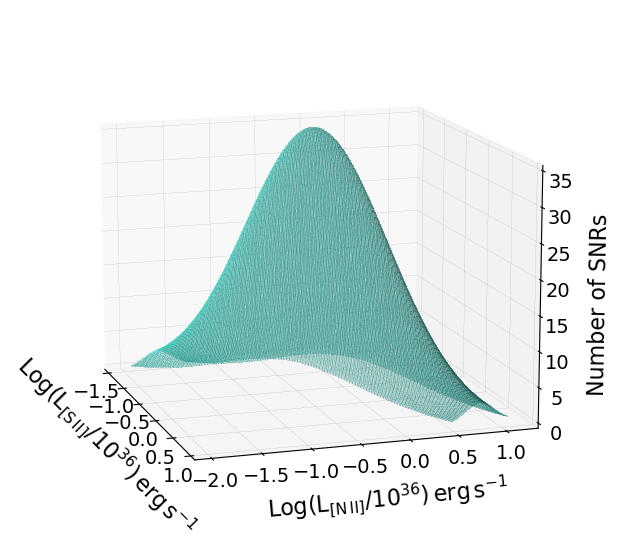}
  \caption{Same as \autoref{fig:SIIHa_LF} but for the joint \nii\ - \sii\ luminosity function.}\label{fig:NIISII_LF}
\endminipage
\end{figure*}

\begin{figure*}
\minipage{0.9\textwidth}
  \includegraphics[width=1\linewidth]{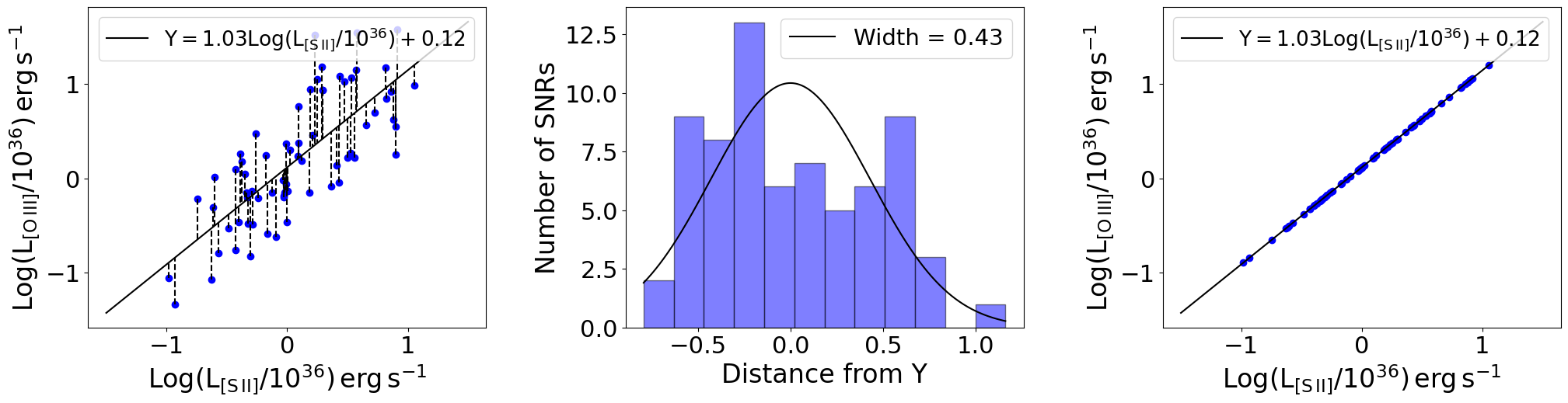}
  \includegraphics[width=0.5\linewidth]{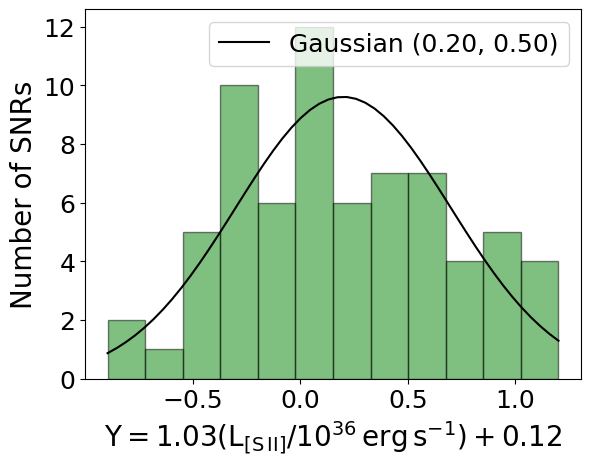}
\hfil
  \includegraphics[width=0.5\linewidth]{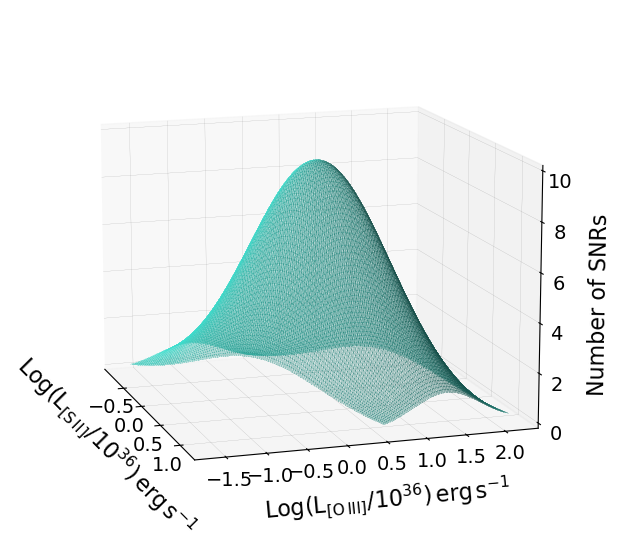}
  \caption{Same as \autoref{fig:SIIHa_LF} but for the joint \oiii\ - \sii\ luminosity function.}\label{fig:OIIISII_LF}
\endminipage
\end{figure*}

\begin{figure*}
\minipage{0.9\textwidth}
  \includegraphics[width=1\linewidth]{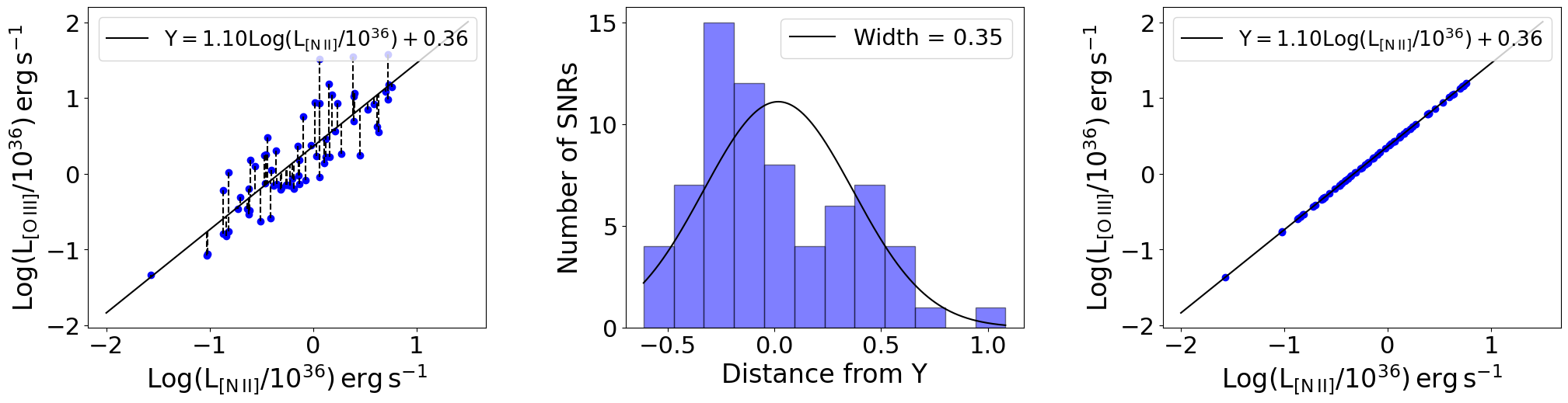}
  \includegraphics[width=0.5\linewidth]{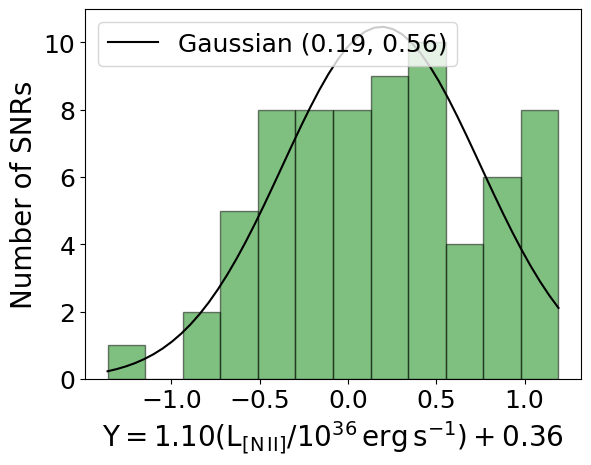}
\hfil
  \includegraphics[width=0.5\linewidth]{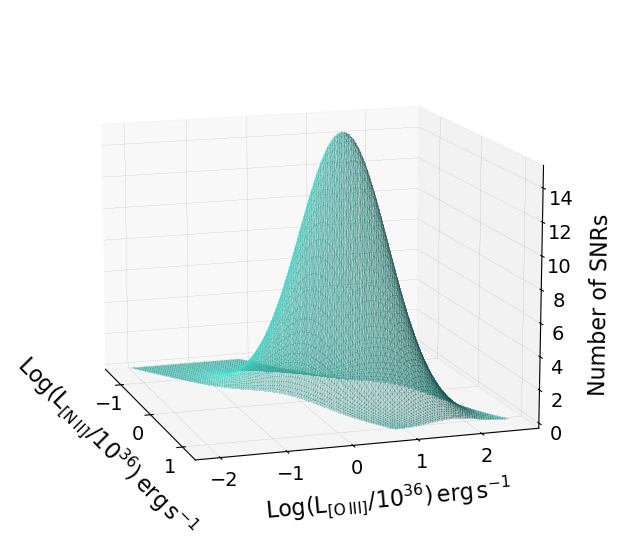}
  \caption{Same as \autoref{fig:SIIHa_LF} but for the joint \oiii\ - \nii\ luminosity function.}\label{fig:OIIINII_LF}
\endminipage
\end{figure*}

\begin{table}
	\centering
		\caption{Best - fit parameters ($\rm \mu$, $\rm\sigma$) of the Gaussian distributions for the 1D luminosity functions. The luminosity L is in units of $\rm L/10^{36}\, erg\, s^{-1}$} 
	\begin{tabular}{lll} 
		\hline
		LF & $\rm \mu (log(L))$  & $\rm \sigma$ (log(L))  \\
		\hline
	     	\ha    & $\, \, \, \, 0.38^{+0.03}_{-0.03}$   & $0.50^{+0.03}_{-0.01}$ \\
            \nii    & $-0.24^{+0.03}_{-0.03}$  & $0.54^{+0.02}_{-0.03}$ \\
            \sii    & $-0.08^{+0.02}_{-0.04}$  & $0.49^{+0.03}_{-0.01}$ \\
            \oiii   & $\, \, \, \, 0.19^{+0.09}_{-0.07}$  & $0.67^{+0.08}_{-0.04}$  \\

		\hline
	\end{tabular}
	\label{table:1d_LF_params}
	
\end{table}

\begin{table}
	\centering
		\caption{Best - fit parameters ($\rm \mu$, $\rm\sigma$) of the Gaussian distributions for the 2D luminosity functions. The luminosity L is in units of $\rm L/10^{36}\, erg\, s^{-1}$. The $\rm \mu$  and $\rm \sigma$ are the mean and the sigma values of the log(L) of  the Gaussian that give the shape of the 2D LFs, the 'Width' indicates the width of the 2D LFs along the best-fit lines that are given in the last column as 'Best-fit Line'.} 
	\begin{tabular}{ l@{\hspace*{-0.01cm}} l  l l@{\hspace*{-0.02
em}}r} 
		\hline
		LF & $\,\,\,\,\rm \mu$  & $\rm \sigma$   &  Width & Best-fit Line \\
		\hline
		\sii\ - \ha      & $\,\,\,\,-0.08^{+0.03}_{-0.03}$  &  $0.46^{+0.03}_{-0.01}$  & 0.18 &  
  $\rm Y = 0.91(L_{H\alpha}) - 0.44$\\
            \nii\ - \ha      & $\,\,\,\,-0.25^{+0.04}_{-0.03}$  &  $0.51^{+0.03}_{-0.02}$  & 0.14 &  
  $\rm Y = 1.01(L_{H\alpha}) - 0.63$\\
            \nii\ - \sii     & $\,\,\,\,-0.24^{+0.03}_{-0.03}$  &  $0.51^{+0.02}_{-0.02}$  & 0.15 &  
  $\rm Y = 1.03(L_{[S\, II]}) - 0.15$\\
            \oiii\ - \sii    & $\,\,\,\,\,\,\,\,0.20^{+0.06}_{-0.06}$  &  $0.50^{+0.05}_{-0.04}$  & 0.43 &  
  $\rm Y = 1.03(L_{[S\, II]}) + 0.12$\\
            \oiii\ - \nii    & $\,\,\,\,\,\,\,\,0.19^{+0.07}_{-0.07}$  &  $0.56^{+0.046}_{-0.04}$  & 0.35 &  
  $\rm Y = 1.10(L_{[N\, II]}) + 0.36$\\

		\hline
	\end{tabular}
	\label{table:2d_LF_params}
	
\end{table}

\subsubsection{Comparison of Luminosity Functions with other studies}
We compare the \ha\ and the joint \sii-\ha\ LFs of this study with those in \citet{2021MNRAS.507.6020K}. The basic difference between the two studies is that in the older one the authors have used the \sii/\ha\ > 0.4 criterion on imaging data, while in this study we use the multi-line diagnostics from \citet{2020MNRAS.491..889K} along with IFS. In the left panel of \autoref{fig:HaLF comp} we show the \ha\ LF of this study (black line), the study of \citet{2021MNRAS.507.6020K} for SNRs for which \sii/\ha\ - 0.4  >  $\rm 3\sigma$ (purple-dashed line) and for SNRs with the \sii/\ha\ - 0.4  >  $\rm 2\sigma$, sample that contains fainter SNRs (green-dotted line). We have also added the distribution of the SNR candidates of our sample that satisfy the  \sii/\ha\ > 0.4 criterion (grey histogram). All LFs and the distribution are normalized to 1. We see that the \ha\ LF of our study is wider than those of the other study, especially than the sample with the \sii/\ha\ ratio $\rm 3\sigma$ above the 0.4 threshold, which by definition consists of brighter sources. This is most probably the result of using deeper data in combination with the multi-line diagnostics. However, we see a quite good agreement between our sample with \sii/\ha\ > 0.4 and the sample of \citet{2021MNRAS.507.6020K} with \sii/\ha\ - 0.4  >  $\rm 2\sigma$.  This happens because setting a less strict significance limit in  \sii/\ha\ - 0.4 (i.e higher than  $\rm 2\sigma$ and not higher than >  $\rm 3\sigma$), results to the detection of also fainter SNRs. Similar behavior we see in the case of the joint \sii\ - \ha\ LF in the middle panel of \autoref{fig:HaLF comp}. 

The 2D LFs of SNRs in this work and in the study of \citet{2021MNRAS.507.6020K} are presented along the \sii-\ha\ best-fit line (e.g. middle panel of \autoref{fig:HaLF comp}). In the present study this line is $\rm Y=0.91Log(L_{H\alpha}/10^{36})-0.44$ for the entire sample and $\rm Y=1.05Log(L_{H\alpha}/10^{36})-0.23$ for the SNRs  with \sii/\ha\ > 0.4,  $\rm Y=0.88Log(L_{H\alpha}/10^{36})-0.06$  and $\rm Y=0.86Log(L_{H\alpha}/10^{36)}-0.07$ in the study of \citet{2021MNRAS.507.6020K} for SNRs with \sii/\ha\  $\rm 3\sigma$ and $\rm 2\sigma$ above the 0.4 threshold respectively.  We see that in all cases there is a sub-linearity in the $\rm Log(L_{[S\,II]}) - Log(L_{H\alpha} )$ relation. As it is also mentioned by \citet{2021MNRAS.507.6020K}, this may happen because SNRs tend to be located in regions with enhanced \ha\ (e.g. close to or even embedded in \ion{H}{II} regions) and hence in our measurements there is also contamination by these regions. However, we see that in the present study the sub-linearity slightly decreases, giving slopes closer to 1.  This probably happens because IFS provides more accurate measurements of the emission lines. For example, the \ha\ emission is not contaminated by \nii\ emission as happens in the imaging process. An underestimated \nii\ contribution would lead to overestimated \ha\ luminosities and hence, underestimated \sii/\ha\ ratios. In this study,  the majority of the SNRs gives \nii\ \lam\lam 6548, 84 / \ha\ ratios from 0.15 to 0.5 with a median value at 0.31 (in the study of \citet{2021MNRAS.507.6020K} a uniform \nii\ \lam\lam 6548, 84 / \ha\ ratio of 0.27 was assumed).

We also compare our \ha\ LF with the one from the work of \citet{2023A&A...672A.148C}. The authors have detected among others, shock excited regions based on the shock models of \citet{2008ApJS..178...20A}. This method is the most similar to our method for the identification of SNRs. \citet{2023A&A...672A.148C} present four samples: i) the 'all' sample which is the full sample and  no significance in the line detection is required; ii) the '\ha ' sample for which they require a $\rm 3\sigma$ detection of \ha\; iii) the '\oiii ' sample for which they require a $\rm 3\sigma$ detection of \oiii\; iv) the 'BPT' sample that requires a  $\rm 3\sigma$ detection of the lines that are used to built the BPT diagram (i.e of the \ha, \hb, \oi, \nii, \sii, and \oiii\ lines; \citealt{1981PASP...93....5B}; \citealt{1987ApJS...63..295V}). In the top left panel of their Figure 17 they present the distribution of the \ha\ luminosity of the sources that have identified as shock excited for all of their samples. The distribution of the samples 'all', \ha\, and \oiii\ present similar distributions while the BPT sample slightly different. For this reason we compare the \ha\ distribution of our sample with their 'all' and  BPT samples (right panel in \autoref{fig:HaLF comp}).

There is more agreement between our sample and their 'BPT' sample than their 'all' sample. This probably happens because in our sample we have also considered a significance of 3\sig\ in the emission lines. We can also notice a slight extent of the 'BPT' sample to lower luminosities. A possible explanation for this could be the fact that \citet{2023A&A...672A.148C} used a wide sample of galaxies, and data that fully cover each galaxy. This can result to identification of SNRs with different properties (e.g. different metallicities and densities) and hence, different luminosity distributions.  

In addition, we compare our \ha\ LF with the one from SNRs in NGC 4030, identified by \citet{2021MNRAS.502.1386C} using the \nii\ - \sii\ diagnostics from \citet{2020MNRAS.491..889K}, assuming a distance of 10.6 Mpc. The distance of this galaxy is quite uncertain, from 10.6 Mpc to 29.9 Mpc  (e.g. \citealt{2013ApJ...771...88L}, \citealt{2013AJ....146...86T}). As it is expected, in that distant galaxies, the faintest limit cannot be at very low luminosities and hence, the \ha\ LF occupies the brighter part of the \ha\ LF of this work and of the work of \citeauthor{2023A&A...672A.148C} (\citeyear{2023A&A...672A.148C}; \autoref{fig:HaLF comp}). However, if we assume a distance of 29 Mpc, as \citet{2021MNRAS.502.1386C} do, the \ha\ LF is shifted to brighter luminosities by a factor of $\sim$1.5 (in logarithmic scale). We should mention here, that as was indicated by the authors after private communication, in NGC 4030 the calculated fluxes may suffer from contamination by \ion{H}{II} regions, and this would give overestimated \ha\ luminosities.

\begin{figure*}
\minipage{\textwidth}
  \includegraphics[width=0.34\linewidth]{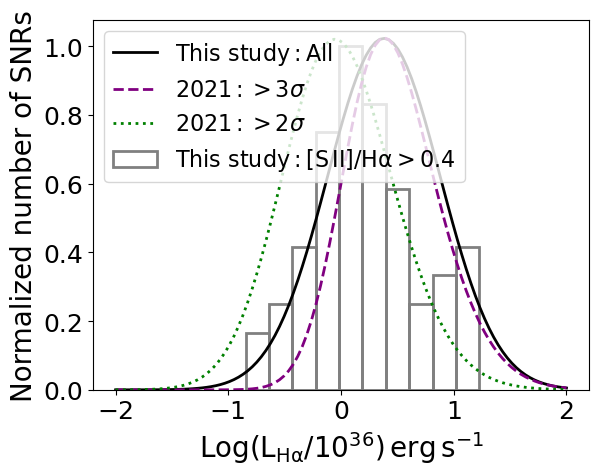}
\hfil
  \includegraphics[width=0.325\linewidth]{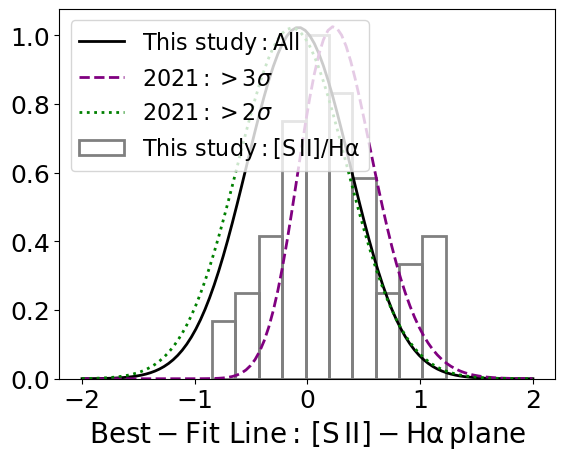}
\hfil
  \includegraphics[width=0.325\linewidth]{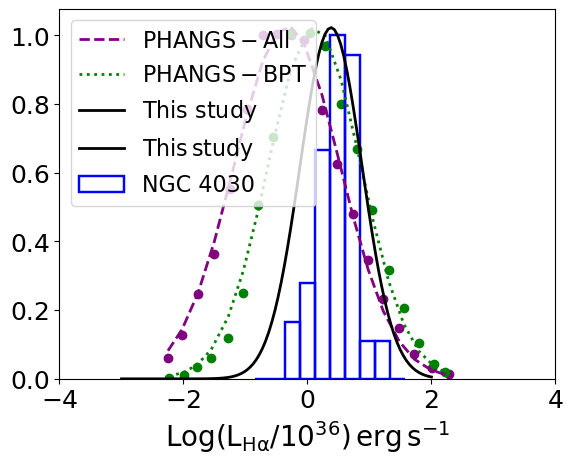}
  \caption{Comparison of Luminosity functions. {\textit{Left}}: We compare our \ha\ LF for the SNRs identified using the multi-line diagnostics (black line; best-fit Gaussian to the distribution of \ha\ luminosities), the \ha\ luminosity distribution of SNRs of these study that satisfy the \sii/\ha\ > 0.4 (grey histogram), the SNRs from the work of \citet{2021MNRAS.507.6020K} based on the \sii/\ha\ > 0.4 criterion with this ratio 3\sig\ above the 0.4 threshold (purple-dashed histogram) and with the \sii/\ha\ ratio 2\sig\ above the 0.4 threshold (green-dotted line). {\textit{Middle}}: Same as in the left panel but now for the joint \sii\ - \ha\ LF. {\textit{Right}}: Luminosity function of this study (black line), the luminosity functions of \citet{2023A&A...672A.148C} for their 'all' sample (purple-dashed line) and their 'BPT' sample (green-dotted line), and the distribution of the \ha\ of SNRs in NGC 4030 assuming a distance of 10 Mpc (blue histogram; \citealt{2021MNRAS.502.1386C})}\label{fig:HaLF comp}
\endminipage
\end{figure*}

\subsection{Shock velocities and densities of SNRs} \label{velocities}
Having multi-line information for the identified SNRs we can take advantage of theoretical models and estimate some of their physical parameters.  We use the shock models from MAPPINGS III \citep{2008ApJS..178...20A} in order to estimate the shock velocity of the SNRs. These models predict emission line ratios for shock excited regions for a wide range of parameters such as: i) shock velocity (100 - 1000 $\rm km\, s^{-1}$), ii) pre-shock density (0.001 - 1000 $\rm cm ^{-3}$), and iii) magnetic parameter $\rm \mu = B/\sqrt{n}$, where B ($\rm \mu \, G$) the magnetic field and n ($\rm cm ^{-3}$) the pre-shock density ($\rm \mu = 10^{-4} - 10\, \mu G\, cm^{3/2}$). In \autoref{fig:4dmod} we see these models in the \sii/\ha\ - \nii/\ha\ - \oiii/\hb\ - Shock Velocity space. The dimensions are the emission-line ratios and the colorbar the shock velocity. These are models that contain both shock and precursor and for solar  metallicities.  We use our emission-line ratios \sii/\ha, \nii/\ha, and \oiii/\hb\ to interpolate in this 4 dimensional space and we estimate the shock velocity. The distribution of the shock velocity of the SNRs is presented in \autoref{fig:vel}. We note here that for the calculation of  shock velocities, it is necessary the emission in all the emission lines (and in higher than a 3\sig\ significance). This means that sources for which we do not detect for example \oiii\ or \ha\ emission, the shock velocity cannot be calculated. In addition, even if we have emission in all wavelengths,   some of the sources may fall out of the interpolation limits, presenting lower emission line ratios and hence indicating velocities lower than 100 $\rm km\, s^{-1}$ (grey points in \autoref{fig:diagnostics}). Those sources are not presented in  \autoref{fig:vel}. Hence finally, shock velocities have been calculated only for the $\sim$ 34\% of the total sample. As can be seen in \autoref{fig:4dmod}, three emission-line ratios are enough in order to not have degeneracy in the models because of different densities and magnetic parameters.

The $\sim 65\%$ of the SNRs (of which we can calculate velocities) have velocities between 100 and 300 $\rm km\, s^{-1}$. These  are expected  for optical SNRs. Usually, the optical emission comes from regions where the shock velocity has decreased, and so the temperature, allowing recombination processes to take place, e.g. of \ha, \nii, \sii, \oiii\ (e.g \citealt {2022hxga.book...77B}). The rest $\sim 35\%$ presents velocities between 300 and 700  $\rm km\, s^{-1}$. These are sources with strong \oiii\ emission, probably indicating SNRs in highly non-uniform environments, where some parts of their shock are moving with low velocities, while others still with high velocities resulting in strong \oiii\ emission. The most interesting and extremest case, with a velocity of 700 $\rm km\, s^{-1}$, is the NGC7793\_SNR\_91.  This  has been identified by all the diagnostics and it presents a  \oiii\ flux of $\rm 2.3\times10^{-14}\, erg\,cm^{-2}\, s^{-1}$ . This SNR along with its spectrum can be seen in \autoref{fig:spec}.  Its ring-like structure of a diameter of $\rm \sim 30 pc$, could indicate a relatively old SNR. Apart from a SNR evolving in a non-uniform medium, another possible scenario could be that the strong \oiii\ emission is the result of the photoionization of gas shocked by a pulsar wind nebula in the interior of the SNR (e.g. \citealt{2018ApJ...864L..36M}). As can be seen in \autoref{fig:spec}, this source also presents  emission in \ion{He}{II} $\lambda 4685$, [\ion{Ar}{III}]$\lambda 7135$, [\ion{S}{III}]$\lambda 9068$ and a weaker [\ion{S}{III}]$\lambda 6312$ emission.


We explored possible correlations between emission-line ratios and shock velocities. In \autoref{fig:vel_lines} we see the \sii/\ha\ (left panel), \nii/\ha\ (middle panel), and \oiii/\hb\ (right panel) of the SNRs as function of their shock velocity. In the first two cases, they appear two branches, the  \sii/\ha\ and \nii/\ha\ emission-line ratio of which seems to increase with the shock velocity. This trend is expected since it probably means that the higher the shock velocity is, the stronger the shock excitation appears, represented by higher \nii/\ha\ and \sii/\ha\ ratios. In the case of \oiii\hb\, the trend is more obvious and the \oiii\hb\ emission-line ratio has a very tight correlation with the shock velocity. This behavior is also expected from theoretical studies (e.g. \citealt {2008ApJS..178...20A}; \citealt{1979ApJS...39....1R}) and is evident in observational ones (e.g. \citealt{2022MNRAS.512.1658B}). The \oiii/\hb\ ratio presents a very slight dependence on density (e.g. \citealt{2017ApJS..229...35D}; \citealt{2022MNRAS.tmp.1558P}) and on the abundances (e.g. \citealt{2008ApJS..178...20A}) and hence is affected more directly by the temperature which increases with the shock velocity.

We also use the \sii 6716/31 ratio in order to estimate the density of the candidate SNRs, which refers to the post-shock density and it depends on the pre-shock density of their ambient ISM. This is possible when we have lines from the same ions that have more or less the same excitation energy and hence, the collisional excitation will be proportional to the collision rates (e.g. \citealt{2006agna.book.....O}; \citealt{2011piim.book.....D}).  The relation between  the \sii 6716/31 ratio and the density for different temperatures is shown in  \autoref{fig:den_rat_theor}. We used the python tool {\texttt{PYNEB}} (\citealt{2015A&A...573A..42L}) in order to estimate the densities for the SNRs. In our case, we give as input the \sii 6716/31 and the temperatures $\rm 5\times 10 ^ 3\, K, 10 ^ 4\, K, 3\times 10 ^ 4\, K,$ and $\rm 5\times 10 ^ 4\, K$, which are more or less the temperatures that we expect from SNRs in the radiative phase, and it gives as output the density in $\rm cm^{-3}$. Hence, for each temperature, we have the density distribution for our SNRs (\autoref{fig:den}). The median values of the density in our sample are: $\rm \sim 70\, cm^{-3},  \sim 80\, cm^{-3}, \sim 160\, cm^{-3}, and \sim 240\, cm^{-3}$, for temperatures $\rm 5\times 10 ^ 3\, K, 10 ^ 4\, K, 3\times 10 ^ 4\, K,$ and $\rm 5\times 10 ^ 4\, K$  respectively. We note that the {\texttt{PYNEB}} cannot recover densities lower than 1 that correspond to  \sii 6716/31 ratios higher than $\sim$ 1.45 as can be seen in \autoref{fig:den_rat_theor}. Hence, we have not calculated the density of SNRs with \sii 6716/31 ratios from $\sim$ 1.45 to $\sim$ 2.22 (the majority of these sources have \sii 6716/31 ratio around 1.5), and this is the $\sim 45\%$ of our sample.

An estimation of the distribution of physical parameters of SNRs, such as density and velocity can also be used as input in theoretical analysis for the development of population synthesis models (e.g. \citealt{2022MNRAS.514.3260K}).

\begin{figure}
\centering
 \includegraphics[width=0.5\textwidth]{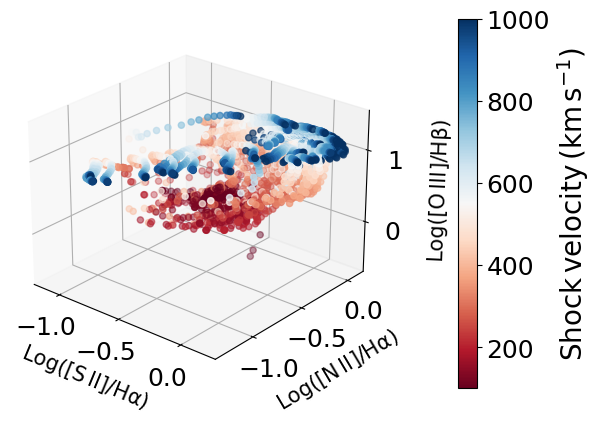}
\caption{\label{fig:4dmod} The shock models of \citet{2008ApJS..178...20A} in the \sii/\ha\ - \nii/\ha\ - \oiii/\hb\ space. The colorbar indicates the shock velocity.}
\end{figure}

\begin{figure}
\centering
 \includegraphics[width=0.45\textwidth]{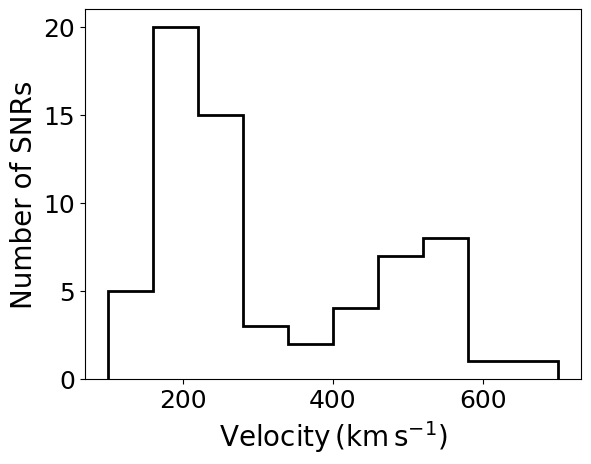}
\caption{\label{fig:vel} The shock velocity distribution of the candidate SNRs identified in this study. We calculate the shock velocity interpolating in the \sii/\ha\ - \nii/\ha\ - \oiii/\hb\ - Shock velocity space of the shock models of \citet{2008ApJS..178...20A} as it seems in \autoref{fig:4dmod}.}
\end{figure}

\begin{figure}
  \includegraphics[width=\linewidth]{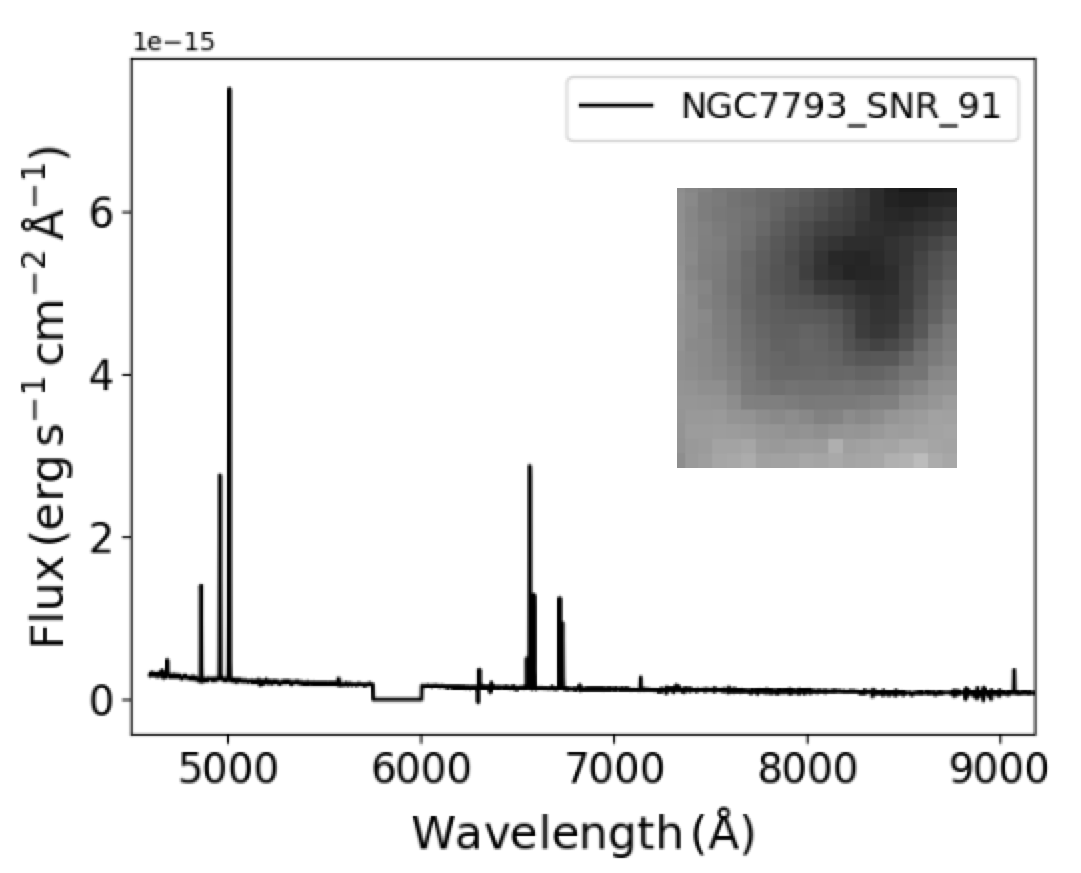}
\caption{The SNR NGC7793\_SNR\_91 in a box of size $\rm 4\times4\, arcsec^2$ and its spectrum. }\label{fig:spec}
\end{figure} 

\begin{figure*}
\minipage{0.95\textwidth}
  \includegraphics[width=0.33\linewidth]{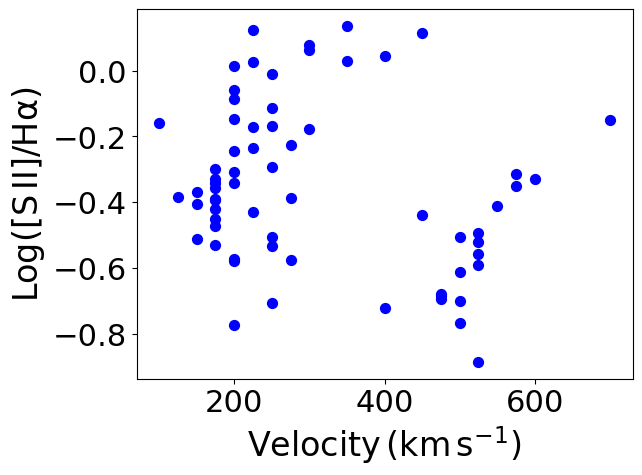}
\hfil
  \includegraphics[width=0.33\linewidth]{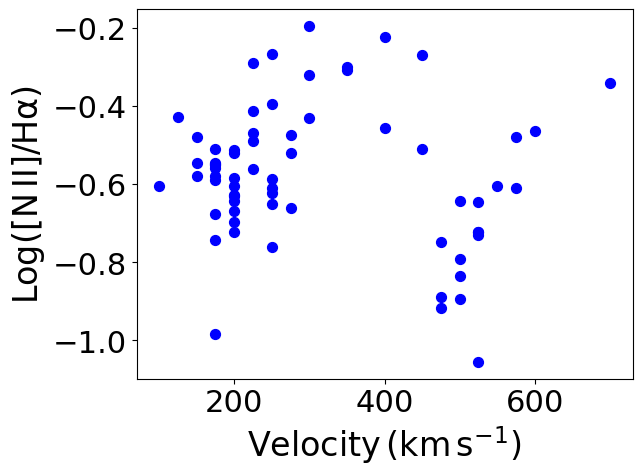}
\hfil
  \includegraphics[width=0.33\linewidth]{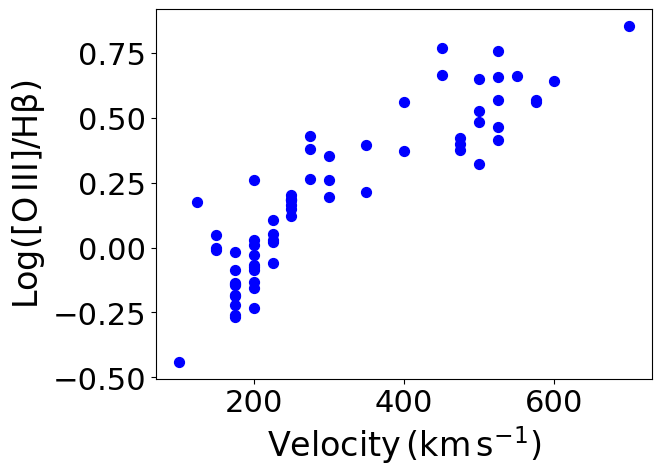}
  \caption{The logarithmic emission line ratios log(\sii/\ha) (left), log(\nii/\ha) (middle), and log(\oiii/\hb) (right) of the SNR candidates identified in this work, as function of their shock velocity.}\label{fig:vel_lines}
\endminipage
\end{figure*}

\begin{figure}
\includegraphics[width=0.45\textwidth]{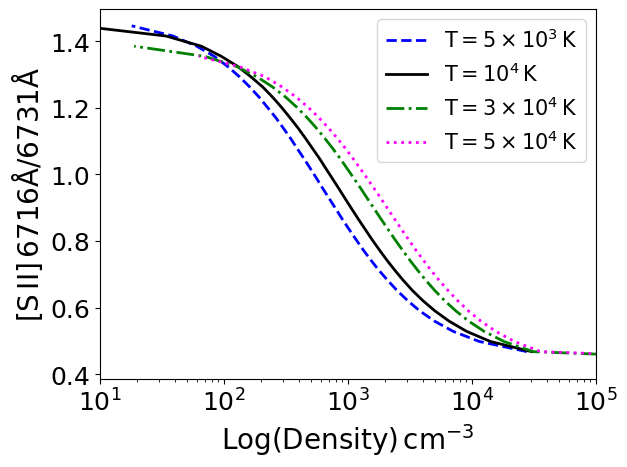}
\caption{\label{fig:den_rat_theor} The dependence of the \sii 6716/31 ratio on the pre-shock density for different temperatures.}
\end{figure}\hfil
\begin{figure}
\includegraphics[width=0.45\textwidth]{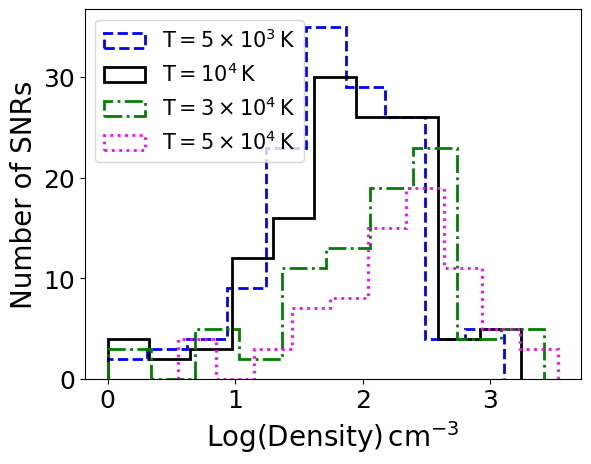}
\caption{\label{fig:den} The distribution of the densities of the candidate SNRs, identified in this study, for different temperatures.}
\end{figure}

\subsection{Investigation of SNRs' ambient environment via IFS - Correlation between SNRs and ISM properties.}
In an attempt to look for correlations between the SNR properties and the properties of the ISM, we construct a density map of the available fields of the galaxy, using the \sii 6716/31 ratio as described in \S \ref{velocities}. In order to do that, we first bin the \sii\ $\rm \lambda$6717 map (which is the weaker line between the two \sii\ lines) so that every bin to have S/N > 20. We reconstruct also the \sii\ $\rm \lambda$6731 map using the same binning and we create the the \sii 6716/31 ratio map. We do the binning and the reconstruction using the ADABIN code\footnote{https://github.com/zidianjun/adabin} (adaptive binning algorithm for 2D maps; \citealt{2023MNRAS.518..286L}, \citealt{2023arXiv231114254G}). Then, we make contours on the map of regions that have \ha\ flux higher than $\rm 10^{-17}\, erg\, s^{-1}\, cm^{-2}$ and set the rest equal to zero. We do that for the purpose of minimizing the presence of background on the map. Of course, by setting this cutoff, we may miss regions that contain SNRs. The final map is presented in \autoref{fig:density_map} with the colorbar showing the density. On the map we have added the identified SNRs and we have categorized them in those for which  \sii/\ha\ < 0.4 (magenta triangles) and \sii/\ha\ > 0.4 (black circles). Aiming to make possible correlations more obvious, we also construct the density histogram of the SNRs with \sii/\ha\ < 0.4 (magenta-dashed histogram) and \sii/\ha\ > 0.4 (black histogram). The density map and the densities used in the histograms, have been calculated assuming a  temperature of $\rm 10^4 \, K$, a typical temperature for SNRs (e.g. \citealt{2022MNRAS.512.1658B}). We do not see any specific correlation between density and the two categories of SNRs. We also calculated the density in an annulus of a width of 0.4" and for a radius 1.5 times the radius we used for the photometry of each source, which represents the pre-shock density of the SNRs' ambient, but again, no correlation was apparent.  This could be an indication that the properties of SNRs (for example the shock excitation) are affected by the density of the very local environment, that we cannot study in that distant galaxies. Another possible explanation could be that the density currently observed  in the surrounding ISM (absolute values and also structures) is not correlated to that of the time of the SN explosion and the first years of the SNR evolution, and hence, no correlations should be expected. In any case, we must keep in mind that here we explore more central regions of the galaxy. A more complete coverage of the galaxy could present higher variations in density and probably a different trend in the densities of the two categories (or in general the different \sii/\ha\ ratios).

\begin{figure*}
\minipage{\textwidth}
  \includegraphics[width=0.95\linewidth]{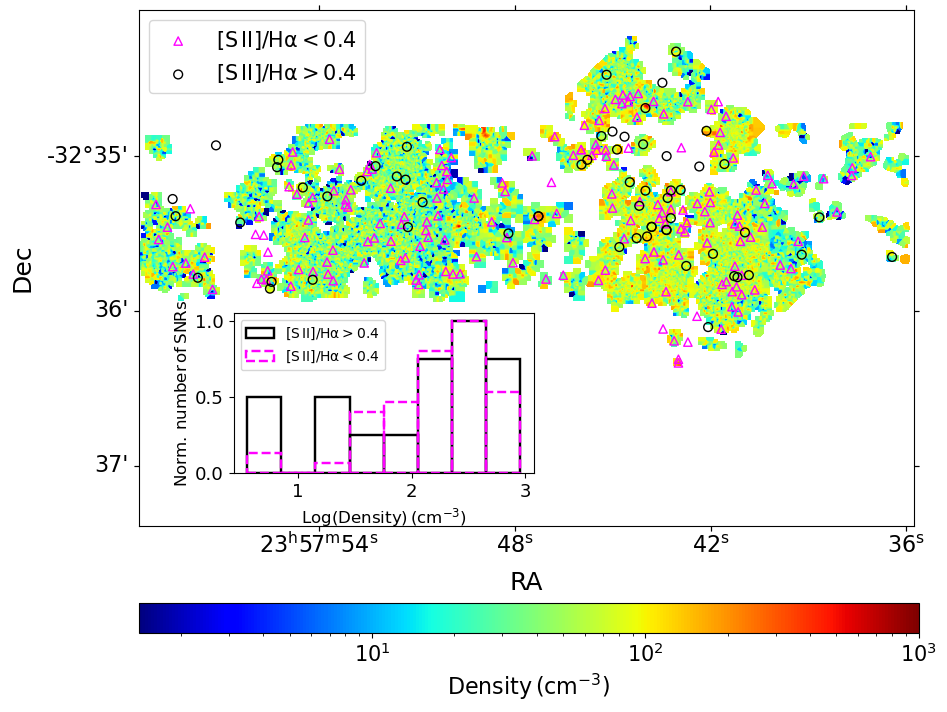}
  \caption{The density map of the available fields of NGC 7793, using the  \sii 6716/31 ratio. The apparent regions are those for which the \ha\ flux is higher than $\rm 10^{-17}\, erg\, s^{-1}\, cm^{-2}$. The black circles indicate the position of candidate SNRs with \sii/\ha\ > 0.4 and the magenta triangles those with \sii/\ha\ < 0.4. The black histogram in the bottom-left panel shows the density distribution of SNRs with \sii/\ha\ > 0.4 while the magenta-dashed histogram the density distribution for those with \sii/\ha\ < 0.4. The densities in this plot have been calculating assuming a temperature of $\rm 10^4 \, K$.}\label{fig:density_map}
\endminipage
\end{figure*}

We also explore if there is any correlation between density and shock velocity. In \autoref{fig:denvel} we plot the density of SNRs as a function of their shock velocity (for those that we could calculate both densities and shock velocities). Again, we do not see any tight correlation between density and shock velocity. One could say that there is a lack of high-velocity SNRs evolving in lower density environments. This would probably mean that because of the low density and the relatively high velocity that has as a result higher temperatures, the shock excitation in the optical wavelengths is still very weak to be detected, or even  absent. However, the statistics is very poor to come to an accurate conclusion. In both cases (Figures \ref{fig:density_map} and \ref{fig:denvel}), it is evident the need of larger samples by having a full coverage of the galaxy and/or similar studies in other galaxies.
\begin{figure}
\centering
 \includegraphics[width=0.45\textwidth]{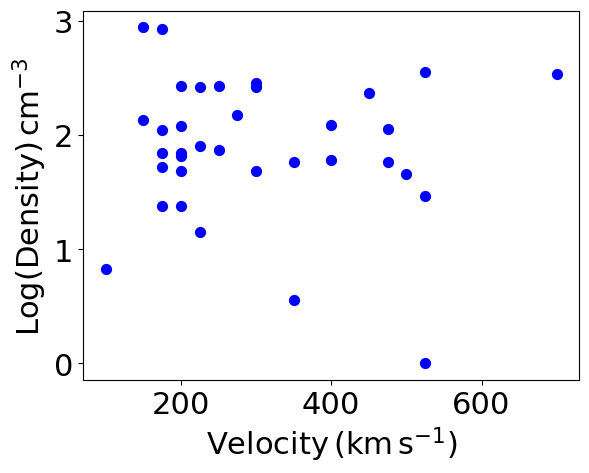}
\caption{\label{fig:denvel} The density of the candidate SNRs (calculated using the \siiratio\ ratio and assuming a temperature of $\rm 10^4 \, K$) as function of their shock velocity estimated by using shock  models of \citet{2008ApJS..178...20A}.}
\end{figure}



\section{Conclusions} \label{conclusions}
In this study, we use MUSE IFS of the galaxy NGC 7793 and we apply the multi-line diagnostics developed by \citet{2020MNRAS.491..889K} for the identification of SNRs. We find in total 238 candidate SNRs, 225 of which, new identifications, down to a \ha\ luminosity of $\rm 9\times 10 ^ {-17}\, erg\, cm^{-2}\, s^{-1}$ and $\sim 72\%$ more than those that would be identified by using the traditional diagnostic tool \sii/\ha\ > 0.4. Our sample consists of sources for which the \ha, \hb, \nii, \sii, and \oiii\ emission lines, and the  \nii/\ha, \sii/\ha, and \oiii/\hb\ emission line ratios, have a significance higher than 3\sig. The \ha\ velocity dispersion of the candidate SNRs was also calculated giving values between 15 $\rm km\, s^{-1}$ and 70 $\rm km\, s^{-1}$, most of them in agreement with those suggested in \citet{2019ApJ...887...66P} for SNRs.

We construct the luminosity functions  of the \ha, \sii, and for the first time of the \nii\ and \oiii\ emission lines. We fit Gaussian distributions, the best parameters of which are calculated by using a Markov Chain Monte Carlo (MCMC) method, and are presented in \autoref{table:1d_LF_params}. We also create the joint \sii\ - \ha, and for the first time the \nii\ - \ha, \nii\ - \sii, \oiii\ - \sii, and \oiii\ - \nii\ luminosity functions. Although there is a relatively tight correlation for the majority of the emission line ratios (i.e. small scatter when plotting SNRs in a 2D emission-line ratio plane), we see a large scatter in relations that contain the \oiii\ emission line (Figures \ref{fig:SIIHa_LF} - \ref{fig:OIIINII_LF}). This is probably an indication that the \oiii\ emission comes from a different region of the shock than the \sii\ or \nii\ emissions. For the joint luminosity functions we also fit Gaussian distributions, the best-fit parameters of which are shown in \autoref{table:2d_LF_params}.

We use the observed \nii/\ha, \sii/\ha, and \oiii/\hb\ emission line ratios of the SNRs to calculate their shock velocity. In order to do that, we interpolate  in the \sii/\ha\ - \nii/\ha\ - \oiii/\hb\ - Shock Velocity space of the shock  models of \citet{2008ApJS..178...20A}. We find that the $\sim 65\%$ of these SNRs has shock velocities lower than 300 $\rm km\, s^{-1}$,  as expected for optical SNRs. We also see that the SNRs that are missed by the traditional criterion, are those with lower shock velocities (most of them lower than 100  $\rm km\, s^{-1}$), or high enough to give strong optical emission lines (because higher velocities result to higher temperatures where recombination processes of \sii\ for example are less probable to happen). Moreover, we see a clear correlation between \oiii/\hb\ ratio and shock velocity, while less clear but still evident for the \sii/\ha\ and \nii/\ha\ emission line ratios.

In addition, we use the \sii 6716/31 ratio and assuming different temperatures we find the density of the candidate SNRs. The median value of the density of the SNRs in our sample for a temperature of $\rm T = 10^4\, K$, is $\rm \sim 80\, cm^{-3}$. Velocity and density distribution, as well as, luminosity functions of various emission lines can be used in theoretical analyses for the development of population synthesis models.

We find no evident correlation between density and SNRs with \sii/\ha\ > 0.4 and those with \sii/\ha\ < 0.4. However, this could be result of the fact that we have only central regions of the galaxy available and not a full coverage. We do not find any correlation between density and shock velocity either, which could also be result of poor statistics. It is very probable that no trend between density and other properties of SNRs should be expected, since the medium could have been very different at the time of the SN explosion and during the first stages of the SNR evolution. However, for more accurate conclusions we need larger samples of the same but also of different galaxies. 
\clearpage
\section*{Acknowledgements}

We thank the anonymous referee for the thorough review and the very useful comments that helped to improve the clarity of the paper. We thank Andreas Zezas for the useful discussion about densities. The SNICE research group acknowledges financial support from the program Unidad de Excelencia Mar\'{i}a de Maeztu CEX2020-001058-M, Spanish Ministerio de Ciencia e Innovaci\'{o}n (MCIN), the Agencia Estatal de Investigaci\'{o}n (AEI) 10.13039/501100011033,  the European Social Fund (ESF) “Investing in your future” under the 2019 Ram\'{o}n y Cajal program RYC2019-027683-I, and the PID2020-115253GA-I00 HOSTFLOWS project, from Centro Superior de Investigaciones Cient\'{i}ficas (CSIC) under the PIE project 20215AT016. 

\section*{Data Availability}
For this study we have used data available on the ESO Archive Science Portal.







\newpage
\onecolumn
\appendix

\section{Candidate SNRs in NGC 7793}
\begin{ThreePartTable}
\begin{longtable}{l@{\hspace*{0.5em}}l@{\hspace*{0.2em}}l@{\hspace*{1em}}l@{\hspace*{0.5em}}l@{\hspace*{0.5em}}l@{\hspace*{0.5em}}l@{\hspace*{0.5em}}l@{\hspace*{1em}}l@{\hspace*{-0.4em}}l@{\hspace*{0.3em}}l@{\hspace*{0.3em}}l@{\hspace*{0.3em}}l@{\hspace*{0.3em}}l@{\hspace*{0.8em}}lll}
\caption{Information of the candidate SNRs. Col.9 corresponds to the radius of the aperture used for the photometry. Col.10, col.11, col.12, col.13, and col.14 indicate if each source has been identified (y) or not (n) with the diagnostics 1: \sii/\ha, 2: \sii/\ha\ - \nii/\ha, 3: \sii/\ha\ - \oiii/\hb, 4: \nii/\ha\ - \oiii/\hb, 5: \nii/\ha\ - \sii/\ha\ - \oiii/\hb. The last column shows the \ha\ velocity dispersion.}
\label{table:opt_SNRs}\\
\hline
ID & RA &   DEC &  $\rm F_{H\alpha} \times 10^{-16} $ & $\rm F_{H\beta} \times 10^{-16} $ & $\rm Log(\frac{[\ion{S}{II}]}{H\alpha}) $ & $\rm Log(\frac{[\ion{N}{II}]}{H\alpha}) $ & $\rm Log(\frac{[\ion{O}{III}]}{H\beta}) $ & aper &  1 & 2 & 3 & 4 & 5 & $\rm V(\sigma)$\\
   NGC7793 & (J2000) & (J2000) & ($\rm erg\, cm^{-2}\, s^{-1}$) & ($\rm erg\, cm^{-2}\, s^{-1}$) & & & & (arcsec) & & & & & & $\rm km\, s^{-1}$\\

\hline
SNR\_1 & 23:57:36.4 & -32:35:39.3 &  7.0 $\pm$  0.2 & 2.47 $\pm$ 0.11 & -0.06 $\pm$ 0.03 & -0.70 $\pm$ 0.04 & -0.08 $\pm$ 0.06 & 1.86 & y & y & y & n & y & 37.04 $\pm$ 1.12 \\ 
SNR\_2 & 23:57:37.1 & -32:35:00.4 & 128.3 $\pm$  0.6 & 59.78 $\pm$ 0.40 & -0.51 $\pm$ 0.01 & -0.59 $\pm$ 0.01 & 0.18 $\pm$ 0.01 & 3.25 & n & y & y & y & y & 24.52 $\pm$ 0.55 \\ 
SNR\_3 & 23:57:37.6 & -32:35:04.7 & 13.7 $\pm$  0.2 & 5.41 $\pm$ 0.11 & -0.61 $\pm$ 0.02 & -0.79 $\pm$ 0.02 & 0.53 $\pm$ 0.03 & 1.30 & n & y & y & y & y & 21.79 $\pm$ 0.50 \\ 
SNR\_4 & 23:57:37.6 & -32:35:07.7 & 11.6 $\pm$  0.1 & $\cdots$ & -0.48 $\pm$ 0.01 & -0.65 $\pm$ 0.02 & $\cdots$ & 1.00 & n & y & n & n & n & 24.90 $\pm$ 0.55 \\ 
SNR\_5 & 23:57:38.1 & -32:35:21.8 & 34.1 $\pm$  0.4 & 14.56 $\pm$ 0.29 & -0.56 $\pm$ 0.02 & -0.73 $\pm$ 0.02 & 0.57 $\pm$ 0.03 & 2.98 & n & y & y & y & y & 23.39 $\pm$ 0.54 \\ 
SNR\_6 & 23:57:38.5 & -32:35:08.9 &  7.4 $\pm$  0.1 & $\cdots$ & -0.49 $\pm$ 0.02 & -0.70 $\pm$ 0.02 & $\cdots$ & 1.21 & n & y & n & n & n & 20.49 $\pm$ 0.54 \\ 
SNR\_7 & 23:57:38.6 & -32:35:23.9 &  8.4 $\pm$  0.2 & 2.94 $\pm$ 0.12 & -0.32 $\pm$ 0.03 & -0.61 $\pm$ 0.04 & 0.56 $\pm$ 0.05 & 1.78 & y & y & y & y & y & 26.89 $\pm$ 0.68 \\ 
SNR\_8 & 23:57:39.1 & -32:35:08.0 & 11.0 $\pm$  0.3 & $\cdots$ & -0.61 $\pm$ 0.03 & -0.79 $\pm$ 0.04 & $\cdots$ & 1.28 & n & y & n & n & n & 24.56 $\pm$ 0.56 \\ 
SNR\_9 & 23:57:39.2 & -32:35:38.4 & 70.2 $\pm$  0.4 & 29.84 $\pm$ 0.26 & -0.16 $\pm$ 0.01 & -0.61 $\pm$ 0.01 & -0.44 $\pm$ 0.01 & 2.27 & y & y & y & n & y & 34.96 $\pm$ 0.90 \\ 
SNR\_10 & 23:57:39.3 & -32:35:33.2 & 25.0 $\pm$  0.3 & $\cdots$ & -0.62 $\pm$ 0.01 & -0.68 $\pm$ 0.01 & $\cdots$ & 1.69 & n & y & n & n & n & 23.72 $\pm$ 0.54 \\ 
SNR\_11 & 23:57:39.4 & -32:35:11.0 & 10.9 $\pm$  0.2 & $\cdots$ & -0.65 $\pm$ 0.02 & -0.61 $\pm$ 0.02 & $\cdots$ & 1.00 & n & y & n & n & n & 19.03 $\pm$ 0.45 \\ 
SNR\_12 & 23:57:39.5 & -32:35:43.8 & 17.9 $\pm$  0.2 & $\cdots$ & -0.61 $\pm$ 0.02 & -0.77 $\pm$ 0.02 & $\cdots$ & 1.66 & n & y & n & n & n & 26.61 $\pm$ 0.54 \\ 
SNR\_13 & 23:57:39.8 & -32:35:42.4 & 34.0 $\pm$  0.2 & $\cdots$ & -0.52 $\pm$ 0.01 & -0.67 $\pm$ 0.01 & $\cdots$ & 1.66 & n & y & n & n & n & 30.47 $\pm$ 0.58 \\ 
SNR\_14 & 23:57:40.1 & -32:35:10.9 &  4.6 $\pm$  0.1 & $\cdots$ & -0.43 $\pm$ 0.03 & -0.89 $\pm$ 0.04 & $\cdots$ & 1.26 & n & y & n & n & n & 20.90 $\pm$ 0.48 \\ 
SNR\_15 & 23:57:40.2 & -32:35:07.6 &  7.2 $\pm$  0.2 & $\cdots$ & -0.46 $\pm$ 0.03 & -0.77 $\pm$ 0.03 & $\cdots$ & 1.51 & n & y & n & n & n & 24.61 $\pm$ 0.61 \\ 
SNR\_16 & 23:57:40.3 & -32:35:18.5 & 32.1 $\pm$  0.4 & $\cdots$ & -0.48 $\pm$ 0.01 & -0.70 $\pm$ 0.02 & $\cdots$ & 1.76 & n & y & n & n & n & 22.59 $\pm$ 0.44 \\ 
SNR\_17 & 23:57:40.4 & -32:35:27.7 & 115.6 $\pm$  0.4 & 47.86 $\pm$ 0.26 & -0.70 $\pm$ 0.00 & -0.89 $\pm$ 0.01 & 0.65 $\pm$ 0.01 & 2.36 & n & n & y & y & y & 27.21 $\pm$ 0.55 \\ 
SNR\_18 & 23:57:40.6 & -32:35:13.3 & 24.0 $\pm$  0.3 & $\cdots$ & -0.58 $\pm$ 0.02 & -0.88 $\pm$ 0.02 & $\cdots$ & 1.57 & n & y & n & n & n & 26.04 $\pm$ 0.61 \\ 
SNR\_19 & 23:57:40.6 & -32:35:52.0 & 21.6 $\pm$  0.2 & $\cdots$ & -0.60 $\pm$ 0.01 & -0.84 $\pm$ 0.01 & $\cdots$ & 0.80 & n & y & n & n & n & 38.47 $\pm$ 0.78 \\ 
SNR\_20 & 23:57:40.7 & -32:35:31.6 & 45.3 $\pm$  0.2 & $\cdots$ & -0.60 $\pm$ 0.00 & -0.67 $\pm$ 0.00 & $\cdots$ & 0.80 & n & y & n & n & n & 22.98 $\pm$ 0.49 \\ 
SNR\_21 & 23:57:40.8 & -32:35:46.3 & 23.0 $\pm$  0.4 & 7.92 $\pm$ 0.24 & -0.15 $\pm$ 0.02 & -0.51 $\pm$ 0.02 & -0.16 $\pm$ 0.04 & 1.17 & y & y & y & n & y & 37.49 $\pm$ 0.85 \\ 
SNR\_22 & 23:57:40.9 & -32:35:29.8 &  9.2 $\pm$  0.1 & $\cdots$ & -0.21 $\pm$ 0.02 & -0.71 $\pm$ 0.02 & $\cdots$ & 0.99 & y & y & n & n & n & 28.42 $\pm$ 0.58 \\ 
SNR\_23 & 23:57:41.0 & -32:35:53.9 & 135.6 $\pm$  0.5 & $\cdots$ & -0.62 $\pm$ 0.00 & -0.68 $\pm$ 0.01 & $\cdots$ & 0.80 & n & y & n & n & n & 24.45 $\pm$ 0.45 \\ 
SNR\_24 & 23:57:41.0 & -32:35:46.3 & 61.4 $\pm$  0.2 & $\cdots$ & -0.44 $\pm$ 0.00 & -0.60 $\pm$ 0.01 & $\cdots$ & 1.10 & n & y & n & n & n & 27.80 $\pm$ 0.46 \\ 
SNR\_25 & 23:57:41.1 & -32:35:27.3 &  1.9 $\pm$  0.1 & 0.65 $\pm$ 0.07 & -0.47 $\pm$ 0.08 & -0.51 $\pm$ 0.09 & -0.09 $\pm$ 0.15 & 0.80 & n & y & y & n & y & 21.61 $\pm$ 0.45 \\ 
SNR\_26 & 23:57:41.1 & -32:36:00.5 & 19.2 $\pm$  0.2 & 6.69 $\pm$ 0.12 & -0.51 $\pm$ 0.01 & -0.64 $\pm$ 0.02 & 0.32 $\pm$ 0.02 & 1.43 & n & y & y & y & y & 28.54 $\pm$ 0.60 \\ 
SNR\_27 & 23:57:41.1 & -32:35:47.1 &  9.2 $\pm$  0.2 & $\cdots$ & -0.25 $\pm$ 0.03 & -0.60 $\pm$ 0.03 & $\cdots$ & 0.80 & y & y & n & n & n & 30.00 $\pm$ 0.48 \\ 
SNR\_28 & 23:57:41.1 & -32:35:23.1 & 15.6 $\pm$  0.2 & $\cdots$ & -0.48 $\pm$ 0.02 & -0.66 $\pm$ 0.02 & $\cdots$ & 1.34 & n & y & n & n & n & 26.86 $\pm$ 0.45 \\ 
SNR\_29 & 23:57:41.2 & -32:35:50.8 & 58.1 $\pm$  1.3 & 19.80 $\pm$ 0.95 & -0.68 $\pm$ 0.03 & -0.92 $\pm$ 0.03 & 0.42 $\pm$ 0.07 & 0.77 & n & n & y & y & y & 25.89 $\pm$ 0.46 \\ 
SNR\_30 & 23:57:41.2 & -32:35:42.9 & 11.4 $\pm$  0.1 & $\cdots$ & -0.56 $\pm$ 0.01 & -0.65 $\pm$ 0.02 & $\cdots$ & 0.80 & n & y & n & n & n & 28.53 $\pm$ 0.46 \\ 
SNR\_31 & 23:57:41.2 & -32:35:26.4 &  2.5 $\pm$  0.1 & $\cdots$ & -0.64 $\pm$ 0.04 & -0.73 $\pm$ 0.05 & $\cdots$ & 0.80 & n & y & n & n & n & 27.90 $\pm$ 0.55 \\ 
SNR\_32 & 23:57:41.2 & -32:35:46.7 & 30.5 $\pm$  0.8 & $\cdots$ & -0.39 $\pm$ 0.03 & -0.70 $\pm$ 0.04 & $\cdots$ & 0.80 & y & y & n & n & n & 29.44 $\pm$ 0.44 \\ 
SNR\_33 & 23:57:41.3 & -32:35:00.9 &  8.0 $\pm$  0.1 & 3.54 $\pm$ 0.07 & -0.49 $\pm$ 0.02 & -0.72 $\pm$ 0.02 & 0.41 $\pm$ 0.03 & 1.29 & n & y & y & y & y & 25.32 $\pm$ 0.55 \\ 
SNR\_34 & 23:57:41.3 & -32:35:52.7 & 32.6 $\pm$  0.3 & $\cdots$ & -0.55 $\pm$ 0.01 & -0.66 $\pm$ 0.01 & $\cdots$ & 0.80 & n & y & n & n & n & 27.20 $\pm$ 0.56 \\ 
SNR\_35 & 23:57:41.3 & -32:35:58.5 &  6.1 $\pm$  0.1 & $\cdots$ & -0.56 $\pm$ 0.02 & -0.84 $\pm$ 0.03 & $\cdots$ & 0.80 & n & y & n & n & n & 28.94 $\pm$ 0.70 \\ 
SNR\_36 & 23:57:41.5 & -32:35:48.8 & 80.3 $\pm$  0.4 & 34.80 $\pm$ 0.24 & -0.89 $\pm$ 0.01 & -1.06 $\pm$ 0.01 & 0.76 $\pm$ 0.01 & 0.93 & n & n & y & y & y & 25.66 $\pm$ 0.38 \\ 
SNR\_37 & 23:57:41.5 & -32:34:44.8 &  6.3 $\pm$  0.1 & $\cdots$ & -0.51 $\pm$ 0.03 & -0.69 $\pm$ 0.03 & $\cdots$ & 1.27 & n & y & n & n & n & 23.48 $\pm$ 0.67 \\ 
SNR\_38 & 23:57:41.5 & -32:35:03.3 & 40.7 $\pm$  0.3 & 16.86 $\pm$ 0.19 & -0.17 $\pm$ 0.01 & -0.61 $\pm$ 0.01 & 0.12 $\pm$ 0.02 & 2.45 & y & y & y & y & y & 31.01 $\pm$ 0.71 \\ 
SNR\_39 & 23:57:41.6 & -32:35:49.9 & 49.8 $\pm$  0.3 & $\cdots$ & -0.52 $\pm$ 0.01 & -0.63 $\pm$ 0.01 & $\cdots$ & 0.91 & n & y & n & n & n & 26.64 $\pm$ 0.43 \\ 
SNR\_40 & 23:57:41.6 & -32:36:07.2 & 14.7 $\pm$  0.4 & $\cdots$ & -0.56 $\pm$ 0.03 & -0.87 $\pm$ 0.03 & $\cdots$ & 1.36 & n & y & n & n & n & 33.78 $\pm$ 0.94 \\ 
SNR\_41 & 23:57:41.7 & -32:34:51.0 & 43.1 $\pm$  0.2 & $\cdots$ & -0.63 $\pm$ 0.01 & -0.69 $\pm$ 0.01 & $\cdots$ & 0.80 & n & y & n & n & n & 17.20 $\pm$ 0.36 \\ 
SNR\_42 & 23:57:41.7 & -32:34:55.9 & 20.1 $\pm$  0.5 & 6.91 $\pm$ 0.33 & -0.53 $\pm$ 0.03 & -0.98 $\pm$ 0.03 & -0.18 $\pm$ 0.06 & 1.97 & n & y & n & n & y & 24.49 $\pm$ 0.68 \\ 
SNR\_43 & 23:57:41.7 & -32:34:39.1 &  2.4 $\pm$  0.1 & $\cdots$ & -0.51 $\pm$ 0.03 & -0.68 $\pm$ 0.04 & $\cdots$ & 0.80 & n & y & n & n & n & 23.35 $\pm$ 0.73 \\ 
SNR\_44 & 23:57:41.9 & -32:34:58.6 &  6.3 $\pm$  0.1 & $\cdots$ & -0.51 $\pm$ 0.02 & -0.62 $\pm$ 0.02 & $\cdots$ & 1.15 & n & y & n & n & n & 23.99 $\pm$ 0.62 \\ 
SNR\_45 & 23:57:41.9 & -32:35:38.1 & 13.0 $\pm$  0.2 & $\cdots$ & -0.27 $\pm$ 0.02 & -0.53 $\pm$ 0.03 & $\cdots$ & 1.53 & y & y & n & n & n & 27.94 $\pm$ 0.55 \\ 
SNR\_46 & 23:57:41.9 & -32:34:42.1 & 26.0 $\pm$  0.3 & $\cdots$ & -0.50 $\pm$ 0.01 & -0.89 $\pm$ 0.02 & $\cdots$ & 2.62 & n & y & n & n & n & 24.53 $\pm$ 0.67 \\ 
SNR\_47 & 23:57:42.0 & -32:35:18.2 &  3.2 $\pm$  0.1 & $\cdots$ & -0.46 $\pm$ 0.03 & -0.51 $\pm$ 0.04 & $\cdots$ & 0.80 & n & y & n & n & n & 25.15 $\pm$ 0.46 \\ 
SNR\_48 & 23:57:42.0 & -32:35:26.0 & 10.6 $\pm$  0.1 & $\cdots$ & -0.62 $\pm$ 0.01 & -0.64 $\pm$ 0.01 & $\cdots$ & 0.80 & n & y & n & n & n & 22.49 $\pm$ 0.50 \\ 
SNR\_49 & 23:57:42.0 & -32:35:14.0 & 45.2 $\pm$  0.2 & $\cdots$ & -0.65 $\pm$ 0.00 & -0.65 $\pm$ 0.01 & $\cdots$ & 1.85 & n & y & n & n & n & 20.55 $\pm$ 0.52 \\ 
SNR\_50 & 23:57:42.0 & -32:36:06.4 &  1.3 $\pm$  0.1 & $\cdots$ & -0.37 $\pm$ 0.06 & -0.72 $\pm$ 0.07 & $\cdots$ & 0.80 & y & y & n & n & n & 27.52 $\pm$ 0.70 \\ 
SNR\_51 & 23:57:42.1 & -32:35:33.8 & 15.4 $\pm$  0.2 & $\cdots$ & -0.62 $\pm$ 0.02 & -0.67 $\pm$ 0.02 & $\cdots$ & 0.80 & n & y & n & n & n & 21.22 $\pm$ 0.42 \\ 
SNR\_52 & 23:57:42.1 & -32:34:50.4 &  2.4 $\pm$  0.1 & $\cdots$ & -0.31 $\pm$ 0.04 & -0.71 $\pm$ 0.05 & $\cdots$ & 0.80 & y & y & n & n & n & 22.21 $\pm$ 0.60 \\ 
SNR\_53 & 23:57:42.1 & -32:35:21.7 & 22.2 $\pm$  0.2 & $\cdots$ & -0.60 $\pm$ 0.01 & -0.65 $\pm$ 0.01 & $\cdots$ & 1.15 & n & y & n & n & n & 25.93 $\pm$ 0.48 \\ 
SNR\_54 & 23:57:42.3 & -32:35:04.2 &  4.2 $\pm$  0.1 & 1.86 $\pm$ 0.07 & -0.24 $\pm$ 0.03 & -0.56 $\pm$ 0.03 & 0.05 $\pm$ 0.05 & 1.48 & y & y & y & n & y & 25.57 $\pm$ 0.68 \\ 
SNR\_55 & 23:57:42.4 & -32:35:18.5 & 92.0 $\pm$  0.3 & $\cdots$ & -0.64 $\pm$ 0.00 & -0.79 $\pm$ 0.00 & $\cdots$ & 2.50 & n & y & n & n & n & 20.74 $\pm$ 0.47 \\ 
SNR\_56 & 23:57:42.4 & -32:36:02.2 &  2.3 $\pm$  0.1 & $\cdots$ & -0.51 $\pm$ 0.03 & -0.72 $\pm$ 0.04 & $\cdots$ & 0.80 & n & y & n & n & n & 22.49 $\pm$ 0.54 \\ 
SNR\_57 & 23:57:42.7 & -32:36:12.2 &  2.4 $\pm$  0.1 & $\cdots$ & -0.57 $\pm$ 0.03 & -0.61 $\pm$ 0.03 & $\cdots$ & 0.80 & n & y & n & n & n & 19.49 $\pm$ 0.50 \\ 
SNR\_58 & 23:57:42.7 & -32:34:39.1 & 230.1 $\pm$  0.5 & $\cdots$ & -0.59 $\pm$ 0.00 & -0.69 $\pm$ 0.00 & $\cdots$ & 1.71 & n & y & n & n & n & 23.76 $\pm$ 0.53 \\ 
SNR\_59 & 23:57:42.7 & -32:35:26.0 &  3.7 $\pm$  0.1 & $\cdots$ & -0.45 $\pm$ 0.02 & -0.85 $\pm$ 0.03 & $\cdots$ & 0.80 & n & y & n & n & n & 27.73 $\pm$ 0.58 \\ 
SNR\_60 & 23:57:42.7 & -32:35:42.8 & 15.0 $\pm$  0.2 & $\cdots$ & -0.35 $\pm$ 0.01 & -0.48 $\pm$ 0.02 & $\cdots$ & 1.42 & y & y & n & n & n & 25.36 $\pm$ 0.54 \\ 
SNR\_61 & 23:57:42.8 & -32:35:29.5 & 17.2 $\pm$  0.1 & $\cdots$ & -0.62 $\pm$ 0.01 & -0.58 $\pm$ 0.01 & $\cdots$ & 1.07 & n & y & n & n & n & 25.53 $\pm$ 0.50 \\ 
SNR\_62 & 23:57:42.8 & -32:35:21.1 &  4.6 $\pm$  0.2 & $\cdots$ & -0.54 $\pm$ 0.06 & -0.45 $\pm$ 0.07 & $\cdots$ & 0.80 & n & y & n & n & n & 28.06 $\pm$ 0.66 \\ 
SNR\_63 & 23:57:42.9 & -32:34:56.9 &  2.7 $\pm$  0.1 & $\cdots$ & -0.61 $\pm$ 0.04 & -0.82 $\pm$ 0.04 & $\cdots$ & 0.80 & n & y & n & n & n & 24.11 $\pm$ 0.70 \\ 
SNR\_64 & 23:57:42.9 & -32:35:28.4 &  5.1 $\pm$  0.1 & 1.83 $\pm$ 0.05 & -0.41 $\pm$ 0.02 & -0.55 $\pm$ 0.03 & -0.01 $\pm$ 0.04 & 1.01 & n & y & y & n & y & 24.65 $\pm$ 0.52 \\ 
SNR\_65 & 23:57:42.9 & -32:35:13.3 &  3.3 $\pm$  0.1 & 1.33 $\pm$ 0.05 & -0.31 $\pm$ 0.03 & -0.60 $\pm$ 0.04 & -0.13 $\pm$ 0.05 & 1.02 & y & y & y & n & y & 26.17 $\pm$ 0.77 \\ 
SNR\_66 & 23:57:42.9 & -32:35:38.5 & 17.5 $\pm$  0.2 & $\cdots$ & -0.47 $\pm$ 0.01 & -0.52 $\pm$ 0.01 & $\cdots$ & 1.46 & n & y & n & n & n & 24.16 $\pm$ 0.53 \\ 
SNR\_67 & 23:57:43.0 & -32:36:18.9 &  3.6 $\pm$  0.1 & 1.27 $\pm$ 0.04 & -0.52 $\pm$ 0.02 & -0.65 $\pm$ 0.03 & 0.46 $\pm$ 0.04 & 0.80 & n & y & y & y & y & 29.88 $\pm$ 0.76 \\ 
SNR\_68 & 23:57:43.0 & -32:36:20.3 &  7.4 $\pm$  0.1 & $\cdots$ & -0.47 $\pm$ 0.02 & -0.63 $\pm$ 0.02 & $\cdots$ & 1.08 & n & y & n & n & n & 27.53 $\pm$ 0.70 \\ 
SNR\_69 & 23:57:43.0 & -32:34:19.7 & 11.1 $\pm$  0.3 & $\cdots$ & -0.28 $\pm$ 0.03 & -0.59 $\pm$ 0.04 & $\cdots$ & 1.85 & y & y & n & n & n & 27.45 $\pm$ 0.73 \\ 
SNR\_70 & 23:57:43.1 & -32:36:11.8 & 46.0 $\pm$  0.2 & 20.51 $\pm$ 0.14 & -0.59 $\pm$ 0.01 & -0.72 $\pm$ 0.01 & 0.66 $\pm$ 0.01 & 1.42 & n & y & y & y & y & 24.78 $\pm$ 0.52 \\ 
SNR\_71 & 23:57:43.2 & -32:35:13.5 & 14.0 $\pm$  0.1 & 5.92 $\pm$ 0.09 & -0.39 $\pm$ 0.01 & -0.55 $\pm$ 0.01 & -0.19 $\pm$ 0.02 & 1.47 & y & y & y & n & y & 30.62 $\pm$ 0.89 \\ 
SNR\_72 & 23:57:43.2 & -32:35:20.8 &  4.8 $\pm$  0.1 & $\cdots$ & -0.50 $\pm$ 0.02 & -0.74 $\pm$ 0.03 & $\cdots$ & 1.18 & n & y & n & n & n & 22.88 $\pm$ 0.56 \\ 
SNR\_73 & 23:57:43.2 & -32:35:24.3 &  0.9 $\pm$  0.1 & 0.33 $\pm$ 0.03 & -0.08 $\pm$ 0.08 & -0.72 $\pm$ 0.10 & -0.07 $\pm$ 0.14 & 0.80 & y & y & y & n & y & 22.95 $\pm$ 0.52 \\ 
SNR\_74 & 23:57:43.3 & -32:35:16.4 &  4.4 $\pm$  0.5 & $\cdots$ & -0.15 $\pm$ 0.15 & -0.45 $\pm$ 0.17 & $\cdots$ & 0.80 & y & y & n & n & n & 24.69 $\pm$ 0.64 \\ 
SNR\_75 & 23:57:43.3 & -32:35:00.2 &  2.5 $\pm$  0.1 & 0.88 $\pm$ 0.03 & -0.24 $\pm$ 0.03 & -0.64 $\pm$ 0.03 & -0.23 $\pm$ 0.05 & 1.23 & y & y & y & n & y & 27.05 $\pm$ 0.89 \\ 
SNR\_76 & 23:57:43.3 & -32:35:28.8 & 21.0 $\pm$  0.4 & 7.26 $\pm$ 0.27 & -0.01 $\pm$ 0.02 & -0.27 $\pm$ 0.03 & 0.19 $\pm$ 0.05 & 1.90 & y & y & y & y & y & 35.40 $\pm$ 0.78 \\ 
SNR\_77 & 23:57:43.3 & -32:35:52.7 & 98.3 $\pm$  0.3 & $\cdots$ & -0.64 $\pm$ 0.00 & -0.70 $\pm$ 0.00 & $\cdots$ & 1.02 & n & y & n & n & n & 23.34 $\pm$ 0.42 \\ 
SNR\_78 & 23:57:43.4 & -32:35:26.4 & 14.2 $\pm$  0.4 & $\cdots$ & -0.43 $\pm$ 0.04 & -0.64 $\pm$ 0.04 & $\cdots$ & 1.17 & n & y & n & n & n & 29.33 $\pm$ 0.60 \\ 
SNR\_79 & 23:57:43.4 & -32:34:43.8 & 19.7 $\pm$  0.2 & 7.68 $\pm$ 0.13 & -0.41 $\pm$ 0.01 & -0.60 $\pm$ 0.02 & 0.66 $\pm$ 0.02 & 1.48 & n & y & y & y & y & 19.73 $\pm$ 0.52 \\ 
SNR\_80 & 23:57:43.4 & -32:36:07.1 &  4.2 $\pm$  0.3 & $\cdots$ & -0.51 $\pm$ 0.07 & -0.69 $\pm$ 0.09 & $\cdots$ & 0.80 & n & y & n & n & n & 18.83 $\pm$ 0.59 \\ 
SNR\_81 & 23:57:43.4 & -32:35:40.4 &  3.7 $\pm$  0.1 & $\cdots$ & -0.42 $\pm$ 0.03 & -0.66 $\pm$ 0.03 & $\cdots$ & 0.80 & n & y & n & n & n & 22.31 $\pm$ 0.51 \\ 
SNR\_82 & 23:57:43.4 & -32:34:31.8 &  1.6 $\pm$  0.1 & $\cdots$ & -0.35 $\pm$ 0.04 & -0.68 $\pm$ 0.04 & $\cdots$ & 0.80 & y & y & n & n & n & 21.70 $\pm$ 0.83 \\ 
SNR\_83 & 23:57:43.5 & -32:35:22.2 & 81.7 $\pm$  0.3 & $\cdots$ & -0.66 $\pm$ 0.00 & -0.66 $\pm$ 0.01 & $\cdots$ & 1.33 & n & y & n & n & n & 21.16 $\pm$ 0.59 \\ 
SNR\_84 & 23:57:43.5 & -32:35:24.3 & 12.8 $\pm$  0.1 & $\cdots$ & -0.57 $\pm$ 0.01 & -0.75 $\pm$ 0.01 & $\cdots$ & 0.80 & n & y & n & n & n & 27.81 $\pm$ 0.56 \\ 
SNR\_85 & 23:57:43.6 & -32:35:19.0 & 42.7 $\pm$  0.2 & $\cdots$ & -0.58 $\pm$ 0.00 & -0.72 $\pm$ 0.01 & $\cdots$ & 0.94 & n & y & n & n & n & 24.71 $\pm$ 0.67 \\ 
SNR\_86 & 23:57:43.7 & -32:35:37.3 &  4.9 $\pm$  0.1 & $\cdots$ & -0.40 $\pm$ 0.03 & -0.44 $\pm$ 0.04 & $\cdots$ & 1.07 & n & y & n & n & n & 21.38 $\pm$ 0.48 \\ 
SNR\_87 & 23:57:43.7 & -32:34:39.0 & 63.6 $\pm$  0.2 & 22.30 $\pm$ 0.11 & -0.77 $\pm$ 0.00 & -0.84 $\pm$ 0.00 & 0.48 $\pm$ 0.01 & 0.80 & n & n & y & y & y & 14.80 $\pm$ 0.35 \\ 
SNR\_88 & 23:57:43.8 & -32:35:57.2 &  3.0 $\pm$  0.1 & $\cdots$ & -0.51 $\pm$ 0.03 & -0.61 $\pm$ 0.04 & $\cdots$ & 0.80 & n & y & n & n & n & 27.59 $\pm$ 0.80 \\ 
SNR\_89 & 23:57:43.8 & -32:35:27.6 & 39.7 $\pm$  0.3 & 13.88 $\pm$ 0.15 & 0.05 $\pm$ 0.01 & -0.22 $\pm$ 0.01 & 0.56 $\pm$ 0.01 & 1.73 & y & y & y & y & y & 55.88 $\pm$ 2.42 \\ 
SNR\_90 & 23:57:43.9 & -32:35:31.3 &  6.8 $\pm$  0.1 & 2.64 $\pm$ 0.08 & 0.14 $\pm$ 0.02 & -0.31 $\pm$ 0.03 & 0.21 $\pm$ 0.04 & 1.27 & y & y & y & y & y & 46.66 $\pm$ 2.36 \\ 
SNR\_91 & 23:57:44.0 & -32:34:41.6 & 71.0 $\pm$  0.3 & 32.48 $\pm$ 0.23 & -0.15 $\pm$ 0.01 & -0.34 $\pm$ 0.01 & 0.85 $\pm$ 0.01 & 1.56 & y & y & y & y & y & 34.96 $\pm$ 1.37  \\ 
SNR\_92 & 23:57:44.0 & -32:34:55.6 & 46.5 $\pm$  0.3 & 19.08 $\pm$ 0.21 & -0.35 $\pm$ 0.01 & -0.48 $\pm$ 0.01 & 0.57 $\pm$ 0.02 & 2.74 & y & y & y & y & y & 28.06 $\pm$ 0.78  \\ 
SNR\_93 & 23:57:44.0 & -32:35:13.6 &  5.1 $\pm$  0.2 & 1.75 $\pm$ 0.18 & -0.36 $\pm$ 0.06 & -0.74 $\pm$ 0.07 & -0.22 $\pm$ 0.14 & 0.80 & y & y & y & n & y & 25.16 $\pm$ 0.55 \\ 
SNR\_94 & 23:57:44.0 & -32:35:39.9 & 18.5 $\pm$  0.2 & $\cdots$ & -0.46 $\pm$ 0.01 & -0.53 $\pm$ 0.01 & $\cdots$ & 1.31 & n & y & n & n & n &  21.67 $\pm$ 0.61\\ 
SNR\_95 & 23:57:44.1 & -32:35:17.8 & 13.0 $\pm$  0.2 & 4.48 $\pm$ 0.12 & -0.45 $\pm$ 0.02 & -0.55 $\pm$ 0.02 & -0.02 $\pm$ 0.04 & 0.80 & n & y & y & n & y & 30.44 $\pm$ 0.88 \\ 
SNR\_96 & 23:57:44.2 & -32:34:35.9 & 14.9 $\pm$  0.1 & $\cdots$ & -0.56 $\pm$ 0.01 & -0.66 $\pm$ 0.01 & $\cdots$ & 0.80 & n & y & n & n & n & 22.16 $\pm$ 0.55 \\ 
SNR\_97 & 23:57:44.2 & -32:34:45.1 & 27.5 $\pm$  0.2 & $\cdots$ & -0.43 $\pm$ 0.01 & -0.68 $\pm$ 0.01 & $\cdots$ & 1.91 & n & y & n & n & n & 22.66 $\pm$ 0.59 \\ 
SNR\_98 & 23:57:44.2 & -32:35:32.0 &  3.7 $\pm$  0.1 & 1.66 $\pm$ 0.05 & 0.12 $\pm$ 0.02 & -0.29 $\pm$ 0.03 & -0.06 $\pm$ 0.04 & 0.80 & y & y & y & y & y & 42.81 $\pm$ 2.07 \\ 
SNR\_99 & 23:57:44.4 & -32:35:25.1 & 79.4 $\pm$  0.4 & $\cdots$ & -0.62 $\pm$ 0.01 & -0.57 $\pm$ 0.01 & $\cdots$ & 2.61 & n & y & n & n & n & 23.64 $\pm$ 0.51 \\ 
SNR\_100 & 23:57:44.4 & -32:34:37.1 & 73.3 $\pm$  0.3 & $\cdots$ & -0.55 $\pm$ 0.00 & -0.74 $\pm$ 0.01 & $\cdots$ & 1.44 & n & y & n & n & n & 20.60 $\pm$ 0.53 \\ 
SNR\_101 & 23:57:44.4 & -32:35:10.3 &  8.3 $\pm$  0.1 & 2.88 $\pm$ 0.08 & -0.30 $\pm$ 0.02 & -0.55 $\pm$ 0.02 & -0.26 $\pm$ 0.04 & 0.94 & y & y & y & n & y & 25.30 $\pm$ 0.74 \\ 
SNR\_102 & 23:57:44.5 & -32:34:39.1 &  7.6 $\pm$  0.2 & $\cdots$ & -0.61 $\pm$ 0.03 & -0.57 $\pm$ 0.04 & $\cdots$ & 0.80 & n & y & n & n & n & 22.15 $\pm$ 0.56 \\ 
SNR\_103 & 23:57:44.5 & -32:34:57.1 &  3.4 $\pm$  0.1 & $\cdots$ & -0.51 $\pm$ 0.02 & -0.71 $\pm$ 0.03 & $\cdots$ & 0.80 & n & y & n & n & n & 23.71 $\pm$ 0.64 \\ 
SNR\_104 & 23:57:44.6 & -32:34:52.7 &  6.3 $\pm$  0.4 & 2.15 $\pm$ 0.31 & -0.34 $\pm$ 0.08 & -0.63 $\pm$ 0.09 & -0.03 $\pm$ 0.20 & 1.44 & y & y & y & n & y & 21.89 $\pm$ 0.70 \\ 
SNR\_105 & 23:57:44.6 & -32:35:32.2 & 37.7 $\pm$  0.2 & 16.22 $\pm$ 0.15 & -0.58 $\pm$ 0.01 & -0.67 $\pm$ 0.01 & 0.03 $\pm$ 0.01 & 1.97 & n & y & y & n & n & 23.12 $\pm$ 0.52 \\ 
SNR\_106 & 23:57:44.7 & -32:34:36.5 & 11.3 $\pm$  0.7 & 3.82 $\pm$ 0.55 & -0.51 $\pm$ 0.07 & -0.58 $\pm$ 0.08 & -0.00 $\pm$ 0.19 & 0.80 & n & y & y & n & y & 26.11 $\pm$ 0.66 \\ 
SNR\_107 & 23:57:44.7 & -32:34:39.9 & 55.5 $\pm$  0.4 & $\cdots$ & -0.60 $\pm$ 0.01 & -0.67 $\pm$ 0.01 & $\cdots$ & 1.30 & n & y & n & n & n & 20.06 $\pm$ 0.48 \\ 
SNR\_108 & 23:57:44.8 & -32:35:35.4 & 14.1 $\pm$  0.1 & $\cdots$ & -0.10 $\pm$ 0.01 & -0.54 $\pm$ 0.01 & $\cdots$ & 1.28 & y & y & n & n & n & 32.75 $\pm$ 0.65 \\ 
SNR\_109 & 23:57:44.8 & -32:34:57.7 &  6.7 $\pm$  0.1 & $\cdots$ & -0.36 $\pm$ 0.02 & -0.60 $\pm$ 0.03 & $\cdots$ & 1.10 & y & y & n & n & n & 25.99 $\pm$ 0.72 \\ 
SNR\_110 & 23:57:44.9 & -32:34:38.2 & 30.7 $\pm$  0.2 & $\cdots$ & -0.53 $\pm$ 0.01 & -0.67 $\pm$ 0.01 & $\cdots$ & 0.80 & n & y & n & n & n & 22.88 $\pm$ 0.52 \\ 
SNR\_111 & 23:57:44.9 & -32:35:03.6 & 33.9 $\pm$  0.2 & $\cdots$ & -0.47 $\pm$ 0.01 & -0.62 $\pm$ 0.01 & $\cdots$ & 2.01 & n & y & n & n & n & 20.43 $\pm$ 0.57 \\ 
SNR\_112 & 23:57:45.0 & -32:34:50.7 &  1.7 $\pm$  0.1 & $\cdots$ & -0.38 $\pm$ 0.04 & -0.57 $\pm$ 0.05 & $\cdots$ & 0.80 & y & y & n & n & n & 21.33 $\pm$ 0.68 \\ 
SNR\_113 & 23:57:45.0 & -32:35:20.3 &  6.6 $\pm$  0.1 & $\cdots$ & -0.44 $\pm$ 0.02 & -0.76 $\pm$ 0.02 & $\cdots$ & 0.80 & n & y & n & n & n & 25.98 $\pm$ 0.56 \\ 
SNR\_114 & 23:57:45.0 & -32:35:14.8 & 50.4 $\pm$  0.2 & $\cdots$ & -0.40 $\pm$ 0.01 & -0.60 $\pm$ 0.01 & $\cdots$ & 1.68 & n & y & n & n & n & 26.48 $\pm$ 0.54 \\ 
SNR\_115 & 23:57:45.2 & -32:34:28.7 & 14.7 $\pm$  0.3 & 7.32 $\pm$ 0.22 & -0.39 $\pm$ 0.02 & -0.58 $\pm$ 0.03 & -0.14 $\pm$ 0.04 & 1.77 & y & y & y & n & y & 26.62 $\pm$ 0.72 \\ 
SNR\_116 & 23:57:45.2 & -32:35:00.2 &  6.5 $\pm$  0.1 & $\cdots$ & -0.59 $\pm$ 0.02 & -0.69 $\pm$ 0.02 & $\cdots$ & 1.06 & n & y & n & n & n & 23.36 $\pm$ 0.64 \\ 
SNR\_117 & 23:57:45.2 & -32:34:57.8 & 21.3 $\pm$  0.1 & $\cdots$ & -0.50 $\pm$ 0.01 & -0.57 $\pm$ 0.01 & $\cdots$ & 1.36 & n & y & n & n & n & 24.82 $\pm$ 0.61 \\ 
SNR\_118 & 23:57:45.2 & -32:35:44.4 &  6.9 $\pm$  0.2 & $\cdots$ & -0.59 $\pm$ 0.03 & -0.80 $\pm$ 0.03 & $\cdots$ & 0.80 & n & y & n & n & n & 31.21 $\pm$ 0.67 \\ 
SNR\_119 & 23:57:45.2 & -32:34:41.8 &  4.6 $\pm$  0.1 & $\cdots$ & -0.50 $\pm$ 0.02 & -0.72 $\pm$ 0.03 & $\cdots$ & 0.80 & n & y & n & n & n & 24.48 $\pm$ 0.66 \\ 
SNR\_120 & 23:57:45.3 & -32:34:52.4 &  4.8 $\pm$  0.1 & 2.10 $\pm$ 0.06 & -0.23 $\pm$ 0.02 & -0.52 $\pm$ 0.03 & 0.26 $\pm$ 0.04 & 1.12 & y & y & y & y & y & 29.97 $\pm$ 0.78 \\ 
SNR\_121 & 23:57:45.4 & -32:34:46.3 & 101.4 $\pm$  0.4 & $\cdots$ & -0.64 $\pm$ 0.00 & -0.65 $\pm$ 0.01 & $\cdots$ & 2.45 & n & y & n & n & n & 22.57 $\pm$ 0.59 \\ 
SNR\_122 & 23:57:45.4 & -32:35:48.3 &  9.1 $\pm$  0.1 & $\cdots$ & -0.45 $\pm$ 0.02 & -0.63 $\pm$ 0.02 & $\cdots$ & 1.04 & n & y & n & n & n & 29.36 $\pm$ 0.75 \\ 
SNR\_123 & 23:57:45.5 & -32:34:57.9 &  3.2 $\pm$  0.1 & $\cdots$ & -0.58 $\pm$ 0.05 & -0.62 $\pm$ 0.06 & $\cdots$ & 0.98 & n & y & n & n & n & 23.36 $\pm$ 0.74 \\ 
SNR\_124 & 23:57:45.5 & -32:34:55.7 & 21.2 $\pm$  0.2 & $\cdots$ & -0.56 $\pm$ 0.01 & -0.59 $\pm$ 0.01 & $\cdots$ & 1.34 & n & y & n & n & n & 22.61 $\pm$ 0.65 \\ 
SNR\_125 & 23:57:45.6 & -32:34:59.9 &  5.5 $\pm$  0.1 & $\cdots$ & -0.50 $\pm$ 0.02 & -0.74 $\pm$ 0.02 & $\cdots$ & 0.80 & n & y & n & n & n & 19.98 $\pm$ 0.54 \\ 
SNR\_126 & 23:57:45.7 & -32:35:01.6 & 64.9 $\pm$  0.3 & 23.82 $\pm$ 0.15 & 0.03 $\pm$ 0.00 & -0.30 $\pm$ 0.01 & 0.39 $\pm$ 0.01 & 1.67 & y & y & y & y & y & 58.00 $\pm$ 2.46 \\ 
SNR\_127 & 23:57:45.8 & -32:34:48.2 &  3.3 $\pm$  0.1 & $\cdots$ & -0.62 $\pm$ 0.03 & -0.81 $\pm$ 0.03 & $\cdots$ & 0.80 & n & y & n & n & n & 24.34 $\pm$ 0.63 \\ 
SNR\_128 & 23:57:45.9 & -32:35:03.5 & 12.1 $\pm$  0.3 & 4.18 $\pm$ 0.21 & -0.18 $\pm$ 0.03 & -0.43 $\pm$ 0.04 & 0.35 $\pm$ 0.07 & 1.38 & y & y & y & y & y & 34.23 $\pm$ 0.89 \\ 
SNR\_129 & 23:57:45.9 & -32:34:57.7 &  1.8 $\pm$  0.1 & $\cdots$ & -0.65 $\pm$ 0.06 & -0.57 $\pm$ 0.07 & $\cdots$ & 0.80 & n & y & n & n & n & 30.48 $\pm$ 1.20 \\ 
SNR\_130 & 23:57:46.2 & -32:34:59.9 & 20.7 $\pm$  0.2 & $\cdots$ & -0.61 $\pm$ 0.01 & -0.78 $\pm$ 0.01 & $\cdots$ & 0.80 & n & y & n & n & n & 36.07 $\pm$ 1.33 \\ 
SNR\_131 & 23:57:46.5 & -32:35:46.3 & 12.0 $\pm$  0.4 & $\cdots$ & -0.63 $\pm$ 0.04 & -0.53 $\pm$ 0.05 & $\cdots$ & 1.17 & n & y & n & n & n & 28.06 $\pm$ 0.93 \\ 
SNR\_132 & 23:57:46.7 & -32:35:22.4 & 43.5 $\pm$  0.4 & $\cdots$ & -0.49 $\pm$ 0.01 & -0.56 $\pm$ 0.01 & $\cdots$ & 2.78 & n & y & n & n & n & 23.66 $\pm$ 0.64 \\ 
SNR\_133 & 23:57:46.7 & -32:34:52.6 &  2.5 $\pm$  0.1 & $\cdots$ & -0.40 $\pm$ 0.04 & -0.66 $\pm$ 0.04 & $\cdots$ & 0.80 & n & y & n & n & n & 24.90 $\pm$ 0.67 \\ 
SNR\_134 & 23:57:46.8 & -32:35:10.3 &  8.0 $\pm$  0.2 & $\cdots$ & -0.63 $\pm$ 0.03 & -0.46 $\pm$ 0.03 & $\cdots$ & 1.49 & n & y & n & n & n & 23.25 $\pm$ 0.80 \\ 
SNR\_135 & 23:57:47.0 & -32:35:47.8 & 13.9 $\pm$  0.2 & $\cdots$ & -0.44 $\pm$ 0.02 & -0.47 $\pm$ 0.02 & $\cdots$ & 1.74 & n & y & n & n & n & 27.23 $\pm$ 0.85 \\ 
SNR\_136 & 23:57:47.2 & -32:35:23.5 & 40.0 $\pm$  0.2 & 13.99 $\pm$ 0.11 & 0.06 $\pm$ 0.01 & -0.20 $\pm$ 0.01 & 0.26 $\pm$ 0.01 & 1.01 & y & y & y & y & y & 69.53 $\pm$ 4.53 \\ 
SNR\_137 & 23:57:47.4 & -32:35:24.7 & 29.0 $\pm$  0.2 & $\cdots$ & -0.50 $\pm$ 0.01 & -0.51 $\pm$ 0.01 & $\cdots$ & 1.03 & n & y & n & n & n & 24.50 $\pm$ 0.61 \\ 
SNR\_138 & 23:57:47.4 & -32:35:00.1 & 15.0 $\pm$  0.2 & $\cdots$ & -0.59 $\pm$ 0.01 & -0.51 $\pm$ 0.02 & $\cdots$ & 1.33 & n & y & n & n & n & 24.09 $\pm$ 0.70 \\ 
SNR\_139 & 23:57:48.0 & -32:35:41.5 &  7.1 $\pm$  0.1 & $\cdots$ & -0.46 $\pm$ 0.02 & -0.60 $\pm$ 0.03 & $\cdots$ & 1.20 & n & y & n & n & n & 25.36 $\pm$ 0.75 \\ 
SNR\_140 & 23:57:48.2 & -32:35:30.1 &  1.9 $\pm$  0.1 & $\cdots$ & -0.31 $\pm$ 0.06 & -0.55 $\pm$ 0.07 & $\cdots$ & 0.80 & y & y & n & n & n & 27.26 $\pm$ 0.81 \\ 
SNR\_141 & 23:57:48.2 & -32:35:31.7 &  2.4 $\pm$  0.1 & $\cdots$ & -0.40 $\pm$ 0.05 & -0.67 $\pm$ 0.06 & $\cdots$ & 0.80 & n & y & n & n & n & 24.28 $\pm$ 0.63 \\ 
SNR\_142 & 23:57:48.3 & -32:35:13.9 & 36.6 $\pm$  0.2 & $\cdots$ & -0.68 $\pm$ 0.01 & -0.50 $\pm$ 0.01 & $\cdots$ & 1.63 & n & y & n & n & n & 22.05 $\pm$ 0.58 \\ 
SNR\_143 & 23:57:48.4 & -32:35:10.6 &  7.5 $\pm$  0.2 & $\cdots$ & -0.63 $\pm$ 0.02 & -0.49 $\pm$ 0.03 & $\cdots$ & 1.28 & n & y & n & n & n & 22.04 $\pm$ 0.72 \\ 
SNR\_144 & 23:57:48.7 & -32:35:25.6 & 34.3 $\pm$  0.3 & $\cdots$ & -0.65 $\pm$ 0.01 & -0.62 $\pm$ 0.01 & $\cdots$ & 1.82 & n & y & n & n & n & 20.68 $\pm$ 0.57 \\ 
SNR\_145 & 23:57:49.1 & -32:35:39.1 & 20.8 $\pm$  0.2 & $\cdots$ & -0.55 $\pm$ 0.01 & -0.57 $\pm$ 0.02 & $\cdots$ & 1.75 & n & y & n & n & n & 27.59 $\pm$ 0.78 \\ 
SNR\_146 & 23:57:49.2 & -32:35:16.4 & 33.3 $\pm$  0.3 & $\cdots$ & -0.60 $\pm$ 0.01 & -0.57 $\pm$ 0.01 & $\cdots$ & 1.83 & n & y & n & n & n & 24.84 $\pm$ 0.68 \\ 
SNR\_147 & 23:57:49.6 & -32:35:45.7 & 32.2 $\pm$  0.2 & $\cdots$ & -0.67 $\pm$ 0.01 & -0.54 $\pm$ 0.01 & $\cdots$ & 1.22 & n & y & n & n & n & 28.51 $\pm$ 0.72 \\ 
SNR\_148 & 23:57:49.9 & -32:35:45.9 & 54.6 $\pm$  0.5 & $\cdots$ & -0.61 $\pm$ 0.01 & -0.59 $\pm$ 0.01 & $\cdots$ & 1.38 & n & y & n & n & n & 28.90 $\pm$ 0.76 \\ 
SNR\_149 & 23:57:49.9 & -32:35:00.8 & 38.1 $\pm$  0.3 & $\cdots$ & -0.62 $\pm$ 0.01 & -0.70 $\pm$ 0.01 & $\cdots$ & 1.60 & n & y & n & n & n & 21.68 $\pm$ 0.55 \\ 
SNR\_150 & 23:57:50.0 & -32:35:09.2 & 12.0 $\pm$  0.1 & $\cdots$ & -0.61 $\pm$ 0.01 & -0.52 $\pm$ 0.02 & $\cdots$ & 0.86 & n & y & n & n & n & 27.50 $\pm$ 0.71 \\ 
SNR\_151 & 23:57:50.1 & -32:35:06.1 & 12.5 $\pm$  0.1 & $\cdots$ & -0.56 $\pm$ 0.01 & -0.59 $\pm$ 0.01 & $\cdots$ & 1.28 & n & y & n & n & n & 25.66 $\pm$ 0.63 \\ 
SNR\_152 & 23:57:50.1 & -32:35:44.8 & 151.1 $\pm$  0.4 & $\cdots$ & -0.59 $\pm$ 0.00 & -0.55 $\pm$ 0.00 & $\cdots$ & 1.68 & n & y & n & n & n & 27.78 $\pm$ 0.60 \\ 
SNR\_153 & 23:57:50.1 & -32:35:32.6 & 10.0 $\pm$  0.1 & $\cdots$ & -0.56 $\pm$ 0.01 & -0.63 $\pm$ 0.02 & $\cdots$ & 1.23 & n & y & n & n & n & 23.34 $\pm$ 0.60 \\ 
SNR\_154 & 23:57:50.2 & -32:35:12.9 &  3.9 $\pm$  0.1 & $\cdots$ & -0.54 $\pm$ 0.02 & -0.76 $\pm$ 0.03 & $\cdots$ & 1.01 & n & y & n & n & n & 26.25 $\pm$ 0.69 \\ 
SNR\_155 & 23:57:50.2 & -32:35:03.8 & 28.9 $\pm$  0.2 & $\cdots$ & -0.53 $\pm$ 0.01 & -0.58 $\pm$ 0.01 & $\cdots$ & 1.68 & n & y & n & n & n & 27.94 $\pm$ 0.72 \\ 
SNR\_156 & 23:57:50.3 & -32:34:57.6 &  6.9 $\pm$  0.1 & 2.39 $\pm$ 0.04 & -0.44 $\pm$ 0.01 & -0.51 $\pm$ 0.01 & 0.66 $\pm$ 0.02 & 1.26 & n & y & y & y & y & 25.89 $\pm$ 0.77 \\ 
SNR\_157 & 23:57:50.3 & -32:35:23.3 & 11.0 $\pm$  0.1 & $\cdots$ & -0.62 $\pm$ 0.01 & -0.69 $\pm$ 0.02 & $\cdots$ & 1.51 & n & y & n & n & n & 24.36 $\pm$ 0.63 \\ 
SNR\_158 & 23:57:50.4 & -32:35:16.4 &  4.8 $\pm$  0.1 & $\cdots$ & -0.57 $\pm$ 0.02 & -0.73 $\pm$ 0.02 & $\cdots$ & 1.26 & n & y & n & n & n & 25.99 $\pm$ 0.69 \\ 
SNR\_159 & 23:57:50.4 & -32:35:10.5 & 54.1 $\pm$  0.2 & $\cdots$ & -0.62 $\pm$ 0.00 & -0.64 $\pm$ 0.01 & $\cdots$ & 2.00 & n & y & n & n & n & 22.93 $\pm$ 0.60 \\ 
SNR\_160 & 23:57:50.6 & -32:35:31.6 & 25.9 $\pm$  0.2 & 9.05 $\pm$ 0.10 & -0.53 $\pm$ 0.01 & -0.65 $\pm$ 0.01 & 0.20 $\pm$ 0.02 & 1.53 & n & y & y & y & y & 20.03 $\pm$ 0.47 \\ 
SNR\_161 & 23:57:50.7 & -32:35:25.6 & 209.0 $\pm$  0.3 & $\cdots$ & -0.55 $\pm$ 0.00 & -0.54 $\pm$ 0.00 & $\cdots$ & 2.00 & n & y & n & n & n & 24.30 $\pm$ 0.51 \\ 
SNR\_162 & 23:57:50.7 & -32:35:29.0 & 94.6 $\pm$  0.3 & $\cdots$ & -0.66 $\pm$ 0.00 & -0.56 $\pm$ 0.00 & $\cdots$ & 1.54 & n & y & n & n & n & 21.83 $\pm$ 0.46 \\ 
SNR\_163 & 23:57:50.8 & -32:35:18.0 & 23.1 $\pm$  0.2 & 8.03 $\pm$ 0.14 & -0.17 $\pm$ 0.01 & -0.47 $\pm$ 0.01 & 0.02 $\pm$ 0.02 & 2.22 & y & y & y & y & y & 26.65 $\pm$ 0.64 \\ 
SNR\_164 & 23:57:50.9 & -32:35:35.1 &  8.4 $\pm$  0.2 & 3.49 $\pm$ 0.10 & -0.43 $\pm$ 0.02 & -0.49 $\pm$ 0.03 & 0.11 $\pm$ 0.04 & 2.00 & n & y & y & y & y & 24.67 $\pm$ 0.61 \\ 
SNR\_165 & 23:57:50.9 & -32:35:46.1 & 18.0 $\pm$  0.1 & $\cdots$ & -0.56 $\pm$ 0.01 & -0.71 $\pm$ 0.01 & $\cdots$ & 1.47 & n & y & n & n & n & 24.79 $\pm$ 0.55 \\ 
SNR\_166 & 23:57:50.9 & -32:35:49.9 & 91.4 $\pm$  0.3 & $\cdots$ & -0.66 $\pm$ 0.00 & -0.58 $\pm$ 0.00 & $\cdots$ & 1.85 & n & y & n & n & n & 24.80 $\pm$ 0.53 \\ 
SNR\_167 & 23:57:51.0 & -32:35:48.2 & 13.7 $\pm$  0.1 & $\cdots$ & -0.55 $\pm$ 0.01 & -0.58 $\pm$ 0.01 & $\cdots$ & 0.95 & n & y & n & n & n & 24.23 $\pm$ 0.56 \\ 
SNR\_168 & 23:57:51.2 & -32:35:27.7 & 63.8 $\pm$  3.5 & 20.90 $\pm$ 4.40 & -0.29 $\pm$ 0.07 & -0.62 $\pm$ 0.08 & 0.16 $\pm$ 0.28 & 1.53 & y & y & y & y & y & 27.86 $\pm$ 0.70 \\ 
SNR\_169 & 23:57:51.3 & -32:34:56.6 &  6.8 $\pm$  0.1 & 2.87 $\pm$ 0.07 & -0.39 $\pm$ 0.02 & -0.43 $\pm$ 0.02 & 0.18 $\pm$ 0.03 & 1.28 & y & y & y & y & y & 23.79 $\pm$ 0.55 \\ 
SNR\_170 & 23:57:51.3 & -32:35:09.3 & 38.7 $\pm$  0.2 & $\cdots$ & -0.32 $\pm$ 0.01 & -0.50 $\pm$ 0.01 & $\cdots$ & 2.00 & y & y & n & n & n & 24.56 $\pm$ 0.56 \\ 
SNR\_171 & 23:57:51.3 & -32:35:39.6 & 87.6 $\pm$  0.3 & 31.05 $\pm$ 0.14 & -0.72 $\pm$ 0.00 & -0.46 $\pm$ 0.00 & 0.37 $\pm$ 0.01 & 2.00 & n & n & y & y & y & 24.13 $\pm$ 0.49 \\ 
SNR\_172 & 23:57:51.4 & -32:35:23.6 & 106.5 $\pm$  0.3 & $\cdots$ & -0.47 $\pm$ 0.00 & -0.63 $\pm$ 0.00 & $\cdots$ & 2.58 & n & y & n & n & n & 27.04 $\pm$ 0.58 \\ 
SNR\_173 & 23:57:51.4 & -32:35:26.6 & 17.7 $\pm$  0.1 & 6.16 $\pm$ 0.15 & -0.59 $\pm$ 0.01 & -0.53 $\pm$ 0.01 & -0.01 $\pm$ 0.03 & 0.91 & n & y & n & n & y & 25.64 $\pm$ 0.52 \\ 
SNR\_174 & 23:57:51.5 & -32:35:40.6 & 134.8 $\pm$  0.3 & 46.91 $\pm$ 0.17 & -0.77 $\pm$ 0.00 & -0.59 $\pm$ 0.00 & 0.26 $\pm$ 0.01 & 2.00 & n & n & y & y & n & 24.05 $\pm$ 0.46 \\ 
SNR\_175 & 23:57:51.6 & -32:35:08.1 &  5.1 $\pm$  0.1 & $\cdots$ & -0.37 $\pm$ 0.03 & -0.77 $\pm$ 0.04 & $\cdots$ & 1.16 & y & y & n & n & n & 26.53 $\pm$ 0.63 \\ 
SNR\_176 & 23:57:51.8 & -32:35:33.1 & 15.4 $\pm$  0.2 & $\cdots$ & -0.61 $\pm$ 0.02 & -0.68 $\pm$ 0.02 & $\cdots$ & 2.00 & n & y & n & n & n & 23.01 $\pm$ 0.50 \\ 
SNR\_177 & 23:57:51.8 & -32:35:12.6 & 39.9 $\pm$  0.2 & $\cdots$ & -0.60 $\pm$ 0.01 & -0.74 $\pm$ 0.01 & $\cdots$ & 1.24 & n & y & n & n & n & 24.60 $\pm$ 0.53 \\ 
SNR\_178 & 23:57:52.1 & -32:35:26.2 & 19.4 $\pm$  0.2 & $\cdots$ & -0.48 $\pm$ 0.01 & -0.64 $\pm$ 0.01 & $\cdots$ & 1.52 & n & y & n & n & n & 20.77 $\pm$ 0.43 \\ 
SNR\_179 & 23:57:52.2 & -32:34:58.9 & 15.5 $\pm$  0.1 & $\cdots$ & -0.60 $\pm$ 0.01 & -0.74 $\pm$ 0.01 & $\cdots$ & 2.26 & n & y & n & n & n & 24.69 $\pm$ 0.60 \\ 
SNR\_180 & 23:57:52.2 & -32:35:04.1 & 13.2 $\pm$  0.1 & 5.18 $\pm$ 0.05 & -0.37 $\pm$ 0.01 & -0.48 $\pm$ 0.01 & 0.05 $\pm$ 0.01 & 1.40 & y & y & y & y & y & 26.10 $\pm$ 0.61 \\ 
SNR\_181 & 23:57:52.3 & -32:35:31.9 &  5.7 $\pm$  0.1 & $\cdots$ & -0.54 $\pm$ 0.02 & -0.60 $\pm$ 0.02 & $\cdots$ & 1.04 & n & y & n & n & n & 23.51 $\pm$ 0.52 \\ 
SNR\_182 & 23:57:52.3 & -32:35:05.2 &  5.5 $\pm$  0.1 & $\cdots$ & -0.54 $\pm$ 0.01 & -0.63 $\pm$ 0.02 & $\cdots$ & 1.03 & n & y & n & n & n & 21.25 $\pm$ 0.46 \\ 
SNR\_183 & 23:57:52.4 & -32:35:03.4 &  1.7 $\pm$  0.1 & $\cdots$ & -0.70 $\pm$ 0.04 & -1.31 $\pm$ 0.04 & $\cdots$ & 0.78 & n & y & n & n & n & 33.01 $\pm$ 1.26 \\ 
SNR\_184 & 23:57:52.5 & -32:35:33.3 & 21.6 $\pm$  0.3 & $\cdots$ & -0.54 $\pm$ 0.02 & -0.70 $\pm$ 0.02 & $\cdots$ & 1.51 & n & y & n & n & n & 23.00 $\pm$ 0.49 \\ 
SNR\_185 & 23:57:52.5 & -32:35:06.1 & 13.7 $\pm$  0.1 & $\cdots$ & -0.45 $\pm$ 0.01 & -0.67 $\pm$ 0.01 & $\cdots$ & 1.21 & n & y & n & n & n & 22.34 $\pm$ 0.51 \\ 
SNR\_186 & 23:57:52.6 & -32:35:41.5 &  7.9 $\pm$  0.1 & $\cdots$ & -0.52 $\pm$ 0.01 & -0.75 $\pm$ 0.01 & $\cdots$ & 1.31 & n & y & n & n & n & 26.28 $\pm$ 0.61 \\ 
SNR\_187 & 23:57:52.7 & -32:35:09.7 & 13.3 $\pm$  0.2 & 4.65 $\pm$ 0.10 & 0.02 $\pm$ 0.02 & -0.41 $\pm$ 0.02 & 0.03 $\pm$ 0.03 & 2.49 & y & y & y & y & y & 28.46 $\pm$ 0.71 \\ 
SNR\_188 & 23:57:53.0 & -32:35:13.2 & 166.1 $\pm$  0.2 & $\cdots$ & -0.54 $\pm$ 0.00 & -0.54 $\pm$ 0.00 & $\cdots$ & 2.00 & n & y & n & n & n & 24.84 $\pm$ 0.45 \\ 
SNR\_189 & 23:57:53.1 & -32:35:19.9 & 12.6 $\pm$  0.1 & $\cdots$ & -0.61 $\pm$ 0.01 & -0.59 $\pm$ 0.01 & $\cdots$ & 0.82 & n & y & n & n & n & 18.17 $\pm$ 0.40 \\ 
SNR\_190 & 23:57:53.1 & -32:35:17.0 & 24.3 $\pm$  0.1 & $\cdots$ & -0.60 $\pm$ 0.01 & -0.63 $\pm$ 0.01 & $\cdots$ & 1.26 & n & y & n & n & n & 18.97 $\pm$ 0.42 \\ 
SNR\_191 & 23:57:53.2 & -32:35:19.0 & 15.6 $\pm$  0.1 & $\cdots$ & -0.61 $\pm$ 0.01 & -0.58 $\pm$ 0.01 & $\cdots$ & 0.79 & n & y & n & n & n & 18.38 $\pm$ 0.38 \\ 
SNR\_192 & 23:57:53.3 & -32:35:05.2 & 34.0 $\pm$  0.1 & $\cdots$ & -0.65 $\pm$ 0.00 & -0.58 $\pm$ 0.00 & $\cdots$ & 1.66 & n & y & n & n & n & 23.85 $\pm$ 0.50 \\ 
SNR\_193 & 23:57:53.5 & -32:35:13.5 & 17.8 $\pm$  0.2 & $\cdots$ & -0.55 $\pm$ 0.01 & -0.62 $\pm$ 0.01 & $\cdots$ & 2.00 & n & y & n & n & n & 23.61 $\pm$ 0.54 \\ 
SNR\_194 & 23:57:53.5 & -32:35:48.4 & 27.5 $\pm$  0.2 & $\cdots$ & -0.55 $\pm$ 0.01 & -0.58 $\pm$ 0.01 & $\cdots$ & 1.16 & n & y & n & n & n & 27.05 $\pm$ 0.51 \\ 
SNR\_195 & 23:57:53.5 & -32:35:36.6 & 28.6 $\pm$  0.1 & $\cdots$ & -0.59 $\pm$ 0.01 & -0.57 $\pm$ 0.01 & $\cdots$ & 1.35 & n & y & n & n & n & 26.14 $\pm$ 0.53 \\ 
SNR\_196 & 23:57:53.6 & -32:34:53.6 & 73.2 $\pm$  0.2 & $\cdots$ & -0.58 $\pm$ 0.00 & -0.64 $\pm$ 0.00 & $\cdots$ & 2.03 & n & y & n & n & n & 28.58 $\pm$ 0.60 \\ 
SNR\_197 & 23:57:53.7 & -32:35:45.8 & 57.9 $\pm$  0.2 & $\cdots$ & -0.66 $\pm$ 0.00 & -0.61 $\pm$ 0.01 & $\cdots$ & 0.83 & n & y & n & n & n & 29.41 $\pm$ 0.52 \\ 
SNR\_198 & 23:57:53.7 & -32:35:15.8 & 17.6 $\pm$  0.2 & $\cdots$ & -0.32 $\pm$ 0.01 & -0.50 $\pm$ 0.01 & $\cdots$ & 2.00 & y & y & n & n & n & 27.66 $\pm$ 0.58 \\ 
SNR\_199 & 23:57:53.7 & -32:35:41.2 & 95.4 $\pm$  0.3 & $\cdots$ & -0.48 $\pm$ 0.00 & -0.67 $\pm$ 0.00 & $\cdots$ & 2.00 & n & y & n & n & n & 26.57 $\pm$ 0.49 \\ 
SNR\_200 & 23:57:54.2 & -32:35:44.9 & 30.5 $\pm$  0.2 & $\cdots$ & -0.57 $\pm$ 0.01 & -0.70 $\pm$ 0.01 & $\cdots$ & 1.30 & n & y & n & n & n & 29.31 $\pm$ 0.61 \\ 
SNR\_201 & 23:57:54.2 & -32:35:48.1 &  9.8 $\pm$  0.2 & $\cdots$ & -0.39 $\pm$ 0.02 & -0.33 $\pm$ 0.03 & $\cdots$ & 1.45 & y & y & n & n & n & 32.72 $\pm$ 0.74 \\ 
SNR\_202 & 23:57:54.2 & -32:35:16.2 &  6.3 $\pm$  0.1 & $\cdots$ & -0.51 $\pm$ 0.02 & -0.58 $\pm$ 0.02 & $\cdots$ & 1.36 & n & y & n & n & n & 25.95 $\pm$ 0.53 \\ 
SNR\_203 & 23:57:54.3 & -32:35:18.3 & 17.0 $\pm$  0.1 & $\cdots$ & -0.41 $\pm$ 0.01 & -0.66 $\pm$ 0.01 & $\cdots$ & 2.00 & n & y & n & n & n & 25.45 $\pm$ 0.54 \\ 
SNR\_204 & 23:57:54.3 & -32:35:24.9 & 16.2 $\pm$  0.1 & 6.13 $\pm$ 0.06 & -0.42 $\pm$ 0.01 & -0.56 $\pm$ 0.01 & -0.14 $\pm$ 0.01 & 1.39 & n & y & y & n & y & 23.37 $\pm$ 0.53 \\ 
SNR\_205 & 23:57:54.4 & -32:35:31.7 & 45.2 $\pm$  0.2 & $\cdots$ & -0.54 $\pm$ 0.01 & -0.60 $\pm$ 0.01 & $\cdots$ & 1.40 & n & y & n & n & n & 24.81 $\pm$ 0.46 \\ 
SNR\_206 & 23:57:54.5 & -32:35:12.3 & 18.4 $\pm$  0.1 & 6.48 $\pm$ 0.06 & 0.08 $\pm$ 0.01 & -0.32 $\pm$ 0.01 & 0.20 $\pm$ 0.01 & 1.52 & y & y & y & y & y & 47.66 $\pm$ 5.05 \\ 
SNR\_207 & 23:57:54.6 & -32:35:44.0 & 88.8 $\pm$  0.3 & $\cdots$ & -0.57 $\pm$ 0.00 & -0.65 $\pm$ 0.00 & $\cdots$ & 2.00 & n & y & n & n & n & 31.48 $\pm$ 0.64 \\ 
SNR\_208 & 23:57:54.6 & -32:35:14.9 &  4.6 $\pm$  0.1 & $\cdots$ & -0.57 $\pm$ 0.02 & -0.64 $\pm$ 0.02 & $\cdots$ & 1.11 & n & y & n & n & n & 28.02 $\pm$ 0.61 \\ 
SNR\_209 & 23:57:54.8 & -32:34:58.5 &  4.8 $\pm$  0.1 & $\cdots$ & -0.42 $\pm$ 0.03 & -0.72 $\pm$ 0.03 & $\cdots$ & 1.11 & n & y & n & n & n & 33.27 $\pm$ 0.94 \\ 
SNR\_210 & 23:57:54.8 & -32:35:03.6 & 19.4 $\pm$  0.3 & $\cdots$ & -0.48 $\pm$ 0.02 & -0.86 $\pm$ 0.02 & $\cdots$ & 1.52 & n & y & n & n & n & 31.56 $\pm$ 0.88 \\ 
SNR\_211 & 23:57:54.8 & -32:35:50.5 & 15.1 $\pm$  0.2 & $\cdots$ & -0.41 $\pm$ 0.02 & -0.55 $\pm$ 0.02 & $\cdots$ & 1.67 & n & y & n & n & n & 32.46 $\pm$ 0.81 \\ 
SNR\_212 & 23:57:54.9 & -32:35:12.0 & 13.2 $\pm$  0.1 & $\cdots$ & -0.44 $\pm$ 0.01 & -0.60 $\pm$ 0.01 & $\cdots$ & 1.43 & n & y & n & n & n & 28.14 $\pm$ 0.75 \\ 
SNR\_213 & 23:57:55.2 & -32:35:01.6 &  6.9 $\pm$  0.1 & 2.77 $\pm$ 0.07 & -0.34 $\pm$ 0.02 & -0.68 $\pm$ 0.02 & -0.14 $\pm$ 0.04 & 1.34 & y & y & y & n & y & 29.31 $\pm$ 0.82 \\ 
SNR\_214 & 23:57:55.3 & -32:35:04.4 &  3.7 $\pm$  0.1 & $\cdots$ & -0.33 $\pm$ 0.04 & -0.71 $\pm$ 0.05 & $\cdots$ & 1.30 & y & y & n & n & n & 27.35 $\pm$ 0.71 \\ 
SNR\_215 & 23:57:55.4 & -32:35:48.9 &  2.9 $\pm$  0.1 & 1.11 $\pm$ 0.06 & 0.01 $\pm$ 0.04 & -0.52 $\pm$ 0.05 & -0.09 $\pm$ 0.08 & 1.21 & y & y & y & n & y & 33.64 $\pm$ 1.02 \\ 
SNR\_216 & 23:57:55.5 & -32:35:51.5 &  7.5 $\pm$  0.2 & 2.59 $\pm$ 0.11 & -0.11 $\pm$ 0.03 & -0.40 $\pm$ 0.03 & 0.18 $\pm$ 0.06 & 1.45 & y & y & y & y & y & 37.25 $\pm$ 1.15 \\ 
SNR\_217 & 23:57:55.5 & -32:35:37.3 &  5.5 $\pm$  0.2 & $\cdots$ & -0.61 $\pm$ 0.06 & -0.68 $\pm$ 0.06 & $\cdots$ & 1.25 & n & y & n & n & n & 28.52 $\pm$ 0.65 \\ 
SNR\_218 & 23:57:55.5 & -32:35:44.6 & 69.0 $\pm$  0.3 & 24.12 $\pm$ 0.15 & -0.58 $\pm$ 0.00 & -0.66 $\pm$ 0.01 & 0.43 $\pm$ 0.01 & 2.13 & n & y & y & y & y & 27.50 $\pm$ 0.57 \\ 
SNR\_219 & 23:57:55.6 & -32:35:48.0 &  3.9 $\pm$  0.1 & $\cdots$ & -0.61 $\pm$ 0.04 & -0.66 $\pm$ 0.05 & $\cdots$ & 1.00 & n & y & n & n & n & 26.87 $\pm$ 0.71 \\ 
SNR\_220 & 23:57:55.7 & -32:35:30.7 &  2.0 $\pm$  0.1 & $\cdots$ & -0.49 $\pm$ 0.05 & -0.62 $\pm$ 0.06 & $\cdots$ & 1.00 & n & y & n & n & n & 28.70 $\pm$ 0.81 \\ 
SNR\_221 & 23:57:55.7 & -32:35:47.8 &  8.4 $\pm$  0.3 & $\cdots$ & -0.58 $\pm$ 0.05 & -0.74 $\pm$ 0.05 & $\cdots$ & 1.00 & n & y & n & n & n & 27.99 $\pm$ 0.57 \\ 
SNR\_222 & 23:57:55.8 & -32:35:23.5 & 28.0 $\pm$  0.2 & 10.18 $\pm$ 0.11 & -0.57 $\pm$ 0.01 & -0.63 $\pm$ 0.01 & 0.01 $\pm$ 0.01 & 1.84 & n & y & y & n & n & 25.49 $\pm$ 0.48 \\ 
SNR\_223 & 23:57:55.9 & -32:35:49.4 &  3.5 $\pm$  0.1 & $\cdots$ & -0.62 $\pm$ 0.03 & -0.73 $\pm$ 0.03 & $\cdots$ & 1.00 & n & y & n & n & n & 25.82 $\pm$ 0.61 \\ 
SNR\_224 & 23:57:55.9 & -32:35:30.4 &  5.0 $\pm$  0.2 & $\cdots$ & -0.52 $\pm$ 0.05 & -0.56 $\pm$ 0.06 & $\cdots$ & 1.00 & n & y & n & n & n & 25.19 $\pm$ 0.68 \\ 
SNR\_225 & 23:57:56.4 & -32:35:25.9 & 17.5 $\pm$  0.2 & $\cdots$ & -0.36 $\pm$ 0.01 & -0.65 $\pm$ 0.01 & $\cdots$ & 1.77 & y & y & n & n & n & 22.50 $\pm$ 0.52 \\ 
SNR\_226 & 23:57:57.1 & -32:34:56.0 &  5.0 $\pm$  0.2 & 2.08 $\pm$ 0.10 & 0.11 $\pm$ 0.04 & -0.27 $\pm$ 0.04 & 0.77 $\pm$ 0.06 & 1.75 & y & y & y & y & y & 48.09 $\pm$ 13.84 \\ 
SNR\_227 & 23:57:57.2 & -32:35:51.8 & 40.3 $\pm$  0.2 & $\cdots$ & -0.57 $\pm$ 0.01 & -0.67 $\pm$ 0.01 & $\cdots$ & 1.50 & n & y & n & n & n & 27.15 $\pm$ 0.43 \\ 
SNR\_228 & 23:57:57.5 & -32:35:39.2 & 44.6 $\pm$  0.2 & $\cdots$ & -0.64 $\pm$ 0.01 & -0.62 $\pm$ 0.01 & $\cdots$ & 1.76 & n & y & n & n & n & 27.81 $\pm$ 0.45 \\ 
SNR\_229 & 23:57:57.7 & -32:35:47.2 &  4.9 $\pm$  0.1 & 1.74 $\pm$ 0.06 & -0.33 $\pm$ 0.03 & -0.46 $\pm$ 0.03 & 0.64 $\pm$ 0.04 & 1.00 & y & y & y & y & y & 30.27 $\pm$ 0.56 \\ 
SNR\_230 & 23:57:57.8 & -32:35:45.8 &  3.8 $\pm$  0.1 & $\cdots$ & -0.46 $\pm$ 0.03 & -0.92 $\pm$ 0.03 & $\cdots$ & 0.86 & n & y & n & n & n & 28.98 $\pm$ 0.59 \\ 
SNR\_231 & 23:57:57.9 & -32:35:20.5 &  4.4 $\pm$  0.1 & $\cdots$ & -0.48 $\pm$ 0.02 & -0.65 $\pm$ 0.03 & $\cdots$ & 1.00 & n & y & n & n & n & 23.06 $\pm$ 0.58 \\ 
SNR\_232 & 23:57:58.1 & -32:35:41.5 & 59.1 $\pm$  0.3 & 20.79 $\pm$ 0.16 & -0.69 $\pm$ 0.01 & -0.75 $\pm$ 0.01 & 0.40 $\pm$ 0.01 & 2.01 & n & n & y & y & y & 25.70 $\pm$ 0.43 \\ 
SNR\_233 & 23:57:58.4 & -32:35:23.4 & 102.7 $\pm$  0.3 & 39.96 $\pm$ 0.20 & -0.33 $\pm$ 0.00 & -0.59 $\pm$ 0.00 & -0.27 $\pm$ 0.01 & 2.30 & y & y & y & n & y & 34.09 $\pm$ 0.72 \\ 
SNR\_234 & 23:57:58.5 & -32:35:42.9 & 134.3 $\pm$  0.5 & $\cdots$ & -0.58 $\pm$ 0.00 & -0.65 $\pm$ 0.01 & $\cdots$ & 2.05 & n & y & n & n & n & 26.53 $\pm$ 0.40 \\ 
SNR\_235 & 23:57:58.5 & -32:35:16.7 &  3.6 $\pm$  0.1 & 1.24 $\pm$ 0.08 & -0.39 $\pm$ 0.04 & -0.47 $\pm$ 0.05 & 0.38 $\pm$ 0.09 & 1.00 & y & y & y & y & y & 28.64 $\pm$ 0.91 \\ 
SNR\_236 & 23:57:58.6 & -32:35:27.6 & 22.6 $\pm$  0.2 & $\cdots$ & -0.56 $\pm$ 0.01 & -0.67 $\pm$ 0.01 & $\cdots$ & 1.57 & n & y & n & n & n & 23.79 $\pm$ 0.43 \\ 
SNR\_237 & 23:57:58.9 & -32:35:32.4 & 47.6 $\pm$  0.5 & $\cdots$ & -0.55 $\pm$ 0.01 & -0.74 $\pm$ 0.02 & $\cdots$ & 2.04 & n & y & n & n & n & 29.63 $\pm$ 0.58 \\ 
SNR\_238 & 23:57:59.0 & -32:35:18.8 &  7.2 $\pm$  0.1 & 2.65 $\pm$ 0.06 & -0.68 $\pm$ 0.02 & -0.89 $\pm$ 0.02 & 0.38 $\pm$ 0.03 & 1.00 & n & n & y & y & y & 37.89 $\pm$ 1.13 \\ 

\hline
\end{longtable}
\end{ThreePartTable}

\begin{figure*}
\minipage{1\textwidth}
  \includegraphics[width=0.97\linewidth]{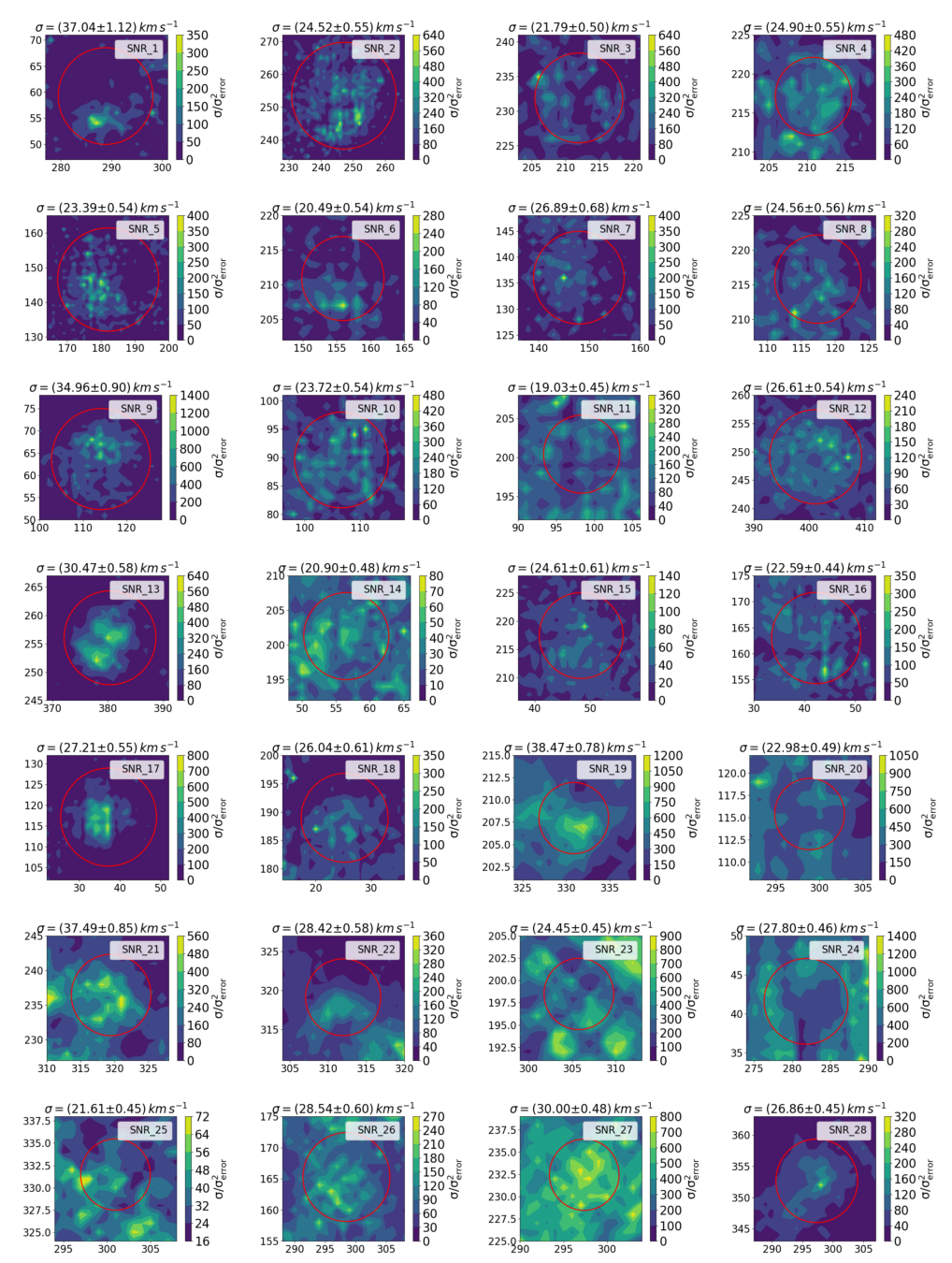}
  \caption{The $\rm \sigma/\sigma_{error}^2$ map of the Gaussian fitted to the \ha\ line of each pixel. The maps have been constructed in the regions of the SNR candidates. The red circles indicate the apertures used for the photometry for each SNR candidate. Above each map the velocity dispersion ($\rm \sigma$) of the integrated \ha\ flux of the candidate SNRs is presented.} \label{fig:VDs_all}
\endminipage
\end{figure*}

\begin{figure*}
\minipage{1\textwidth}
  \includegraphics[width=0.97\linewidth]{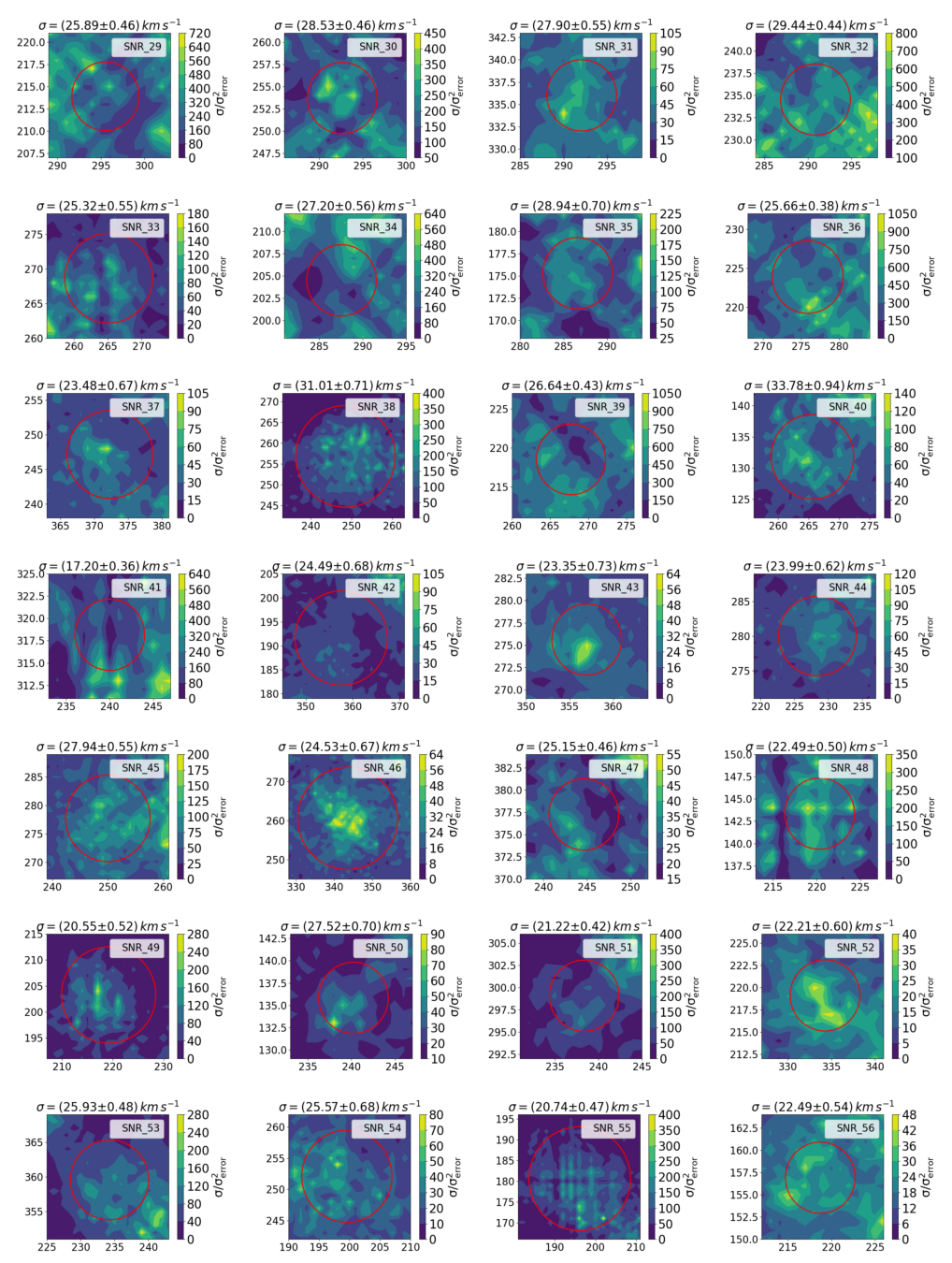}
  \caption*{The $\rm \sigma/\sigma_{error}^2$ map - {\it{continued}}}
\endminipage
\end{figure*}

\begin{figure*}
\minipage{1\textwidth}
  \includegraphics[width=0.97\linewidth]{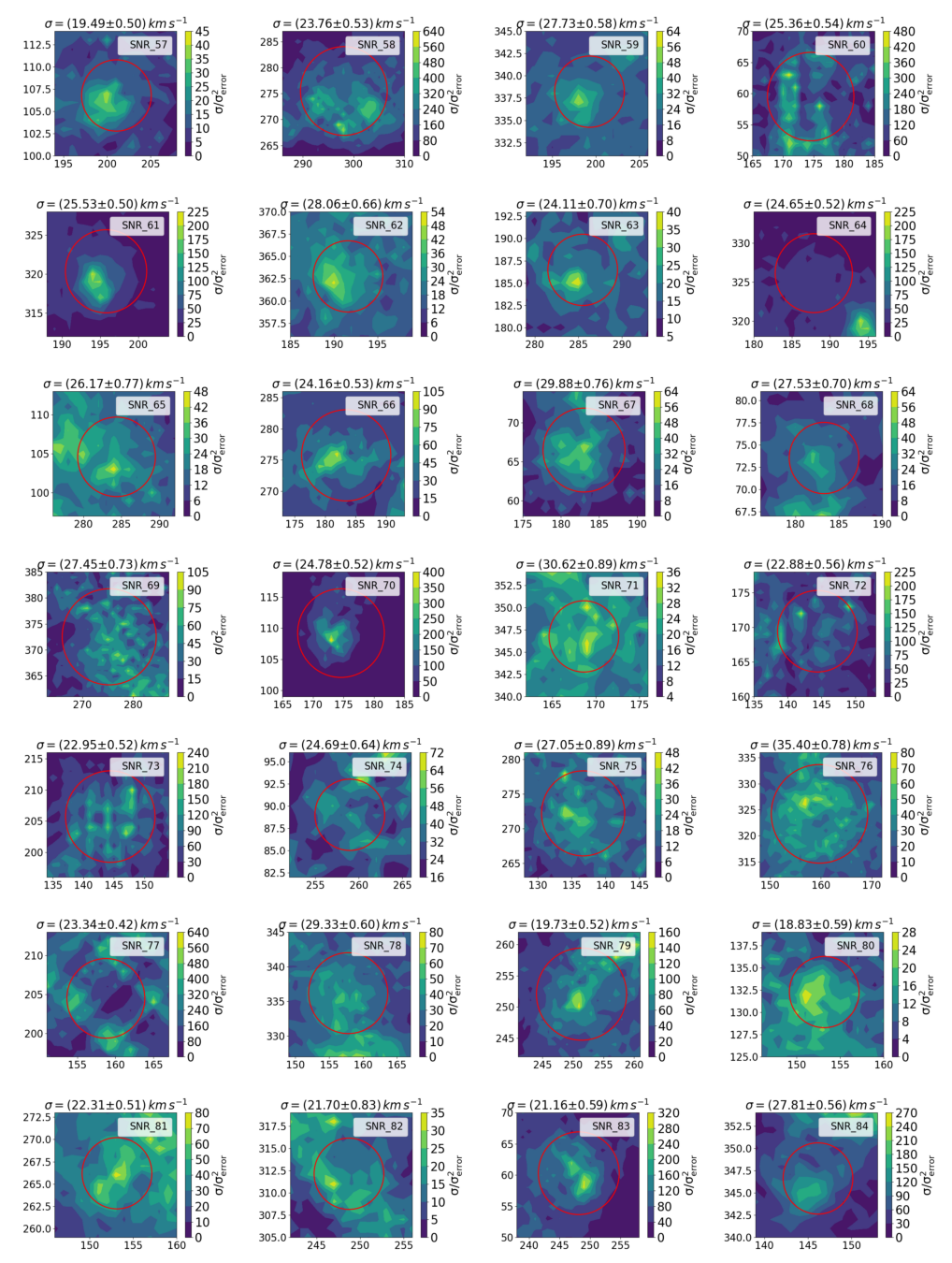}
  \caption*{The $\rm \sigma/\sigma_{error}^2$ map - {\it{continued}}}
\endminipage
\end{figure*}

\begin{figure*}
\minipage{1\textwidth}
  \includegraphics[width=0.97\linewidth]{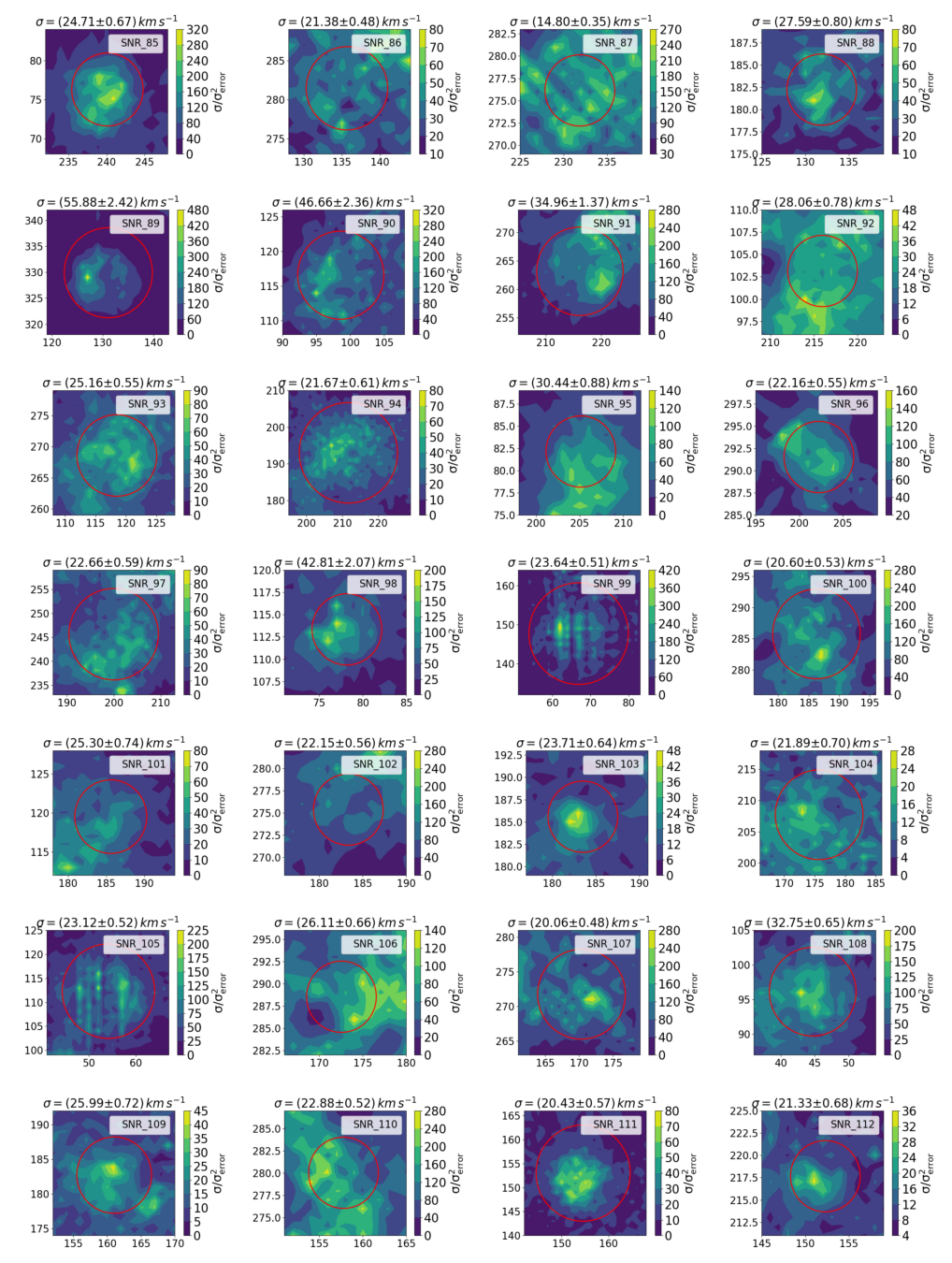}
  \caption*{The $\rm \sigma/\sigma_{error}^2$ map - {\it{continued}}}
\endminipage
\end{figure*}

\begin{figure*}
\minipage{1\textwidth}
  \includegraphics[width=0.97\linewidth]{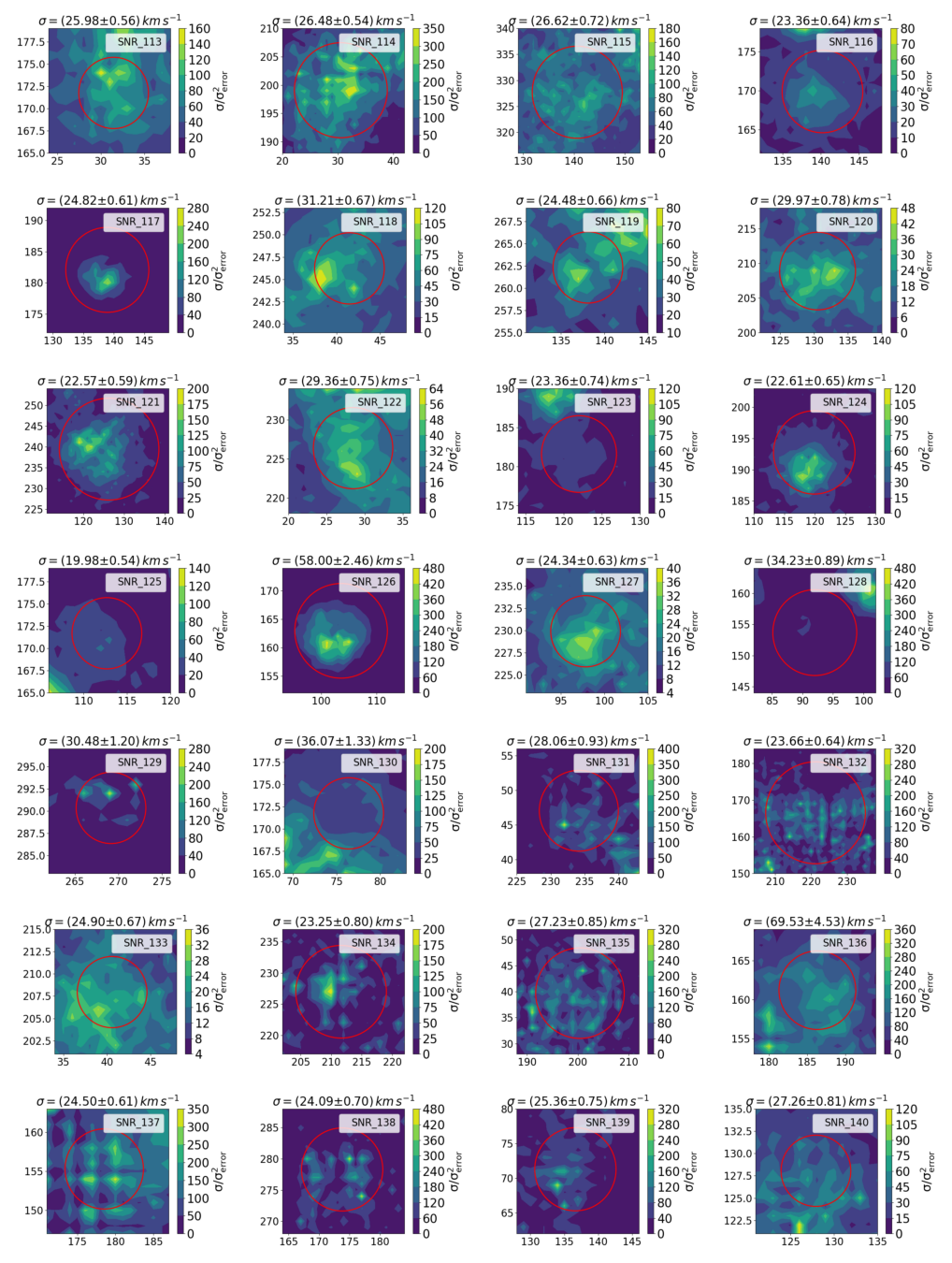}
  \caption*{The $\rm \sigma/\sigma_{error}^2$ map - {\it{continued}}}
\endminipage
\end{figure*}

\begin{figure*}
\minipage{1\textwidth}
  \includegraphics[width=0.97\linewidth]{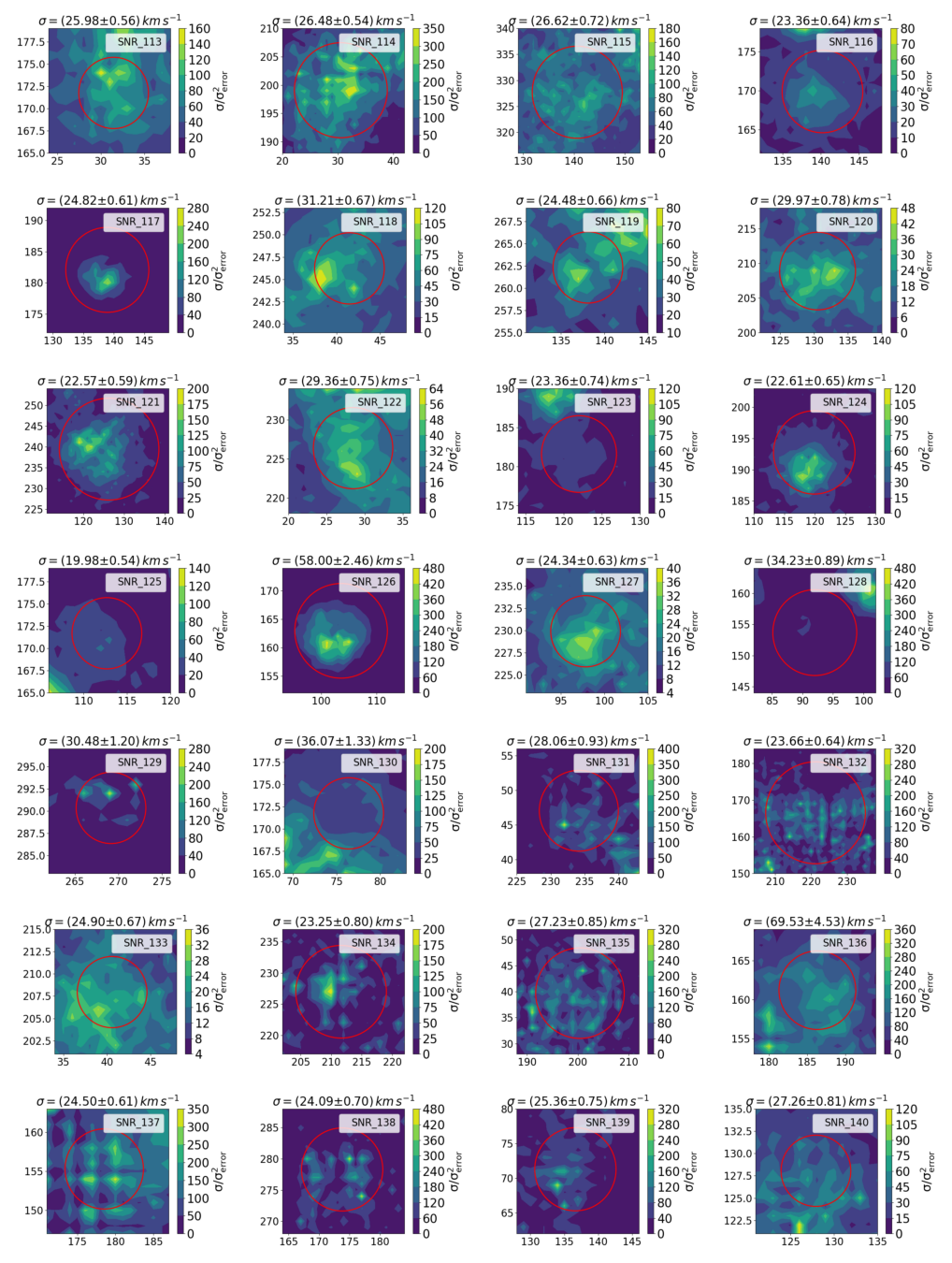}
  \caption*{The $\rm \sigma/\sigma_{error}^2$ map - {\it{continued}}}
\endminipage
\end{figure*}

\begin{figure*}
\minipage{1\textwidth}
  \includegraphics[width=0.97\linewidth]{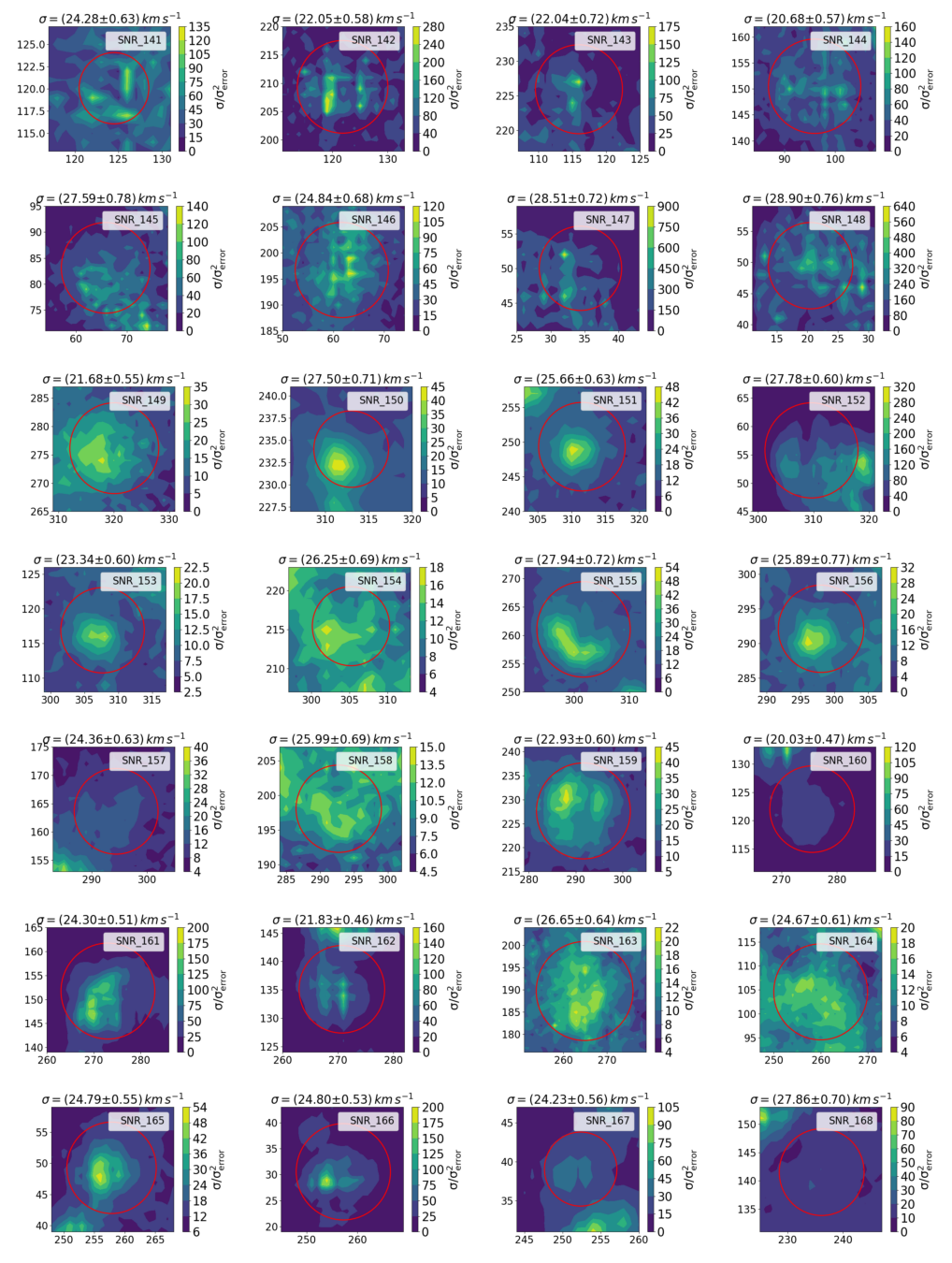}
  \caption*{The $\rm \sigma/\sigma_{error}^2$ map - {\it{continued}}}
\endminipage
\end{figure*}

\begin{figure*}
\minipage{1\textwidth}
  \includegraphics[width=0.97\linewidth]{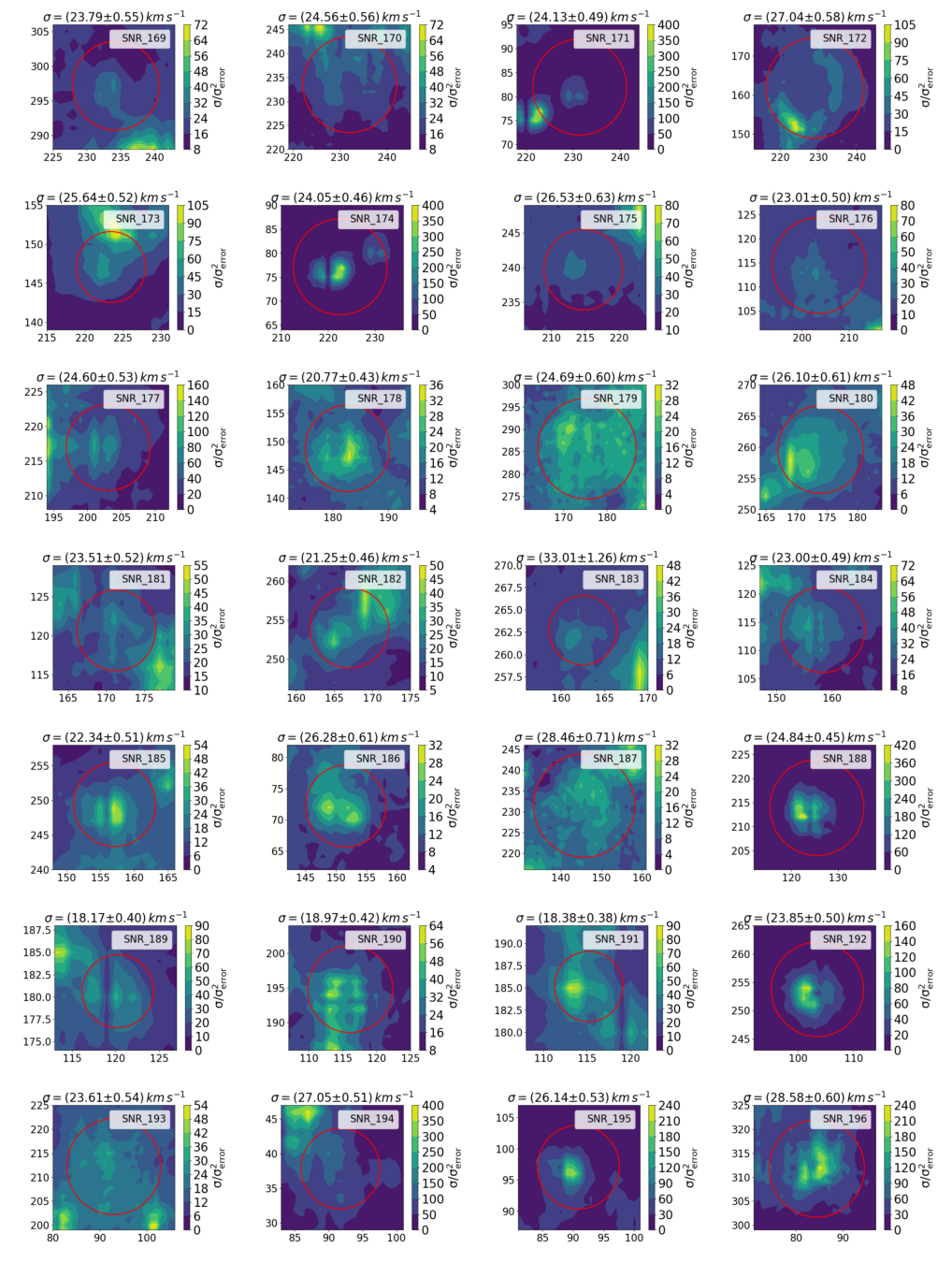}
  \caption*{The $\rm \sigma/\sigma_{error}^2$ map - {\it{continued}}}
\endminipage
\end{figure*}

\begin{figure*}
\minipage{1\textwidth}
  \includegraphics[width=0.97\linewidth]{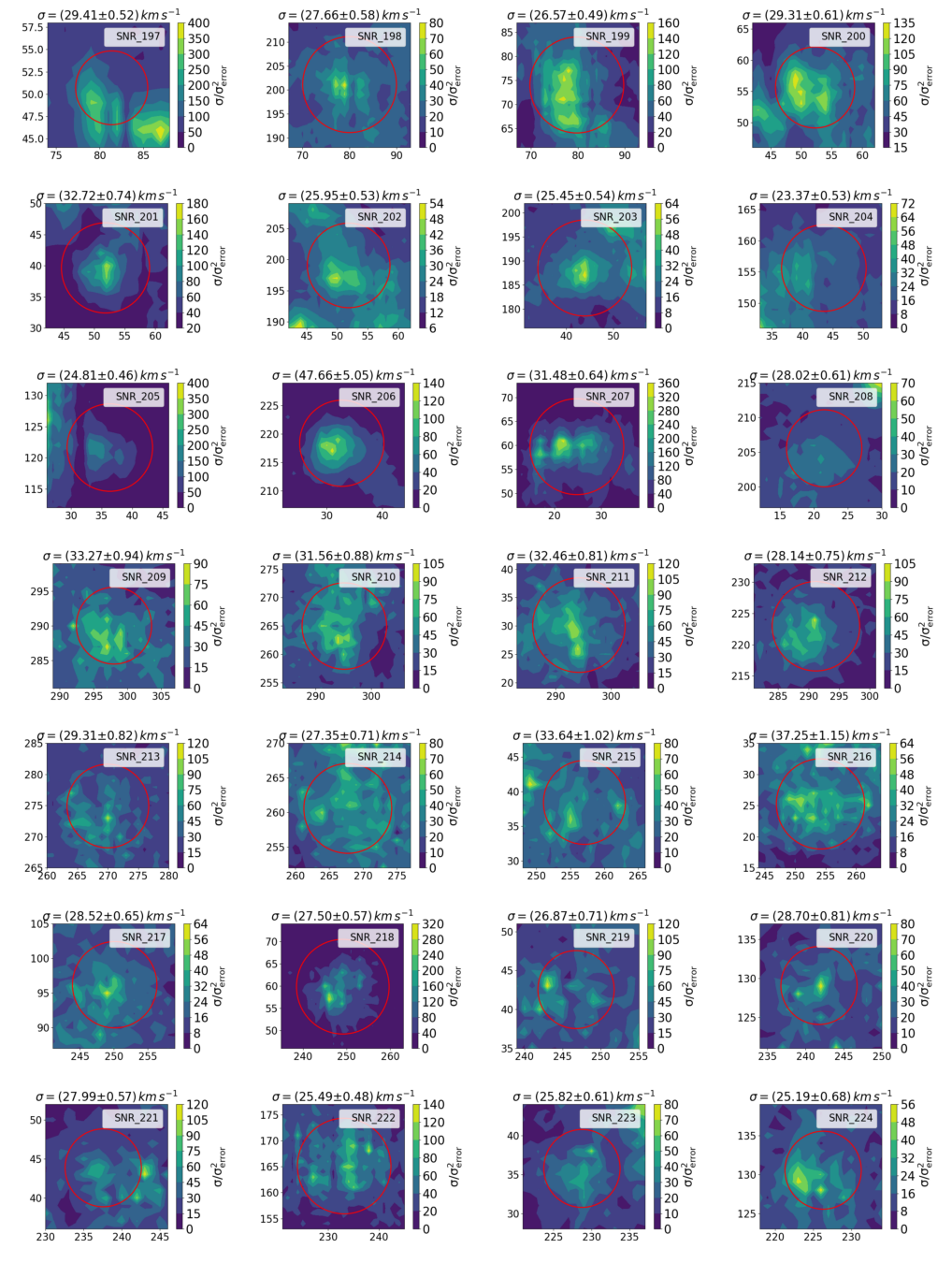}
  \caption*{The $\rm \sigma/\sigma_{error}^2$ map - {\it{continued}}}
\endminipage
\end{figure*}

\begin{figure*}
\minipage{1\textwidth}
  \includegraphics[width=0.97\linewidth]{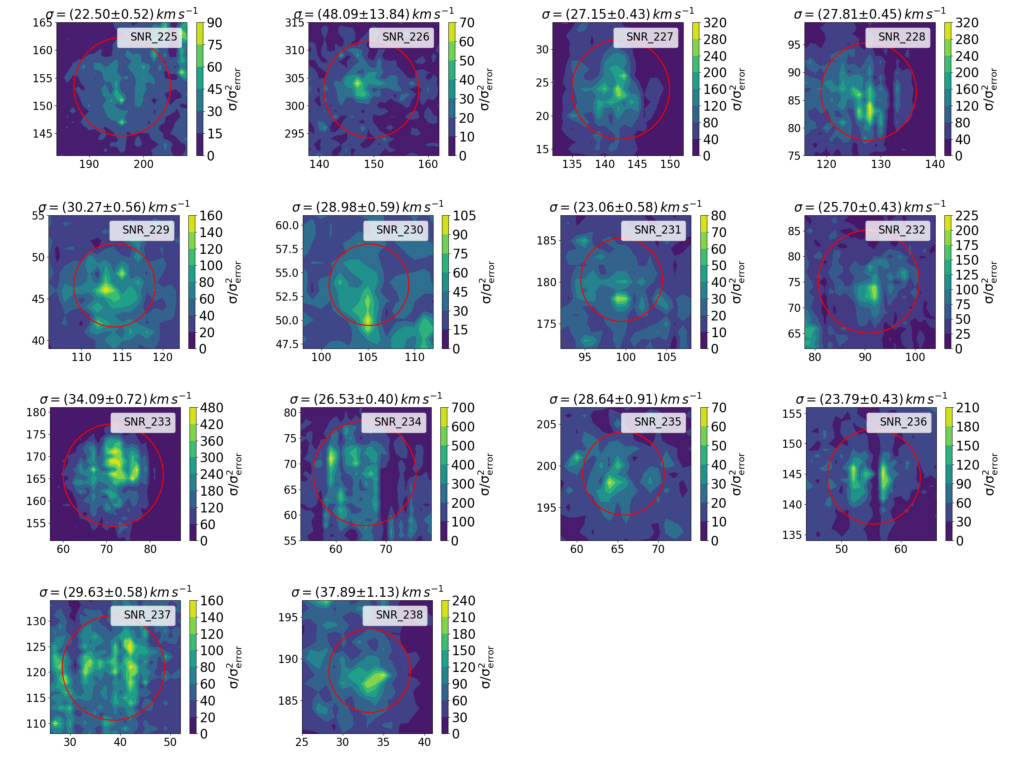}
  \caption*{The $\rm \sigma/\sigma_{error}^2$ map - {\it{continued}}}
\endminipage
\end{figure*}

\begin{figure*}
\minipage{1\textwidth}
  \includegraphics[width=0.97\linewidth]{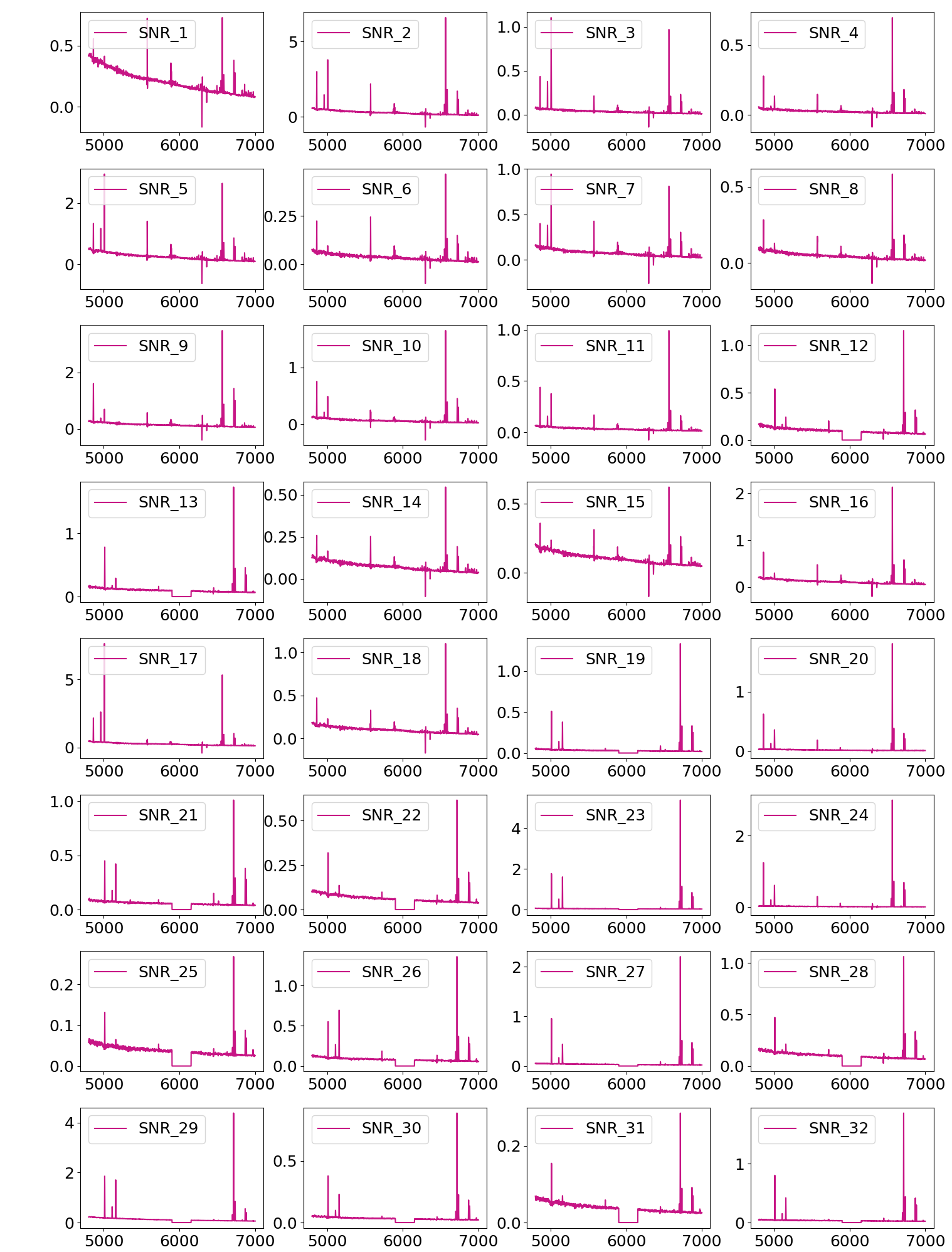}
  \caption{ Spectra of the SNR candidates.} 
  \label{fig:spectra}
\endminipage
\end{figure*}

\begin{figure*}
\minipage{1\textwidth}
  \includegraphics[width=0.97\linewidth]{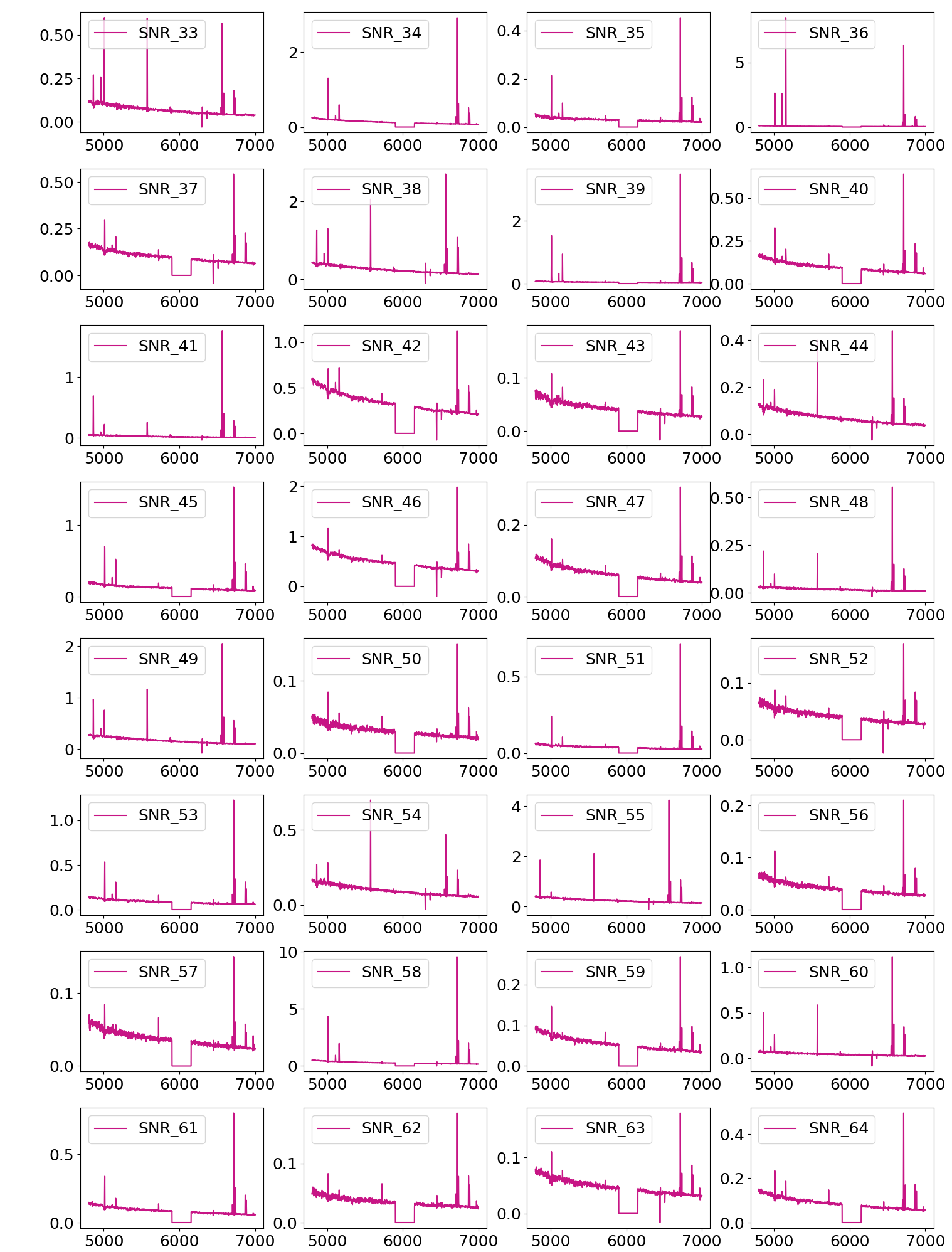}
  \caption*{Spectra - {\it{continued}}}
\endminipage
\end{figure*}

\begin{figure*}
\minipage{1\textwidth}
  \includegraphics[width=0.97\linewidth]{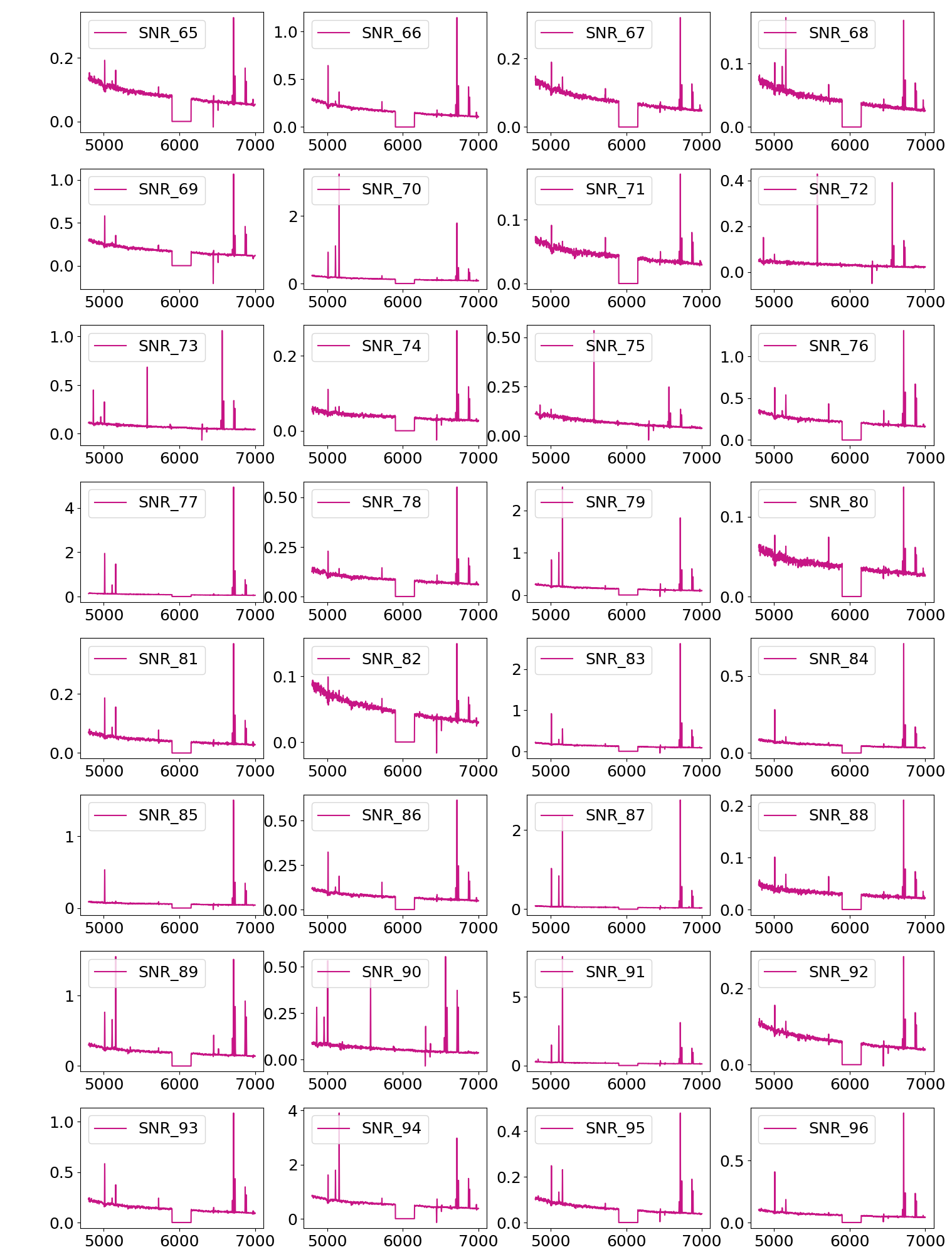}
  \caption*{Spectra - {\it{continued}}}
\endminipage
\end{figure*}

\begin{figure*}
\minipage{1\textwidth}
  \includegraphics[width=0.97\linewidth]{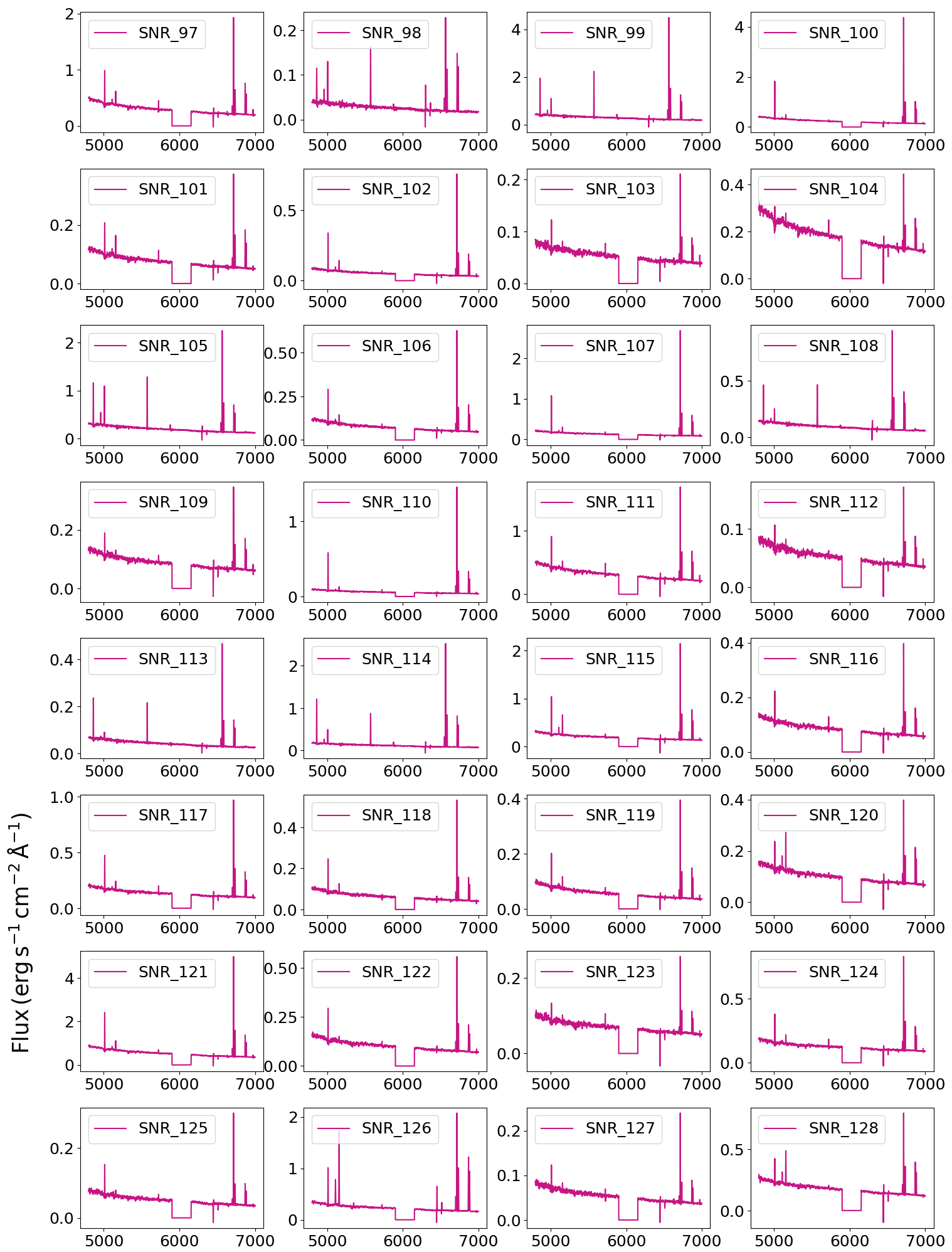}
  \caption*{Spectra - {\it{continued}}}
\endminipage
\end{figure*}

\begin{figure*}
\minipage{1\textwidth}
  \includegraphics[width=0.97\linewidth]{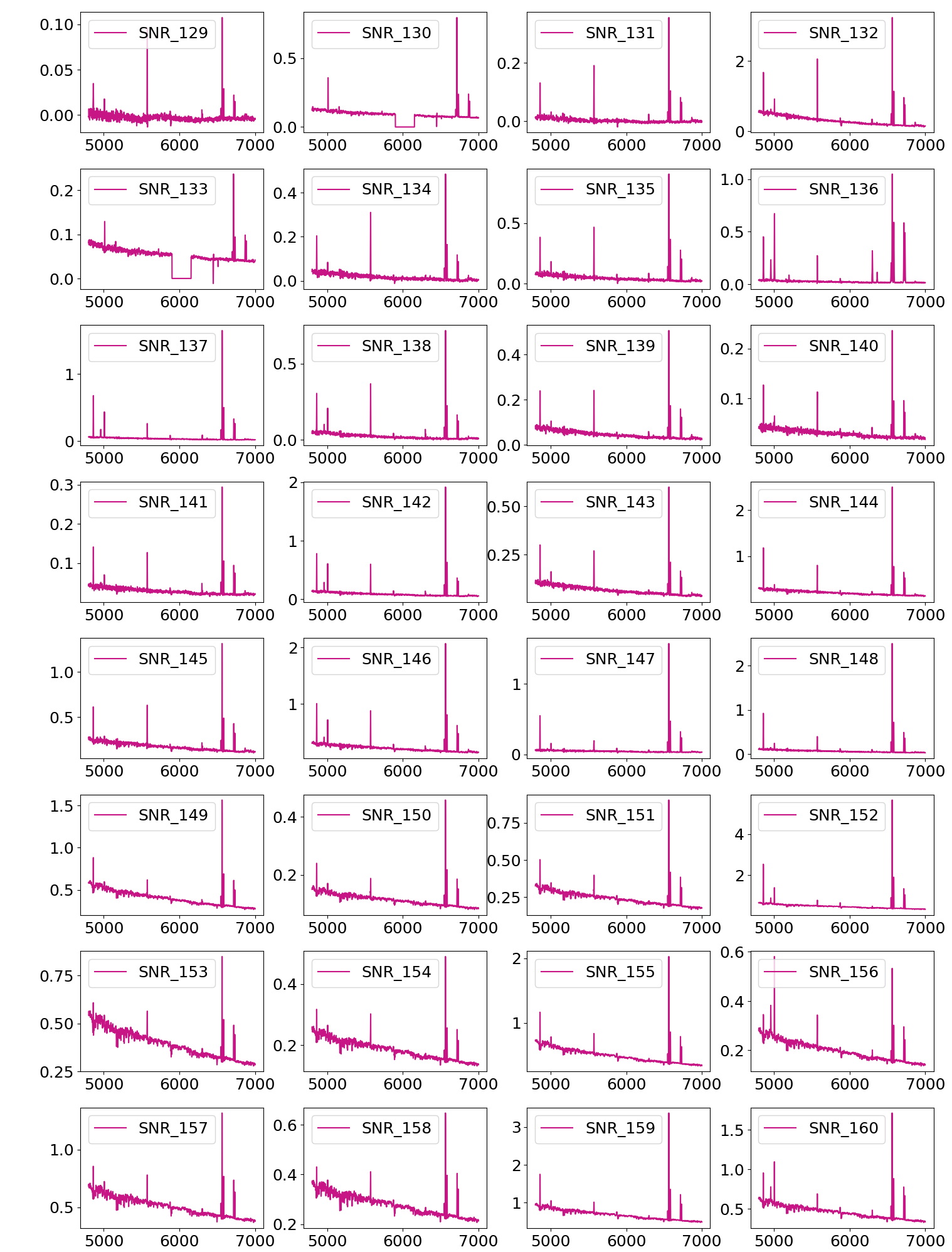}
  \caption*{Spectra - {\it{continued}}}
\endminipage
\end{figure*}

\begin{figure*}
\minipage{1\textwidth}
  \includegraphics[width=0.97\linewidth]{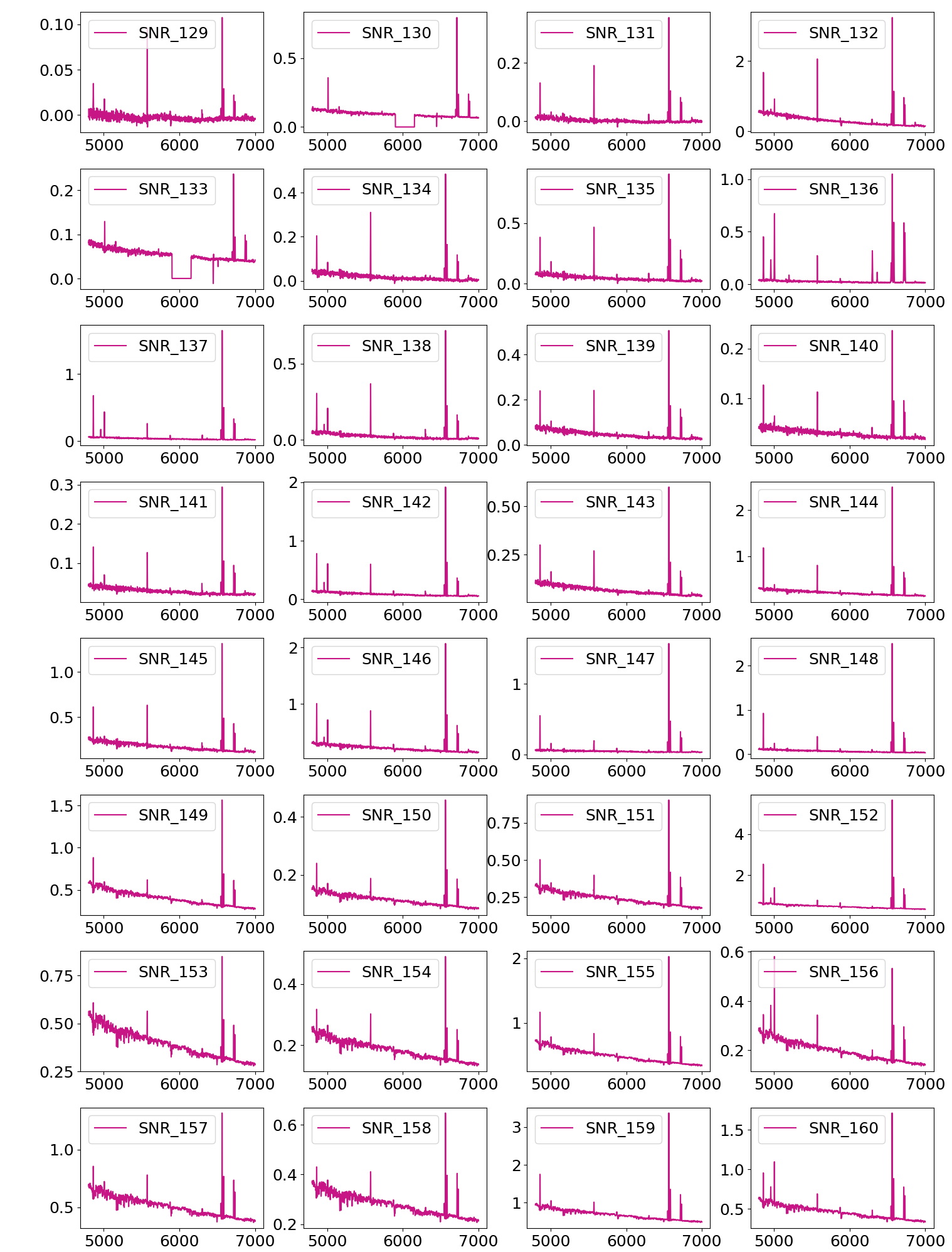}
  \caption*{Spectra - {\it{continued}}}
\endminipage
\end{figure*}

\begin{figure*}
\minipage{1\textwidth}
  \includegraphics[width=0.97\linewidth]{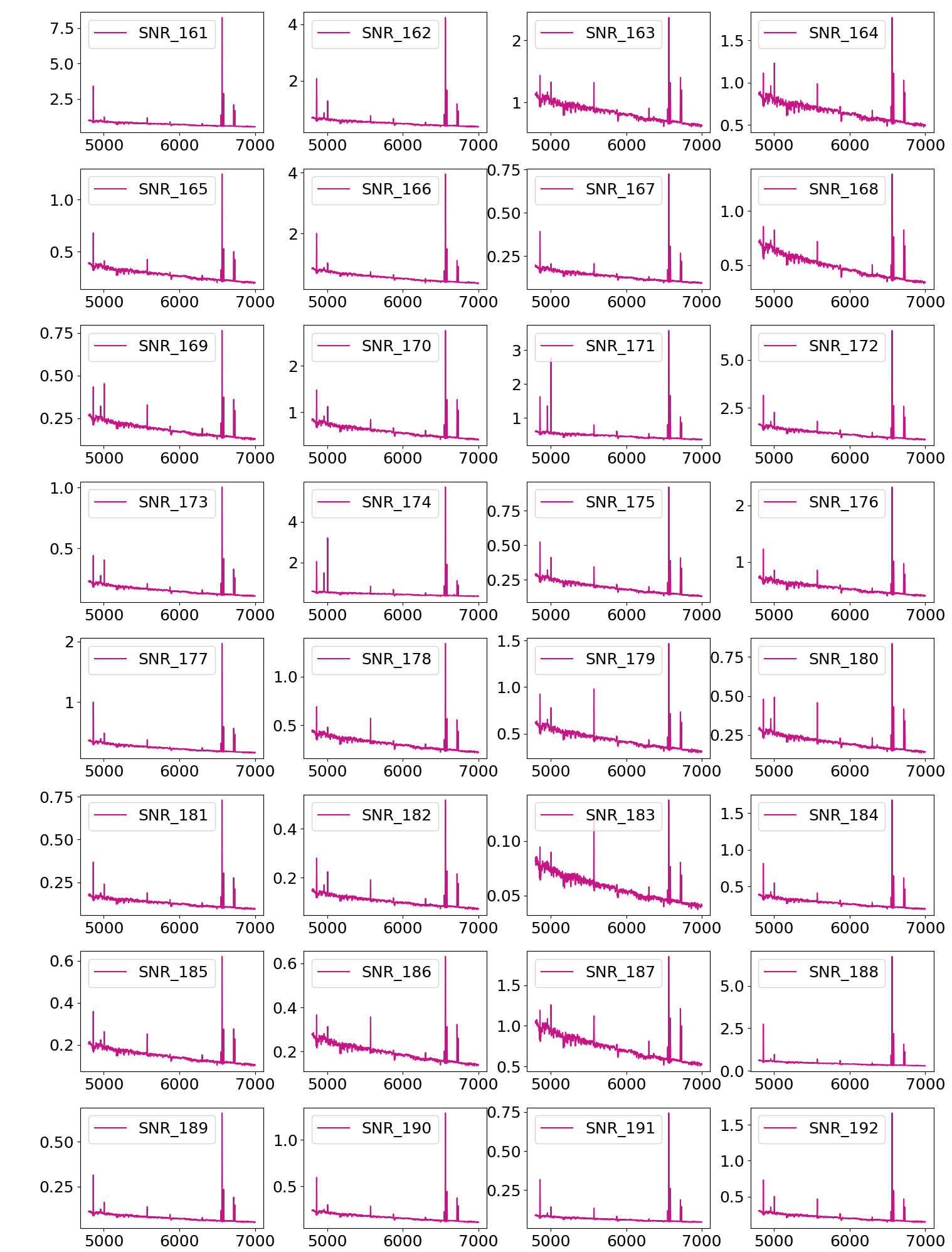}
  \caption*{Spectra - {\it{continued}}}
\endminipage
\end{figure*}

\begin{figure*}
\minipage{1\textwidth}
  \includegraphics[width=0.97\linewidth]{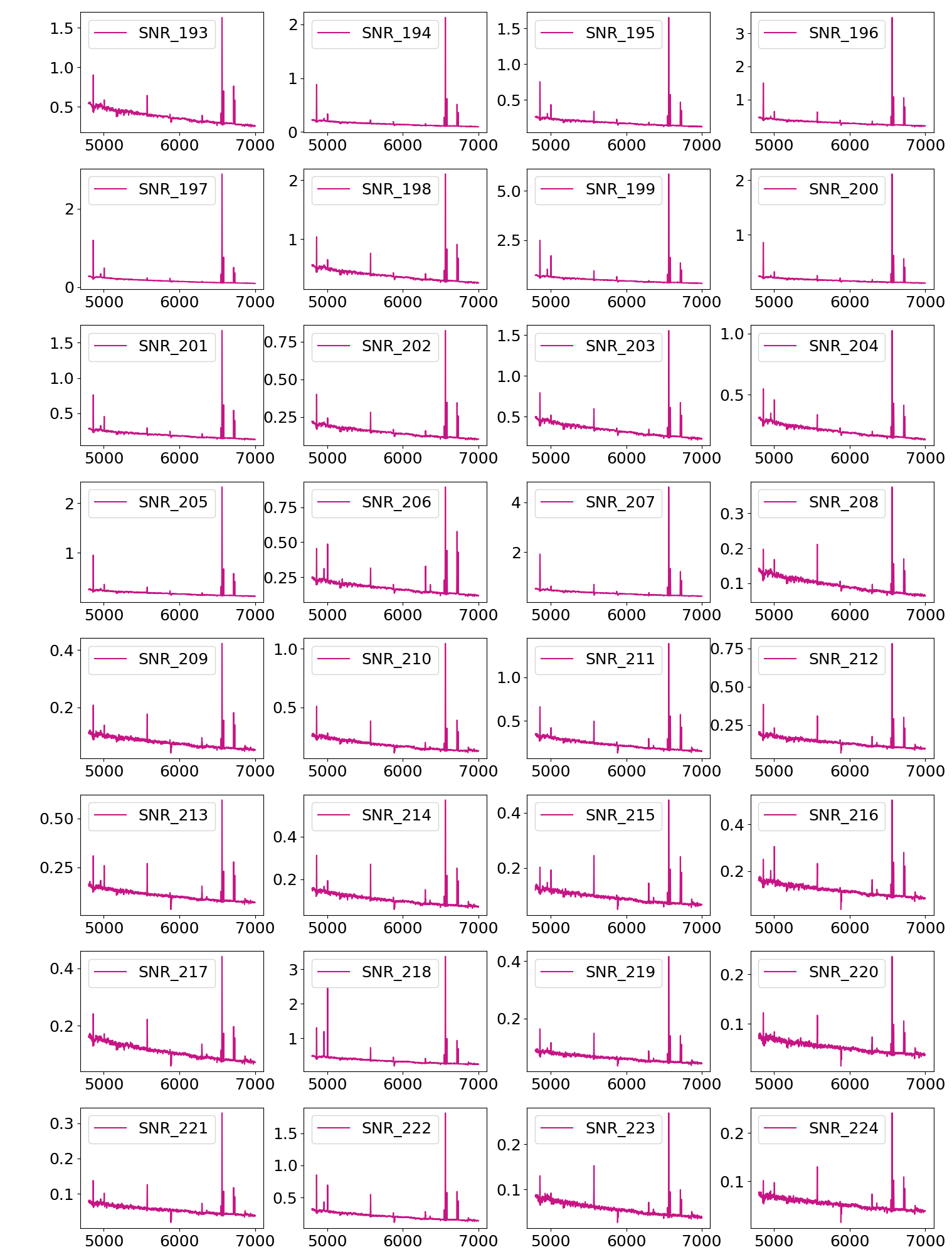}
  \caption*{Spectra - {\it{continued}}}
\endminipage
\end{figure*}

\begin{figure*}
\minipage{1\textwidth}
  \includegraphics[width=0.97\linewidth]{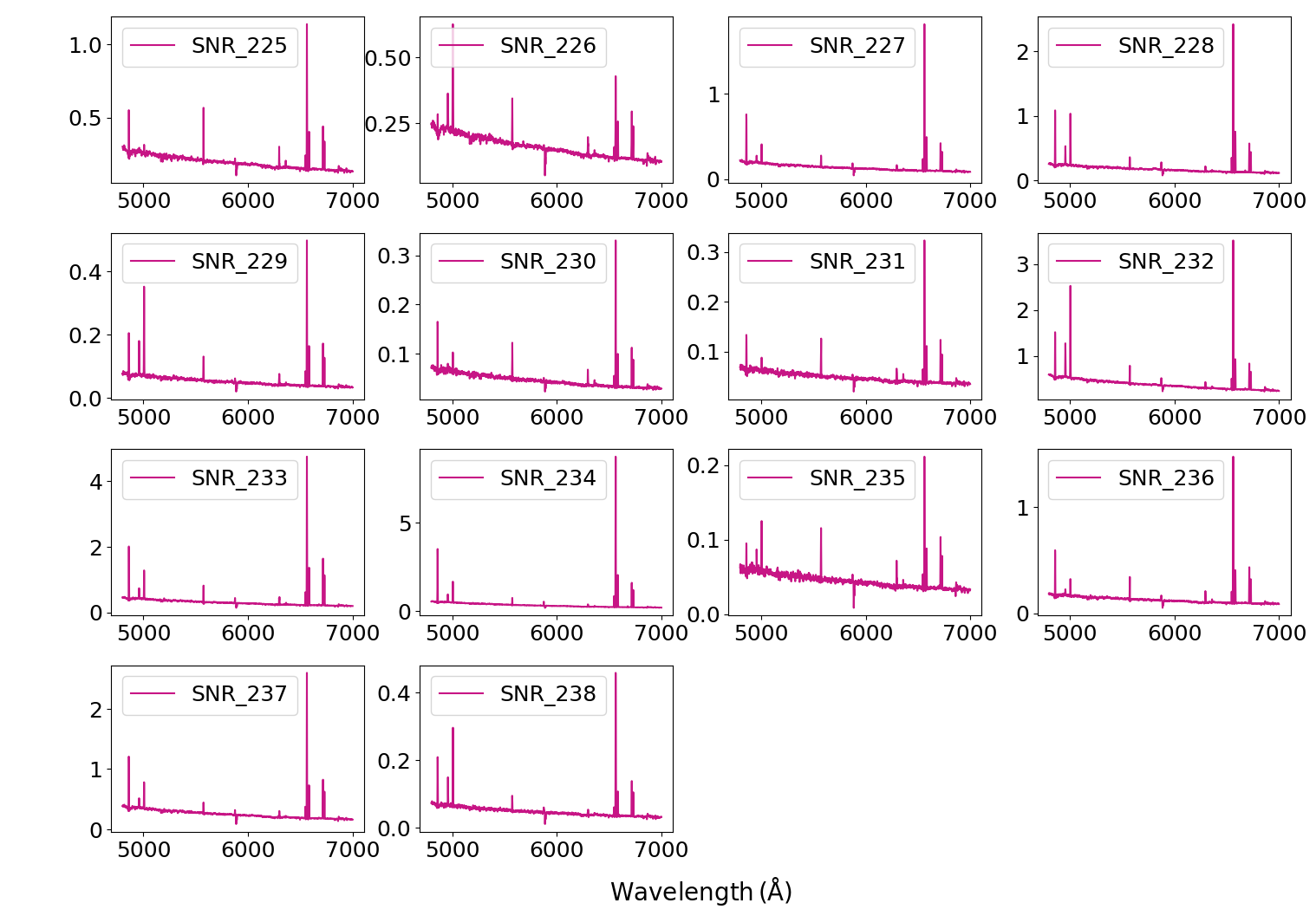}
  \caption*{Spectra - {\it{continued}}}
\endminipage
\end{figure*}


\bsp	
\label{lastpage}
\end{document}